\documentclass[longauth]{aa}

\usepackage{graphicx}
\usepackage{natbib}
\usepackage{scalerel}

\usepackage[table]{xcolor}
\usepackage{txfonts}
\usepackage{hyperref}
\hypersetup{
    colorlinks=true,
    linkcolor=blue,
    filecolor=magenta,      
    urlcolor=blue,
    citecolor=blue
}
\usepackage{tabularray}

\defcitealias{Q1-TP004}{MER25}
\defcitealias{DR1-TP009}{MER26}

\makeatletter
\renewcommand*\aa@pageof{, page \thepage{} of \pageref*{LastPage}}
\makeatother

\usepackage{subcaption}
\usepackage[utf8]{inputenc}
\usepackage{multirow}

\usepackage{euclid}
\usepackage{orcidlink}
\let\orcid\orcidlink

\DeclareSIUnit{\angstrom}{\mathring{A}}
\linenumbers
\renewcommand{\linenumbers}[0]{}

\begin{document} 
\nolinenumbers
\newcommand*{\JWST}{JWST\xspace}
\newcommand*{\sersic}{Sérsic\xspace}
\newcommand*{\zphot}{\texttt{zphot}\xspace}
\newcommand*{\gaia}{\textit{Gaia}\xspace}

\title{Euclid Data Release 1 (DR1)}
\subtitle{Multi-band photometry validation on COSMOS extragalactic sources}    

\author{Euclid Collaboration: T.~Gasparetto\orcid{0000-0002-7913-4866}\thanks{\email{thomas.gasparetto@inaf.it}}\inst{\ref{aff1}}
\and E.~Merlin\orcid{0000-0001-6870-8900}\inst{\ref{aff2}}
\and M.~Castellano\orcid{0000-0001-9875-8263}\inst{\ref{aff1}}
\and E.~Romelli\orcid{0000-0003-3069-9222}\inst{\ref{aff3}}
\and M.~D.~Lepinzan\orcid{0000-0003-1287-9801}\inst{\ref{aff3}}
\and M.~K\"ummel\orcid{0000-0003-2791-2117}\inst{\ref{aff4}}
\and J.~Gracia-Carpio\orcid{0000-0003-4689-3134}\inst{\ref{aff5}}
\and T.~Dumerchat\orcid{0009-0008-0196-2953}\inst{\ref{aff6}}
\and H.~Kong\orcid{0000-0001-8731-1212}\inst{\ref{aff6}}
\and P.~Dimauro\orcid{0000-0001-7399-2854}\inst{\ref{aff7},\ref{aff1}}
\and H.~Dole\orcid{0000-0002-9767-3839}\inst{\ref{aff8},\ref{aff9}}
\and F.~Caro\orcid{0009-0003-1053-0507}\inst{\ref{aff1}}
\and E.~Soubrie\orcid{0000-0001-9295-1863}\inst{\ref{aff8}}
\and M.~Salvato\orcid{0000-0001-7116-9303}\inst{\ref{aff5}}
\and P.~Santini\orcid{0000-0002-9334-8705}\inst{\ref{aff1}}
\and D.~Paris\orcid{0000-0002-7409-8114}\inst{\ref{aff1}}
\and N.~Aghanim\orcid{0000-0002-6688-8992}\inst{\ref{aff8}}
\and B.~Altieri\orcid{0000-0003-3936-0284}\inst{\ref{aff10}}
\and S.~Andreon\orcid{0000-0002-2041-8784}\inst{\ref{aff11}}
\and N.~Auricchio\orcid{0000-0003-4444-8651}\inst{\ref{aff12}}
\and H.~Aussel\orcid{0000-0002-1371-5705}\inst{\ref{aff13}}
\and C.~Baccigalupi\orcid{0000-0002-8211-1630}\inst{\ref{aff14},\ref{aff3},\ref{aff15},\ref{aff16}}
\and M.~Baldi\orcid{0000-0003-4145-1943}\inst{\ref{aff17},\ref{aff12},\ref{aff18}}
\and A.~Balestra\orcid{0000-0002-6967-261X}\inst{\ref{aff2}}
\and S.~Bardelli\orcid{0000-0002-8900-0298}\inst{\ref{aff12}}
\and P.~Battaglia\orcid{0000-0002-7337-5909}\inst{\ref{aff12}}
\and A.~Biviano\orcid{0000-0002-0857-0732}\inst{\ref{aff3},\ref{aff14}}
\and M.~Bolzonella\orcid{0000-0003-3278-4607}\inst{\ref{aff12}}
\and E.~Branchini\orcid{0000-0002-0808-6908}\inst{\ref{aff19},\ref{aff20},\ref{aff11}}
\and M.~Brescia\orcid{0000-0001-9506-5680}\inst{\ref{aff21},\ref{aff22}}
\and J.~Brinchmann\orcid{0000-0003-4359-8797}\inst{\ref{aff23},\ref{aff24},\ref{aff25}}
\and S.~Camera\orcid{0000-0003-3399-3574}\inst{\ref{aff26},\ref{aff27},\ref{aff28}}
\and G.~Ca\~nas-Herrera\orcid{0000-0003-2796-2149}\inst{\ref{aff29}}
\and V.~Capobianco\orcid{0000-0002-3309-7692}\inst{\ref{aff28}}
\and C.~Carbone\orcid{0000-0003-0125-3563}\inst{\ref{aff30}}
\and V.~F.~Cardone\inst{\ref{aff1},\ref{aff31}}
\and J.~Carretero\orcid{0000-0002-3130-0204}\inst{\ref{aff32},\ref{aff33}}
\and F.~J.~Castander\orcid{0000-0001-7316-4573}\inst{\ref{aff34},\ref{aff35}}
\and G.~Castignani\orcid{0000-0001-6831-0687}\inst{\ref{aff12}}
\and S.~Cavuoti\orcid{0000-0002-3787-4196}\inst{\ref{aff22},\ref{aff36}}
\and A.~Cimatti\inst{\ref{aff37}}
\and C.~Colodro-Conde\inst{\ref{aff38}}
\and G.~Congedo\orcid{0000-0003-2508-0046}\inst{\ref{aff39}}
\and C.~J.~Conselice\orcid{0000-0003-1949-7638}\inst{\ref{aff40}}
\and L.~Conversi\orcid{0000-0002-6710-8476}\inst{\ref{aff41},\ref{aff10}}
\and Y.~Copin\orcid{0000-0002-5317-7518}\inst{\ref{aff42}}
\and A.~Costille\inst{\ref{aff43}}
\and H.~M.~Courtois\orcid{0000-0003-0509-1776}\inst{\ref{aff44}}
\and M.~Cropper\orcid{0000-0003-4571-9468}\inst{\ref{aff45}}
\and A.~Da~Silva\orcid{0000-0002-6385-1609}\inst{\ref{aff46},\ref{aff47}}
\and R.~da~Silva\orcid{0000-0003-4788-677X}\inst{\ref{aff1},\ref{aff48}}
\and H.~Degaudenzi\orcid{0000-0002-5887-6799}\inst{\ref{aff49}}
\and G.~De~Lucia\orcid{0000-0002-6220-9104}\inst{\ref{aff3}}
\and F.~Dubath\orcid{0000-0002-6533-2810}\inst{\ref{aff49}}
\and X.~Dupac\inst{\ref{aff10}}
\and S.~Dusini\orcid{0000-0002-1128-0664}\inst{\ref{aff50}}
\and S.~Escoffier\orcid{0000-0002-2847-7498}\inst{\ref{aff51}}
\and M.~Fabricius\orcid{0000-0002-7025-6058}\inst{\ref{aff5},\ref{aff4}}
\and M.~Farina\orcid{0000-0002-3089-7846}\inst{\ref{aff52}}
\and S.~Farrens\orcid{0000-0002-9594-9387}\inst{\ref{aff13}}
\and F.~Faustini\orcid{0000-0001-6274-5145}\inst{\ref{aff1}}
\and S.~Ferriol\inst{\ref{aff42}}
\and S.~Fotopoulou\orcid{0000-0002-9686-254X}\inst{\ref{aff53}}
\and N.~Fourmanoit\orcid{0009-0005-6816-6925}\inst{\ref{aff51}}
\and M.~Frailis\orcid{0000-0002-7400-2135}\inst{\ref{aff3}}
\and E.~Franceschi\orcid{0000-0002-0585-6591}\inst{\ref{aff12}}
\and M.~Fumana\orcid{0000-0001-6787-5950}\inst{\ref{aff30}}
\and L.~Gabarra\orcid{0000-0002-8486-8856}\inst{\ref{aff54}}
\and S.~Galeotta\orcid{0000-0002-3748-5115}\inst{\ref{aff3}}
\and K.~George\orcid{0000-0002-1734-8455}\inst{\ref{aff55}}
\and W.~Gillard\orcid{0000-0003-4744-9748}\inst{\ref{aff51}}
\and C.~Giocoli\orcid{0000-0002-9590-7961}\inst{\ref{aff12},\ref{aff18}}
\and A.~Grazian\orcid{0000-0002-5688-0663}\inst{\ref{aff2}}
\and F.~Grupp\inst{\ref{aff5},\ref{aff4}}
\and L.~Guzzo\orcid{0000-0001-8264-5192}\inst{\ref{aff56},\ref{aff11},\ref{aff57}}
\and W.~G.~Hartley\orcid{0000-0002-4563-5101}\inst{\ref{aff49}}
\and S.~V.~H.~Haugan\orcid{0000-0001-9648-7260}\inst{\ref{aff58}}
\and S.~Hemmati\orcid{0000-0003-2226-5395}\inst{\ref{aff59}}
\and H.~Hoekstra\orcid{0000-0002-0641-3231}\inst{\ref{aff29}}
\and W.~Holmes\orcid{0009-0007-8554-4646}\inst{\ref{aff60}}
\and F.~Hormuth\inst{\ref{aff61}}
\and A.~Hornstrup\orcid{0000-0002-3363-0936}\inst{\ref{aff62},\ref{aff63}}
\and M.~Huertas-Company\orcid{0000-0002-1416-8483}\inst{\ref{aff38},\ref{aff64},\ref{aff65}}
\and K.~Jahnke\orcid{0000-0003-3804-2137}\inst{\ref{aff66}}
\and M.~Jhabvala\inst{\ref{aff67}}
\and S.~Kermiche\orcid{0000-0002-0302-5735}\inst{\ref{aff51}}
\and A.~Kiessling\orcid{0000-0002-2590-1273}\inst{\ref{aff60}}
\and B.~Kubik\orcid{0009-0006-5823-4880}\inst{\ref{aff42}}
\and M.~Kunz\orcid{0000-0002-3052-7394}\inst{\ref{aff68}}
\and H.~Kurki-Suonio\orcid{0000-0002-4618-3063}\inst{\ref{aff69},\ref{aff70}}
\and A.~M.~C.~Le~Brun\orcid{0000-0002-0936-4594}\inst{\ref{aff71}}
\and P.~Liebing\orcid{0009-0005-8043-7150}\inst{\ref{aff45}}
\and S.~Ligori\orcid{0000-0003-4172-4606}\inst{\ref{aff28}}
\and P.~B.~Lilje\orcid{0000-0003-4324-7794}\inst{\ref{aff58}}
\and V.~Lindholm\orcid{0000-0003-2317-5471}\inst{\ref{aff69},\ref{aff70}}
\and I.~Lloro\orcid{0000-0001-5966-1434}\inst{\ref{aff72}}
\and M.~Magliocchetti\orcid{0000-0001-9158-4838}\inst{\ref{aff52}}
\and G.~Mainetti\orcid{0000-0003-2384-2377}\inst{\ref{aff73}}
\and O.~Mansutti\orcid{0000-0001-5758-4658}\inst{\ref{aff3}}
\and S.~Marcin\inst{\ref{aff74}}
\and O.~Marggraf\orcid{0000-0001-7242-3852}\inst{\ref{aff75}}
\and M.~Martinelli\orcid{0000-0002-6943-7732}\inst{\ref{aff1},\ref{aff31}}
\and N.~Martinet\orcid{0000-0003-2786-7790}\inst{\ref{aff43}}
\and F.~Marulli\orcid{0000-0002-8850-0303}\inst{\ref{aff76},\ref{aff12},\ref{aff18}}
\and R.~J.~Massey\orcid{0000-0002-6085-3780}\inst{\ref{aff77}}
\and N.~Mauri\orcid{0000-0001-8196-1548}\inst{\ref{aff37},\ref{aff18}}
\and S.~Mei\orcid{0000-0002-2849-559X}\inst{\ref{aff78},\ref{aff79}}
\and M.~Meneghetti\orcid{0000-0003-1225-7084}\inst{\ref{aff12},\ref{aff18}}
\and P.~Monaco\orcid{0000-0003-2083-7564}\inst{\ref{aff80},\ref{aff3},\ref{aff15},\ref{aff14}}
\and A.~Mora\orcid{0000-0002-1922-8529}\inst{\ref{aff81}}
\and M.~Moresco\orcid{0000-0002-7616-7136}\inst{\ref{aff76},\ref{aff12}}
\and C.~Moretti\orcid{0000-0003-3314-8936}\inst{\ref{aff3},\ref{aff14},\ref{aff15}}
\and L.~Moscardini\orcid{0000-0002-3473-6716}\inst{\ref{aff76},\ref{aff12},\ref{aff18}}
\and R.~Nakajima\orcid{0009-0009-1213-7040}\inst{\ref{aff75}}
\and C.~Neissner\orcid{0000-0001-8524-4968}\inst{\ref{aff6},\ref{aff33}}
\and S.-M.~Niemi\orcid{0009-0005-0247-0086}\inst{\ref{aff82}}
\and C.~Padilla\orcid{0000-0001-7951-0166}\inst{\ref{aff6}}
\and S.~Paltani\orcid{0000-0002-8108-9179}\inst{\ref{aff49}}
\and F.~Pasian\orcid{0000-0002-4869-3227}\inst{\ref{aff3}}
\and W.~J.~Percival\orcid{0000-0002-0644-5727}\inst{\ref{aff83},\ref{aff84},\ref{aff85}}
\and V.~Pettorino\orcid{0000-0002-4203-9320}\inst{\ref{aff82}}
\and A.~Pezzotta\orcid{0000-0003-0726-2268}\inst{\ref{aff11}}
\and S.~Pires\orcid{0000-0002-0249-2104}\inst{\ref{aff13}}
\and G.~Polenta\orcid{0000-0003-4067-9196}\inst{\ref{aff48}}
\and M.~Poncet\inst{\ref{aff86}}
\and L.~A.~Popa\inst{\ref{aff87}}
\and L.~Pozzetti\orcid{0000-0001-7085-0412}\inst{\ref{aff12}}
\and F.~Raison\orcid{0000-0002-7819-6918}\inst{\ref{aff5}}
\and A.~Refregier\inst{\ref{aff88}}
\and A.~Renzi\orcid{0000-0001-9856-1970}\inst{\ref{aff89},\ref{aff50},\ref{aff12}}
\and J.~Rhodes\orcid{0000-0002-4485-8549}\inst{\ref{aff60}}
\and G.~Riccio\inst{\ref{aff22}}
\and I.~Risso\orcid{0000-0003-2525-7761}\inst{\ref{aff19},\ref{aff20},\ref{aff11}}
\and F.~Rizzo\orcid{0000-0002-9407-585X}\inst{\ref{aff3}}
\and C.~Rosset\orcid{0000-0003-0286-2192}\inst{\ref{aff78}}
\and B.~Rusholme\orcid{0000-0001-7648-4142}\inst{\ref{aff59}}
\and R.~Saglia\orcid{0000-0003-0378-7032}\inst{\ref{aff4},\ref{aff5}}
\and Z.~Sakr\orcid{0000-0002-4823-3757}\inst{\ref{aff90},\ref{aff91},\ref{aff92}}
\and D.~Sapone\orcid{0000-0001-7089-4503}\inst{\ref{aff93}}
\and B.~Sartoris\orcid{0000-0003-1337-5269}\inst{\ref{aff94}}
\and M.~Schirmer\orcid{0000-0003-2568-9994}\inst{\ref{aff66}}
\and P.~Schneider\orcid{0000-0001-8561-2679}\inst{\ref{aff75}}
\and T.~Schrabback\orcid{0000-0002-6987-7834}\inst{\ref{aff95}}
\and M.~Scodeggio\inst{\ref{aff30}}
\and A.~Secroun\orcid{0000-0003-0505-3710}\inst{\ref{aff51}}
\and E.~Sihvola\orcid{0000-0003-1804-7715}\inst{\ref{aff69},\ref{aff70}}
\and P.~Simon\inst{\ref{aff75}}
\and C.~Sirignano\orcid{0000-0002-0995-7146}\inst{\ref{aff89},\ref{aff50}}
\and G.~Sirri\orcid{0000-0003-2626-2853}\inst{\ref{aff18}}
\and L.~Stanco\orcid{0000-0002-9706-5104}\inst{\ref{aff50}}
\and P.~Tallada-Cresp\'{i}\orcid{0000-0002-1336-8328}\inst{\ref{aff32},\ref{aff33}}
\and D.~Tavagnacco\orcid{0000-0001-7475-9894}\inst{\ref{aff3}}
\and A.~N.~Taylor\inst{\ref{aff39}}
\and H.~I.~Teplitz\orcid{0000-0002-7064-5424}\inst{\ref{aff96}}
\and I.~Tereno\orcid{0000-0002-4537-6218}\inst{\ref{aff46},\ref{aff47}}
\and N.~Tessore\orcid{0000-0002-9696-7931}\inst{\ref{aff45}}
\and S.~Toft\orcid{0000-0003-3631-7176}\inst{\ref{aff97},\ref{aff98}}
\and R.~Toledo-Moreo\orcid{0000-0002-2997-4859}\inst{\ref{aff99},\ref{aff100}}
\and F.~Torradeflot\orcid{0000-0003-1160-1517}\inst{\ref{aff33},\ref{aff32}}
\and I.~Tutusaus\orcid{0000-0002-3199-0399}\inst{\ref{aff34},\ref{aff35},\ref{aff91}}
\and J.~Valiviita\orcid{0000-0001-6225-3693}\inst{\ref{aff69},\ref{aff70}}
\and T.~Vassallo\orcid{0000-0001-6512-6358}\inst{\ref{aff3},\ref{aff55}}
\and G.~Verdoes~Kleijn\orcid{0000-0001-5803-2580}\inst{\ref{aff101}}
\and A.~Veropalumbo\orcid{0000-0003-2387-1194}\inst{\ref{aff11},\ref{aff20},\ref{aff19}}
\and Y.~Wang\orcid{0000-0002-4749-2984}\inst{\ref{aff59}}
\and J.~Weller\orcid{0000-0002-8282-2010}\inst{\ref{aff4},\ref{aff5}}
\and A.~Zacchei\orcid{0000-0003-0396-1192}\inst{\ref{aff3},\ref{aff14}}
\and G.~Zamorani\orcid{0000-0002-2318-301X}\inst{\ref{aff12}}
\and F.~M.~Zerbi\orcid{0000-0002-9996-973X}\inst{\ref{aff11}}
\and E.~Zucca\orcid{0000-0002-5845-8132}\inst{\ref{aff12}}
\and V.~Allevato\orcid{0000-0001-7232-5152}\inst{\ref{aff22}}
\and I.~T.~Andika\orcid{0000-0001-6102-9526}\inst{\ref{aff4}}
\and M.~Ballardini\orcid{0000-0003-4481-3559}\inst{\ref{aff102},\ref{aff103},\ref{aff12}}
\and P.~Bergamini\orcid{0000-0003-1383-9414}\inst{\ref{aff12}}
\and S.~Borgani\orcid{0000-0001-6151-6439}\inst{\ref{aff80},\ref{aff14},\ref{aff3},\ref{aff15},\ref{aff104}}
\and E.~Bozzo\orcid{0000-0002-8201-1525}\inst{\ref{aff49}}
\and C.~Burigana\orcid{0000-0002-3005-5796}\inst{\ref{aff105},\ref{aff106}}
\and R.~Cabanac\orcid{0000-0001-6679-2600}\inst{\ref{aff91}}
\and M.~Calabrese\orcid{0000-0002-2637-2422}\inst{\ref{aff107},\ref{aff30}}
\and A.~Cappi\inst{\ref{aff94},\ref{aff12}}
\and T.~Castro\orcid{0000-0002-6292-3228}\inst{\ref{aff108},\ref{aff3},\ref{aff14}}
\and D.~Di~Ferdinando\inst{\ref{aff18}}
\and J.~A.~Escartin~Vigo\inst{\ref{aff5}}
\and J.~Garc\'ia-Bellido\orcid{0000-0002-9370-8360}\inst{\ref{aff90}}
\and V.~Gautard\inst{\ref{aff109}}
\and C.~M.~Gutierrez\orcid{0000-0001-7854-783X}\inst{\ref{aff38},\ref{aff110}}
\and A.~Hall\orcid{0000-0002-3139-8651}\inst{\ref{aff39}}
\and Y.~Kang\orcid{0009-0000-8588-7250}\inst{\ref{aff95}}
\and A.~Loureiro\orcid{0000-0002-4371-0876}\inst{\ref{aff111},\ref{aff112}}
\and J.~Macias-Perez\orcid{0000-0002-5385-2763}\inst{\ref{aff113}}
\and C.~Mancini\orcid{0000-0002-4297-0561}\inst{\ref{aff30}}
\and R.~Maoli\orcid{0000-0002-6065-3025}\inst{\ref{aff114},\ref{aff1}}
\and R.~B.~Metcalf\orcid{0000-0003-3167-2574}\inst{\ref{aff76},\ref{aff12}}
\and M.~P\"ontinen\orcid{0000-0001-5442-2530}\inst{\ref{aff69}}
\and V.~Scottez\orcid{0009-0008-3864-940X}\inst{\ref{aff115},\ref{aff116}}
\and M.~Sereno\orcid{0000-0003-0302-0325}\inst{\ref{aff12},\ref{aff18}}
\and K.~Tanidis\orcid{0000-0001-9843-5130}\inst{\ref{aff117}}
\and M.~Tenti\orcid{0000-0002-4254-5901}\inst{\ref{aff18}}
\and A.~Troja\orcid{0000-0003-0239-4595}\inst{\ref{aff3}}
\and M.~Tucci\inst{\ref{aff49}}
\and M.~Viel\orcid{0000-0002-2642-5707}\inst{\ref{aff14},\ref{aff3},\ref{aff16},\ref{aff15},\ref{aff104}}
\and M.~Wiesmann\orcid{0009-0000-8199-5860}\inst{\ref{aff58}}
\and Abdurro'uf\orcid{0000-0002-5258-8761}\inst{\ref{aff118}}
\and J.~A.~Acevedo~Barroso\orcid{0000-0002-9654-1711}\inst{\ref{aff60}}
\and Y.~Akrami\orcid{0000-0002-2407-7956}\inst{\ref{aff90},\ref{aff119}}
\and G.~Angora\orcid{0000-0002-0316-6562}\inst{\ref{aff22},\ref{aff102}}
\and S.~Anselmi\orcid{0000-0002-3579-9583}\inst{\ref{aff50},\ref{aff89},\ref{aff120}}
\and M.~Archidiacono\orcid{0000-0003-4952-9012}\inst{\ref{aff56},\ref{aff57}}
\and G.~Aric\`o\orcid{0000-0002-2802-2928}\inst{\ref{aff18}}
\and F.~Atrio-Barandela\orcid{0000-0002-2130-2513}\inst{\ref{aff121}}
\and M.~Aubert\orcid{0009-0002-7667-8814}\inst{\ref{aff42}}
\and M.~Baes\orcid{0000-0002-3930-2757}\inst{\ref{aff122}}
\and N.~Bartolo\inst{\ref{aff89},\ref{aff50},\ref{aff2}}
\and L.~Bazzanini\orcid{0000-0003-0727-0137}\inst{\ref{aff102},\ref{aff12}}
\and J.~R.~Bermejo-Climent\inst{\ref{aff123},\ref{aff124}}
\and D.~Bertacca\orcid{0000-0002-2490-7139}\inst{\ref{aff89},\ref{aff2},\ref{aff50}}
\and M.~Bethermin\orcid{0000-0002-3915-2015}\inst{\ref{aff125}}
\and F.~Beutler\orcid{0000-0003-0467-5438}\inst{\ref{aff39}}
\and A.~Blanchard\orcid{0000-0001-8555-9003}\inst{\ref{aff91}}
\and M.~Bonici\orcid{0000-0002-8430-126X}\inst{\ref{aff83},\ref{aff30}}
\and F.~Bouch\`e\orcid{0000-0002-4663-1786}\inst{\ref{aff126},\ref{aff36}}
\and M.~L.~Brown\orcid{0000-0002-0370-8077}\inst{\ref{aff40}}
\and S.~Bruton\orcid{0000-0002-6503-5218}\inst{\ref{aff127}}
\and A.~Calabro\orcid{0000-0003-2536-1614}\inst{\ref{aff1}}
\and B.~Camacho~Quevedo\orcid{0000-0002-8789-4232}\inst{\ref{aff14},\ref{aff16},\ref{aff3}}
\and C.~S.~Carvalho\inst{\ref{aff128}}
\and F.~Cogato\orcid{0000-0003-4632-6113}\inst{\ref{aff12}}
\and T.~E.~Collett\orcid{0000-0001-5564-3140}\inst{\ref{aff129}}
\and S.~Conseil\orcid{0000-0002-3657-4191}\inst{\ref{aff130},\ref{aff42}}
\and A.~R.~Cooray\orcid{0000-0002-3892-0190}\inst{\ref{aff131}}
\and P.~Corcho-Caballero\orcid{0000-0001-6327-7080}\inst{\ref{aff101}}
\and M.~Correnti\orcid{0000-0001-6464-3257}\inst{\ref{aff1},\ref{aff48}}
\and B.~Csizi\orcid{0000-0003-3227-6581}\inst{\ref{aff95}}
\and O.~Cucciati\orcid{0000-0002-9336-7551}\inst{\ref{aff12}}
\and T.~de~Boer\orcid{0000-0001-5486-2747}\inst{\ref{aff132}}
\and G.~Desprez\orcid{0000-0001-8325-1742}\inst{\ref{aff101}}
\and A.~D\'iaz-S\'anchez\orcid{0000-0003-0748-4768}\inst{\ref{aff133}}
\and S.~Di~Domizio\orcid{0000-0003-2863-5895}\inst{\ref{aff19},\ref{aff20}}
\and J.~M.~Diego\orcid{0000-0001-9065-3926}\inst{\ref{aff134}}
\and V.~Duret\orcid{0009-0009-0383-4960}\inst{\ref{aff135}}
\and A.~Enia\orcid{0000-0002-0200-2857}\inst{\ref{aff109},\ref{aff12}}
\and A.~Farina\orcid{0009-0000-3420-929X}\inst{\ref{aff11},\ref{aff20}}
\and A.~Finoguenov\orcid{0000-0002-4606-5403}\inst{\ref{aff69}}
\and A.~Franco\orcid{0000-0002-4761-366X}\inst{\ref{aff136},\ref{aff137},\ref{aff138}}
\and A.~Fumagalli\orcid{0009-0004-0300-2535}\inst{\ref{aff3}}
\and Y.~Fu\orcid{0000-0002-0759-0504}\inst{\ref{aff29},\ref{aff101}}
\and K.~Ganga\orcid{0000-0001-8159-8208}\inst{\ref{aff78}}
\and R.~Gavazzi\orcid{0000-0002-5540-6935}\inst{\ref{aff43},\ref{aff139}}
\and E.~Gaztanaga\orcid{0000-0001-9632-0815}\inst{\ref{aff34},\ref{aff35},\ref{aff129}}
\and Z.~Ghaffari\orcid{0000-0002-6467-8078}\inst{\ref{aff3},\ref{aff14}}
\and F.~Giacomini\orcid{0000-0002-3129-2814}\inst{\ref{aff18}}
\and F.~Gianotti\orcid{0000-0003-4666-119X}\inst{\ref{aff12}}
\and G.~Gozaliasl\orcid{0000-0002-0236-919X}\inst{\ref{aff140},\ref{aff69}}
\and A.~Gruppuso\orcid{0000-0001-9272-5292}\inst{\ref{aff12},\ref{aff18}}
\and M.~Guidi\orcid{0000-0001-9408-1101}\inst{\ref{aff17},\ref{aff12}}
\and C.~Hern\'andez-Monteagudo\orcid{0000-0001-5471-9166}\inst{\ref{aff110},\ref{aff38}}
\and J.~Hjorth\orcid{0000-0002-4571-2306}\inst{\ref{aff141}}
\and M.~Jauzac\orcid{0000-0003-1974-8732}\inst{\ref{aff142},\ref{aff77},\ref{aff143},\ref{aff144},\ref{aff91}}
\and J.~J.~E.~Kajava\orcid{0000-0002-3010-8333}\inst{\ref{aff145},\ref{aff146},\ref{aff147}}
\and J.~Kim\orcid{0000-0003-2776-2761}\inst{\ref{aff54}}
\and C.~C.~Kirkpatrick\inst{\ref{aff148}}
\and A.~Kov\'acs\orcid{0000-0002-5825-579X}\inst{\ref{aff124},\ref{aff123}}
\and I.~Kova{\v{c}}i{\'{c}}\orcid{0000-0001-6751-3263}\inst{\ref{aff122}}
\and K.~Koyama\orcid{0000-0001-6727-6915}\inst{\ref{aff129}}
\and S.~Kruk\orcid{0000-0001-8010-8879}\inst{\ref{aff10}}
\and M.~C.~Lam\orcid{0000-0002-9347-2298}\inst{\ref{aff39}}
\and M.~Lattanzi\orcid{0000-0003-1059-2532}\inst{\ref{aff103}}
\and S.~Lee\orcid{0000-0002-8289-740X}\inst{\ref{aff149},\ref{aff60}}
\and G.~F.~Lesci\orcid{0000-0002-4607-2830}\inst{\ref{aff105}}
\and J.~Lesgourgues\orcid{0000-0001-7627-353X}\inst{\ref{aff150}}
\and T.~I.~Liaudat\orcid{0000-0002-9104-314X}\inst{\ref{aff151}}
\and L.~Linke\orcid{0000-0002-2622-8113}\inst{\ref{aff95}}
\and S.~J.~Liu\orcid{0000-0001-7680-2139}\inst{\ref{aff52}}
\and F.~Mannucci\orcid{0000-0002-4803-2381}\inst{\ref{aff152}}
\and C.~J.~A.~P.~Martins\orcid{0000-0002-4886-9261}\inst{\ref{aff153},\ref{aff24}}
\and L.~Maurin\orcid{0000-0002-8406-0857}\inst{\ref{aff8}}
\and M.~Migliaccio\inst{\ref{aff154},\ref{aff155}}
\and M.~Miluzio\inst{\ref{aff10},\ref{aff156}}
\and G.~Morgante\inst{\ref{aff12}}
\and S.~Nadathur\orcid{0000-0001-9070-3102}\inst{\ref{aff129}}
\and K.~Naidoo\orcid{0000-0002-9182-1802}\inst{\ref{aff129},\ref{aff66}}
\and A.~Navarro-Alsina\orcid{0000-0002-3173-2592}\inst{\ref{aff75}}
\and A.~Nersesian\orcid{0000-0001-6843-409X}\inst{\ref{aff157},\ref{aff122}}
\and S.~Nesseris\orcid{0000-0002-0567-0324}\inst{\ref{aff90}}
\and F.~Oppizzi\orcid{0000-0003-3904-8370}\inst{\ref{aff20},\ref{aff50},\ref{aff89}}
\and F.~Pace\orcid{0000-0001-8039-0480}\inst{\ref{aff26},\ref{aff27},\ref{aff28},\ref{aff47}}
\and M.~J.~Page\orcid{0000-0002-6689-6271}\inst{\ref{aff45}}
\and D.~Paoletti\orcid{0000-0003-4761-6147}\inst{\ref{aff12},\ref{aff106}}
\and G.~Papini\orcid{0000-0002-5810-318X}\inst{\ref{aff12},\ref{aff76}}
\and F.~Passalacqua\orcid{0000-0002-8606-4093}\inst{\ref{aff50}}
\and K.~Paterson\orcid{0000-0001-8340-3486}\inst{\ref{aff66}}
\and C.~Pattison\orcid{0000-0003-3272-2617}\inst{\ref{aff129}}
\and A.~Pisani\orcid{0000-0002-6146-4437}\inst{\ref{aff51}}
\and L.~Pizzuti\orcid{0000-0001-5654-7580}\inst{\ref{aff158}}
\and J.~J.~Popp\orcid{0000-0002-3724-1727}\inst{\ref{aff159}}
\and D.~Potter\orcid{0000-0002-0757-5195}\inst{\ref{aff160}}
\and M.~Poulain\orcid{0000-0002-7664-4510}\inst{\ref{aff161}}
\and G.~W.~Pratt\inst{\ref{aff13}}
\and S.~Quai\orcid{0000-0002-0449-8163}\inst{\ref{aff76},\ref{aff12}}
\and M.~Radovich\orcid{0000-0002-3585-866X}\inst{\ref{aff2}}
\and F.~Ricci\orcid{0000-0001-5742-5980}\inst{\ref{aff162},\ref{aff1}}
\and G.~Rodighiero\orcid{0000-0002-9415-2296}\inst{\ref{aff89},\ref{aff2}}
\and W.~Roster\orcid{0000-0002-9149-6528}\inst{\ref{aff5}}
\and M.~Sahl\'en\orcid{0000-0003-0973-4804}\inst{\ref{aff163}}
\and D.~B.~Sanders\orcid{0000-0002-1233-9998}\inst{\ref{aff132}}
\and A.~Schneider\orcid{0000-0001-7055-8104}\inst{\ref{aff160}}
\and D.~Sciotti\orcid{0009-0008-4519-2620}\inst{\ref{aff1},\ref{aff31}}
\and E.~Sellentin\orcid{0009-0002-2655-3458}\inst{\ref{aff164},\ref{aff29}}
\and S.~Serjeant\orcid{0000-0002-0517-7943}\inst{\ref{aff159}}
\and L.~C.~Smith\orcid{0000-0002-3259-2771}\inst{\ref{aff165}}
\and J.~G.~Sorce\orcid{0000-0002-2307-2432}\inst{\ref{aff166},\ref{aff8}}
\and I.~Szapudi\orcid{0000-0003-2274-0301}\inst{\ref{aff132}}
\and M.~Talia\orcid{0000-0003-4352-2063}\inst{\ref{aff76},\ref{aff12}}
\and G.~Testera\orcid{0000-0003-2970-766X}\inst{\ref{aff20}}
\and R.~Teyssier\orcid{0000-0001-7689-0933}\inst{\ref{aff167}}
\and S.~Tosi\orcid{0000-0002-7275-9193}\inst{\ref{aff19},\ref{aff11},\ref{aff20}}
\and C.~Uhlemann\orcid{0000-0001-7831-1579}\inst{\ref{aff168}}
\and C.~Valieri\inst{\ref{aff18}}
\and A.~Venhola\orcid{0000-0001-6071-4564}\inst{\ref{aff161}}
\and D.~Vergani\orcid{0000-0003-0898-2216}\inst{\ref{aff12}}
\and P.~Vielzeuf\orcid{0000-0003-2035-9339}\inst{\ref{aff51}}
\and E.~Vilenius\orcid{0000-0002-6184-7681}\inst{\ref{aff45}}
\and S.~Vinciguerra\orcid{0009-0005-4018-3184}\inst{\ref{aff43}}
\and M.~von~Wietersheim-Kramsta\orcid{0000-0003-4986-5091}\inst{\ref{aff77},\ref{aff142}}
\and N.~A.~Walton\orcid{0000-0003-3983-8778}\inst{\ref{aff165}}
\and A.~H.~Wright\orcid{0000-0001-7363-7932}\inst{\ref{aff169}}
\and D.~Scott\orcid{0000-0002-6878-9840}\inst{\ref{aff170}}}
										   
\institute{INAF-Osservatorio Astronomico di Roma, Via Frascati 33, 00078 Monteporzio Catone, Italy\label{aff1}
\and
INAF-Osservatorio Astronomico di Padova, Via dell'Osservatorio 5, 35122 Padova, Italy\label{aff2}
\and
INAF-Osservatorio Astronomico di Trieste, Via G. B. Tiepolo 11, 34143 Trieste, Italy\label{aff3}
\and
Universit\"ats-Sternwarte M\"unchen, Fakult\"at f\"ur Physik, Ludwig-Maximilians-Universit\"at M\"unchen, Scheinerstr.~1, 81679 M\"unchen, Germany\label{aff4}
\and
Max Planck Institute for Extraterrestrial Physics, Giessenbachstr. 1, 85748 Garching, Germany\label{aff5}
\and
Institut de F\'{i}sica d'Altes Energies (IFAE), The Barcelona Institute of Science and Technology, Campus UAB, 08193 Bellaterra (Barcelona), Spain\label{aff6}
\and
Observatorio Nacional, Rua General Jose Cristino, 77-Bairro Imperial de Sao Cristovao, Rio de Janeiro, 20921-400, Brazil\label{aff7}
\and
Universit\'e Paris-Saclay, CNRS, Institut d'astrophysique spatiale, 91405, Orsay, France\label{aff8}
\and
Institut universitaire de France (IUF), 1 rue Descartes, 75231 PARIS CEDEX 05, France\label{aff9}
\and
ESAC/ESA, Camino Bajo del Castillo, s/n., Urb. Villafranca del Castillo, 28692 Villanueva de la Ca\~nada, Madrid, Spain\label{aff10}
\and
INAF-Osservatorio Astronomico di Brera, Via Brera 28, 20122 Milano, Italy\label{aff11}
\and
INAF-Osservatorio di Astrofisica e Scienza dello Spazio di Bologna, Via Piero Gobetti 93/3, 40129 Bologna, Italy\label{aff12}
\and
Universit\'e Paris-Saclay, Universit\'e Paris Cit\'e, CEA, CNRS, AIM, 91191, Gif-sur-Yvette, France\label{aff13}
\and
IFPU, Institute for Fundamental Physics of the Universe, via Beirut 2, 34151 Trieste, Italy\label{aff14}
\and
INFN, Sezione di Trieste, Via Valerio 2, 34127 Trieste TS, Italy\label{aff15}
\and
SISSA, International School for Advanced Studies, Via Bonomea 265, 34136 Trieste TS, Italy\label{aff16}
\and
Dipartimento di Fisica e Astronomia, Universit\`a di Bologna, Via Gobetti 93/2, 40129 Bologna, Italy\label{aff17}
\and
INFN-Sezione di Bologna, Viale Berti Pichat 6/2, 40127 Bologna, Italy\label{aff18}
\and
Dipartimento di Fisica, Universit\`a di Genova, Via Dodecaneso 33, 16146, Genova, Italy\label{aff19}
\and
INFN-Sezione di Genova, Via Dodecaneso 33, 16146, Genova, Italy\label{aff20}
\and
Department of Physics "E. Pancini", University Federico II, Via Cinthia 6, 80126, Napoli, Italy\label{aff21}
\and
INAF-Osservatorio Astronomico di Capodimonte, Via Moiariello 16, 80131 Napoli, Italy\label{aff22}
\and
European Southern Observatory, Karl-Schwarzschild-Str.~2, 85748 Garching, Germany\label{aff23}
\and
Instituto de Astrof\'isica e Ci\^encias do Espa\c{c}o, Universidade do Porto, CAUP, Rua das Estrelas, PT4150-762 Porto, Portugal\label{aff24}
\and
Faculdade de Ci\^encias da Universidade do Porto, Rua do Campo de Alegre, 4150-007 Porto, Portugal\label{aff25}
\and
Dipartimento di Fisica, Universit\`a degli Studi di Torino, Via P. Giuria 1, 10125 Torino, Italy\label{aff26}
\and
INFN-Sezione di Torino, Via P. Giuria 1, 10125 Torino, Italy\label{aff27}
\and
INAF-Osservatorio Astrofisico di Torino, Via Osservatorio 20, 10025 Pino Torinese (TO), Italy\label{aff28}
\and
Leiden Observatory, Leiden University, Einsteinweg 55, 2333 CC Leiden, The Netherlands\label{aff29}
\and
INAF-IASF Milano, Via Alfonso Corti 12, 20133 Milano, Italy\label{aff30}
\and
INFN-Sezione di Roma, Piazzale Aldo Moro, 2 - c/o Dipartimento di Fisica, Edificio G. Marconi, 00185 Roma, Italy\label{aff31}
\and
Centro de Investigaciones Energ\'eticas, Medioambientales y Tecnol\'ogicas (CIEMAT), Avenida Complutense 40, 28040 Madrid, Spain\label{aff32}
\and
Port d'Informaci\'{o} Cient\'{i}fica, Campus UAB, C. Albareda s/n, 08193 Bellaterra (Barcelona), Spain\label{aff33}
\and
Institute of Space Sciences (ICE, CSIC), Campus UAB, Carrer de Can Magrans, s/n, 08193 Barcelona, Spain\label{aff34}
\and
Institut d'Estudis Espacials de Catalunya (IEEC),  Edifici RDIT, Campus UPC, 08860 Castelldefels, Barcelona, Spain\label{aff35}
\and
INFN -- Sezione di Napoli, Via Cinthia 6, 80126, Napoli, Italy\label{aff36}
\and
Dipartimento di Fisica e Astronomia "Augusto Righi" - Alma Mater Studiorum Universit\`a di Bologna, Viale Berti Pichat 6/2, 40127 Bologna, Italy\label{aff37}
\and
Instituto de Astrof\'{\i}sica de Canarias, E-38205 La Laguna, Tenerife, Spain\label{aff38}
\and
Institute for Astronomy, University of Edinburgh, Royal Observatory, Blackford Hill, Edinburgh EH9 3HJ, UK\label{aff39}
\and
Jodrell Bank Centre for Astrophysics, Department of Physics and Astronomy, University of Manchester, Oxford Road, Manchester M13 9PL, UK\label{aff40}
\and
European Space Agency/ESRIN, Largo Galileo Galilei 1, 00044 Frascati, Roma, Italy\label{aff41}
\and
Universit\'e Lyon 1, CNRS/IN2P3, IP2I Lyon, Villeurbanne, France\label{aff42}
\and
Aix-Marseille Universit\'e, CNRS, CNES, LAM, Marseille, France\label{aff43}
\and
Universit\'e Lyon 1, CNRS/IN2P3, IUF, IP2I Lyon, Villeurbanne, France \label{aff44}
\and
Mullard Space Science Laboratory, University College London, Holmbury St Mary, Dorking, Surrey RH5 6NT, UK\label{aff45}
\and
Departamento de F\'isica, Faculdade de Ci\^encias, Universidade de Lisboa, Edif\'icio C8, Campo Grande, PT1749-016 Lisboa, Portugal\label{aff46}
\and
Instituto de Astrof\'isica e Ci\^encias do Espa\c{c}o, Faculdade de Ci\^encias, Universidade de Lisboa, Campo Grande, 1749-016 Lisboa, Portugal\label{aff47}
\and
Space Science Data Center, Italian Space Agency, via del Politecnico snc, 00133 Roma, Italy\label{aff48}
\and
Department of Astronomy, University of Geneva, ch. d'Ecogia 16, 1290 Versoix, Switzerland\label{aff49}
\and
INFN-Padova, Via Marzolo 8, 35131 Padova, Italy\label{aff50}
\and
Aix-Marseille Universit\'e, CNRS/IN2P3, CPPM, Marseille, France\label{aff51}
\and
INAF-Istituto di Astrofisica e Planetologia Spaziali, via del Fosso del Cavaliere, 100, 00100 Roma, Italy\label{aff52}
\and
School of Physics, HH Wills Physics Laboratory, University of Bristol, Tyndall Avenue, Bristol, BS8 1TL, UK\label{aff53}
\and
Department of Physics, University of Oxford, Keble Road, Oxford OX1 3RH, UK\label{aff54}
\and
University Observatory, LMU Faculty of Physics, Scheinerstr.~1, 81679 Munich, Germany\label{aff55}
\and
Dipartimento di Fisica "Aldo Pontremoli", Universit\`a degli Studi di Milano, Via Celoria 16, 20133 Milano, Italy\label{aff56}
\and
INFN-Sezione di Milano, Via Celoria 16, 20133 Milano, Italy\label{aff57}
\and
Institute of Theoretical Astrophysics, University of Oslo, P.O. Box 1029 Blindern, 0315 Oslo, Norway\label{aff58}
\and
Caltech/IPAC, 1200 E. California Blvd., Pasadena, CA 91125, USA\label{aff59}
\and
Jet Propulsion Laboratory, California Institute of Technology, 4800 Oak Grove Drive, Pasadena, CA, 91109, USA\label{aff60}
\and
Felix Hormuth Engineering, Goethestr. 17, 69181 Leimen, Germany\label{aff61}
\and
Technical University of Denmark, Elektrovej 327, 2800 Kgs. Lyngby, Denmark\label{aff62}
\and
Cosmic Dawn Center (DAWN), Denmark\label{aff63}
\and
Universit\'e PSL, Observatoire de Paris, Sorbonne Universit\'e, CNRS, LERMA, 75014, Paris, France\label{aff64}
\and
Universit\'e Paris-Cit\'e, 5 Rue Thomas Mann, 75013, Paris, France\label{aff65}
\and
Max-Planck-Institut f\"ur Astronomie, K\"onigstuhl 17, 69117 Heidelberg, Germany\label{aff66}
\and
NASA Goddard Space Flight Center, Greenbelt, MD 20771, USA\label{aff67}
\and
Universit\'e de Gen\`eve, D\'epartement de Physique Th\'eorique and Centre for Astroparticle Physics, 24 quai Ernest-Ansermet, CH-1211 Gen\`eve 4, Switzerland\label{aff68}
\and
Department of Physics, P.O. Box 64, University of Helsinki, 00014 Helsinki, Finland\label{aff69}
\and
Helsinki Institute of Physics, Gustaf H{\"a}llstr{\"o}min katu 2, University of Helsinki, 00014 Helsinki, Finland\label{aff70}
\and
Laboratoire d'etude de l'Univers et des phenomenes eXtremes, Observatoire de Paris, Universit\'e PSL, Sorbonne Universit\'e, CNRS, 92190 Meudon, France\label{aff71}
\and
SKAO, Jodrell Bank, Lower Withington, Macclesfield SK11 9FT, UK\label{aff72}
\and
Centre de Calcul de l'IN2P3/CNRS, 21 avenue Pierre de Coubertin 69627 Villeurbanne Cedex, France\label{aff73}
\and
University of Applied Sciences and Arts of Northwestern Switzerland, School of Computer Science, 5210 Windisch, Switzerland\label{aff74}
\and
Universit\"at Bonn, Argelander-Institut f\"ur Astronomie, Auf dem H\"ugel 71, 53121 Bonn, Germany\label{aff75}
\and
Dipartimento di Fisica e Astronomia "Augusto Righi" - Alma Mater Studiorum Universit\`a di Bologna, via Piero Gobetti 93/2, 40129 Bologna, Italy\label{aff76}
\and
Department of Physics, Institute for Computational Cosmology, Durham University, South Road, Durham, DH1 3LE, UK\label{aff77}
\and
Universit\'e Paris Cit\'e, CNRS, Astroparticule et Cosmologie, 75013 Paris, France\label{aff78}
\and
CNRS-UCB International Research Laboratory, Centre Pierre Bin\'etruy, IRL2007, CPB-IN2P3, Berkeley, USA\label{aff79}
\and
Dipartimento di Fisica - Sezione di Astronomia, Universit\`a di Trieste, Via Tiepolo 11, 34131 Trieste, Italy\label{aff80}
\and
Telespazio UK S.L. for European Space Agency (ESA), Camino bajo del Castillo, s/n, Urbanizacion Villafranca del Castillo, Villanueva de la Ca\~nada, 28692 Madrid, Spain\label{aff81}
\and
European Space Agency/ESTEC, Keplerlaan 1, 2201 AZ Noordwijk, The Netherlands\label{aff82}
\and
Waterloo Centre for Astrophysics, University of Waterloo, Waterloo, Ontario N2L 3G1, Canada\label{aff83}
\and
Department of Physics and Astronomy, University of Waterloo, Waterloo, Ontario N2L 3G1, Canada\label{aff84}
\and
Perimeter Institute for Theoretical Physics, Waterloo, Ontario N2L 2Y5, Canada\label{aff85}
\and
Centre National d'Etudes Spatiales -- Centre spatial de Toulouse, 18 avenue Edouard Belin, 31401 Toulouse Cedex 9, France\label{aff86}
\and
Institute of Space Science, Str. Atomistilor, nr. 409 M\u{a}gurele, Ilfov, 077125, Romania\label{aff87}
\and
Institute for Particle Physics and Astrophysics, Dept. of Physics, ETH Zurich, Wolfgang-Pauli-Strasse 27, 8093 Zurich, Switzerland\label{aff88}
\and
Dipartimento di Fisica e Astronomia "G. Galilei", Universit\`a di Padova, Via Marzolo 8, 35131 Padova, Italy\label{aff89}
\and
Instituto de F\'isica Te\'orica UAM-CSIC, Campus de Cantoblanco, 28049 Madrid, Spain\label{aff90}
\and
Institut de Recherche en Astrophysique et Plan\'etologie (IRAP), Universit\'e de Toulouse, CNRS, UPS, CNES, 14 Av. Edouard Belin, 31400 Toulouse, France\label{aff91}
\and
Universit\'e St Joseph; Faculty of Sciences, Beirut, Lebanon\label{aff92}
\and
Departamento de F\'isica, FCFM, Universidad de Chile, Blanco Encalada 2008, Santiago, Chile\label{aff93}
\and
Universit\'e C\^{o}te d'Azur, Observatoire de la C\^{o}te d'Azur, CNRS, Laboratoire Lagrange, Bd de l'Observatoire, CS 34229, 06304 Nice cedex 4, France\label{aff94}
\and
Universit\"at Innsbruck, Institut f\"ur Astro- und Teilchenphysik, Technikerstr. 25/8, 6020 Innsbruck, Austria\label{aff95}
\and
Infrared Processing and Analysis Center, California Institute of Technology, Pasadena, CA 91125, USA\label{aff96}
\and
Cosmic Dawn Center (DAWN)\label{aff97}
\and
Niels Bohr Institute, University of Copenhagen, Jagtvej 128, 2200 Copenhagen, Denmark\label{aff98}
\and
Universidad Polit\'ecnica de Cartagena, Departamento de Electr\'onica y Tecnolog\'ia de Computadoras,  Plaza del Hospital 1, 30202 Cartagena, Spain\label{aff99}
\and
European University of Technology EUt+, European Union\label{aff100}
\and
Kapteyn Astronomical Institute, University of Groningen, PO Box 800, 9700 AV Groningen, The Netherlands\label{aff101}
\and
Dipartimento di Fisica e Scienze della Terra, Universit\`a degli Studi di Ferrara, Via Giuseppe Saragat 1, 44122 Ferrara, Italy\label{aff102}
\and
Istituto Nazionale di Fisica Nucleare, Sezione di Ferrara, Via Giuseppe Saragat 1, 44122 Ferrara, Italy\label{aff103}
\and
ICSC - Centro Nazionale di Ricerca in High Performance Computing, Big Data e Quantum Computing, Via Magnanelli 2, Bologna, Italy\label{aff104}
\and
INAF, Istituto di Radioastronomia, Via Piero Gobetti 101, 40129 Bologna, Italy\label{aff105}
\and
INFN-Bologna, Via Irnerio 46, 40126 Bologna, Italy\label{aff106}
\and
Astronomical Observatory of the Autonomous Region of the Aosta Valley (OAVdA), Loc. Lignan 39, I-11020, Nus (Aosta Valley), Italy\label{aff107}
\and
Department of Mathematical Physics, Institute of Physics, University of S\~ao Paulo, R. do Mat\~ao 1371, 05508-090, S\~ao Paulo, SP, Brazil\label{aff108}
\and
CEA Saclay, DFR/IRFU, Service d'Astrophysique, Bat. 709, 91191 Gif-sur-Yvette, France\label{aff109}
\and
Universidad de La Laguna, Dpto. Astrof\'\i sica, E-38206 La Laguna, Tenerife, Spain\label{aff110}
\and
Oskar Klein Centre for Cosmoparticle Physics, Department of Physics, Stockholm University, Stockholm, SE-106 91, Sweden\label{aff111}
\and
Astrophysics Group, Blackett Laboratory, Imperial College London, London SW7 2AZ, UK\label{aff112}
\and
Univ. Grenoble Alpes, CNRS, Grenoble INP, LPSC-IN2P3, 53, Avenue des Martyrs, 38000 Grenoble, France\label{aff113}
\and
Dipartimento di Fisica, Sapienza Universit\`a di Roma, Piazzale Aldo Moro 2, 00185 Roma, Italy\label{aff114}
\and
Institut d'Astrophysique de Paris, 98bis Boulevard Arago, 75014, Paris, France\label{aff115}
\and
ICL, Junia, Universit\'e Catholique de Lille, LITL, 59000 Lille, France\label{aff116}
\and
Center for Astrophysics and Cosmology, University of Nova Gorica, Nova Gorica, Slovenia\label{aff117}
\and
Department of Astronomy, Indiana University, 727 East Third Street, Bloomington, IN 47405, USA\label{aff118}
\and
CERCA/ISO, Department of Physics, Case Western Reserve University, 10900 Euclid Avenue, Cleveland, OH 44106, USA\label{aff119}
\and
Laboratoire Univers et Th\'eorie, Observatoire de Paris, Universit\'e PSL, Universit\'e Paris Cit\'e, CNRS, 92190 Meudon, France\label{aff120}
\and
Departamento de F{\'\i}sica Fundamental. Universidad de Salamanca. Plaza de la Merced s/n. 37008 Salamanca, Spain\label{aff121}
\and
Universiteit Gent, Department of Physics and Astronomy, Proeftuinstraat 86 N3, 9000 Ghent, Belgium
\label{aff122}
\and
Konkoly Observatory, HUN-REN CSFK, MTA Centre of Excellence, Budapest, Konkoly Thege Mikl\'os {\'u}t 15-17. H-1121, Hungary\label{aff123}
\and
MTA-CSFK Lend\"ulet Large-Scale Structure Research Group, Konkoly-Thege Mikl\'os \'ut 15-17, H-1121 Budapest, Hungary\label{aff124}
\and
Universit\'e de Strasbourg, CNRS, Observatoire astronomique de Strasbourg, UMR 7550, 67000 Strasbourg, France\label{aff125}
\and
Scuola Superiore Meridionale, Via Mezzocannone 4, 80138, Napoli, Italy\label{aff126}
\and
California Institute of Technology, 1200 E California Blvd, Pasadena, CA 91125, USA\label{aff127}
\and
Instituto de Astrof\'isica e Ci\^encias do Espa\c{c}o, Faculdade de Ci\^encias, Universidade de Lisboa, Tapada da Ajuda, 1349-018 Lisboa, Portugal\label{aff128}
\and
Institute of Cosmology and Gravitation, University of Portsmouth, Portsmouth PO1 3FX, UK\label{aff129}
\and
Centre de Recherche Astrophysique de Lyon, UMR5574, CNRS, Universit\'e Claude Bernard Lyon 1, ENS de Lyon, 69230, Saint-Genis-Laval, France\label{aff130}
\and
Department of Physics \& Astronomy, University of California Irvine, Irvine CA 92697, USA\label{aff131}
\and
Institute for Astronomy, University of Hawaii, 2680 Woodlawn Drive, Honolulu, HI 96822, USA\label{aff132}
\and
Departamento F\'isica Aplicada, Universidad Polit\'ecnica de Cartagena, Campus Muralla del Mar, 30202 Cartagena, Murcia, Spain\label{aff133}
\and
Instituto de F\'isica de Cantabria, Edificio Juan Jord\'a, Avenida de los Castros, 39005 Santander, Spain\label{aff134}
\and
Univ. Grenoble Alpes, CNRS, IPAG, 38000 Grenoble, France\label{aff135}
\and
Department of Mathematics and Physics E. De Giorgi, University of Salento, Via per Arnesano, CP-I93, 73100, Lecce, Italy\label{aff136}
\and
INFN, Sezione di Lecce, Via per Arnesano, CP-193, 73100, Lecce, Italy\label{aff137}
\and
INAF-Sezione di Lecce, c/o Dipartimento Matematica e Fisica, Via per Arnesano, 73100, Lecce, Italy\label{aff138}
\and
Institut d'Astrophysique de Paris, UMR 7095, CNRS, and Sorbonne Universit\'e, 98 bis boulevard Arago, 75014 Paris, France\label{aff139}
\and
Department of Computer Science, Aalto University, PO Box 15400, Espoo, FI-00 076, Finland\label{aff140}
\and
DARK, Niels Bohr Institute, University of Copenhagen, Jagtvej 155, 2200 Copenhagen, Denmark\label{aff141}
\and
Department of Physics, Centre for Extragalactic Astronomy, Durham University, South Road, Durham, DH1 3LE, UK\label{aff142}
\and
Astrophysics Research Centre, University of KwaZulu-Natal, Westville Campus, Durban 4041, South Africa\label{aff143}
\and
School of Mathematics, Statistics \& Computer Science, University of KwaZulu-Natal, Westville Campus, Durban 4041, South Africa\label{aff144}
\and
Department of Physics and Astronomy, Vesilinnantie 5, University of Turku, 20014 Turku, Finland\label{aff145}
\and
Finnish Centre for Astronomy with ESO (FINCA), Quantum, Vesilinnantie 5, University of Turku, 20014 Turku, Finland\label{aff146}
\and
Serco for European Space Agency (ESA), Camino bajo del Castillo, s/n, Urbanizacion Villafranca del Castillo, Villanueva de la Ca\~nada, 28692 Madrid, Spain\label{aff147}
\and
Department of Physics and Helsinki Institute of Physics, Gustaf H\"allstr\"omin katu 2, University of Helsinki, 00014 Helsinki, Finland\label{aff148}
\and
Ohio University, Physics \& Astronomy Department,1 Ohio University, Athens, OH 45701, USA\label{aff149}
\and
Institute for Theoretical Particle Physics and Cosmology (TTK), RWTH Aachen University, 52056 Aachen, Germany\label{aff150}
\and
IRFU, CEA, Universit\'e Paris-Saclay 91191 Gif-sur-Yvette Cedex, France\label{aff151}
\and
INAF-Osservatorio Astrofisico di Arcetri, Largo E. Fermi 5, 50125, Firenze, Italy\label{aff152}
\and
Centro de Astrof\'{\i}sica da Universidade do Porto, Rua das Estrelas, 4150-762 Porto, Portugal\label{aff153}
\and
Dipartimento di Fisica, Universit\`a di Roma Tor Vergata, Via della Ricerca Scientifica 1, Roma, Italy\label{aff154}
\and
INFN, Sezione di Roma 2, Via della Ricerca Scientifica 1, Roma, Italy\label{aff155}
\and
HE Space for European Space Agency (ESA), Camino bajo del Castillo, s/n, Urbanizacion Villafranca del Castillo, Villanueva de la Ca\~nada, 28692 Madrid, Spain\label{aff156}
\and
STAR Institute, University of Li{\`e}ge, Quartier Agora, All\'ee du six Ao\^ut 19c, 4000 Li\`ege, Belgium\label{aff157}
\and
Dipartimento di Fisica ``G. Occhialini", Universit\`a degli Studi di Milano Bicocca, Piazza della Scienza 3, 20126 Milano, Italy\label{aff158}
\and
School of Physical Sciences, The Open University, Milton Keynes, MK7 6AA, UK\label{aff159}
\and
Department of Astrophysics, University of Zurich, Winterthurerstrasse 190, 8057 Zurich, Switzerland\label{aff160}
\and
Space physics and astronomy research unit, University of Oulu, Pentti Kaiteran katu 1, FI-90014 Oulu, Finland\label{aff161}
\and
Department of Mathematics and Physics, Roma Tre University, Via della Vasca Navale 84, 00146 Rome, Italy\label{aff162}
\and
Theoretical astrophysics, Department of Physics and Astronomy, Uppsala University, Box 524, 751 20 Uppsala, Sweden\label{aff163}
\and
Mathematical Institute, University of Leiden, Einsteinweg 55, 2333 CA Leiden, The Netherlands\label{aff164}
\and
Institute of Astronomy, University of Cambridge, Madingley Road, Cambridge CB3 0HA, UK\label{aff165}
\and
Univ. Lille, CNRS, Centrale Lille, UMR 9189 CRIStAL, 59000 Lille, France\label{aff166}
\and
Department of Astrophysical Sciences, Peyton Hall, Princeton University, Princeton, NJ 08544, USA\label{aff167}
\and
Fakult\"at f\"ur Physik, Universit\"at Bielefeld, Postfach 100131, 33501 Bielefeld, Germany\label{aff168}
\and
Ruhr University Bochum, Faculty of Physics and Astronomy, Astronomical Institute (AIRUB), German Centre for Cosmological Lensing (GCCL), 44780 Bochum, Germany\label{aff169}
\and
Department of Physics and Astronomy, University of British Columbia, Vancouver, BC V6T 1Z1, Canada\label{aff170}}    

\date{\today}

\abstract{
The Euclid Data Release 1 (DR1) contains catalogues of sources covering roughly $1900\,{\deg}^2$, including deep coverage of legacy fields that enable an extensive comparison between \Euclid imaging and multiwavelength observations from other instruments. The Cosmic Evolution Survey (COSMOS) field, observed by \Euclid and included in DR1, has been observed with the {\it James Webb} Space Telescope, resulting in the COSMOS2025 catalogue.
We compare the Euclid DR1 photometry with the multiwavelength COSMOS2025 data to assess the accuracy of the photometric measurements produced by the different methods implemented in the OU-MER pipeline. We use data from both the Euclid Wide and Deep Surveys, focussing exclusively on extragalactic sources.
We use the fluxes of COSMOS2025 to predict the expected fluxes in the \Euclid bands via spectral energy distribution fitting. The redshifts of the sources are either spectroscopic, for a sample of brighter sources, or photometric, to include fainter sources. Both the fluxes in the common Hyper Suprime-Cam bands and the SEDs from COSMOS2025 are used to validate the analysis and to characterise any biases or limitations of the method.
The predicted fluxes and the OU-MER measurements are in good agreement, especially for bright and isolated sources. Median offsets depend on band, photometric method, survey depth, and redshift sample. Most are well within 10\%, although the deep photo-z model-fitting measurements show NIR offsets of approximately 12–13\%. However, complementary measurements on \gaia stars show an opposite trend, highlighting that photometric validation approaches may suffer from intrinsic uncertainties that prevent a thorough assessment of systematic effects. We find that the OU-MER fluxes typically have larger offsets for systems that have not been correctly deblended in detection.
Among the three methods, template-fitting photometry generally shows the smallest offset from the SED-based predictions in the NISP bands and toward fainter magnitudes, while aperture photometry and model-fitting are comparable in \IE. 
Future analysis on deeper \Euclid data and on fields at different Galactic latitudes will allow us to strengthen the constraints on the photometric accuracy of the \Euclid surveys.}

\authorrunning{Euclid Collaboration: T. Gasparetto et al.}
\titlerunning{Multi-band photometry validation on COSMOS extragalactic sources}

\maketitle

\section{Introduction}\label{sect:intro}
After its successful launch on 1 July 2023, followed by some months of commissioning and early performance verification, the \Euclid satellite started its nominal six-years-long survey campaign on 14 February 2024. The goal of the \Euclid mission is to perform a photometric and spectroscopic survey at visible and near infrared wavelengths in order to study two primary cosmology probes, namely weak gravitational lensing and galaxy clustering~\citep{Laureijs11}. The target final footprint of the nominal Euclid Wide Survey (EWS) at the end of the mission, following a precise survey plan~\citep{Scaramella-EP1}, will be roughly 14\,000 ${\deg}^2$. Additionally to the EWS, about 12\% of \Euclid's on-sky observations are spent on the Euclid Deep Survey (EDS), observations about 2 magnitudes deeper with respect to the nominal survey that will also serve as monitoring and calibration fields. 

The EDS is further complemented by the Euclid Auxiliary Fields~\citep[EAFs, see sect.\,4.2.2 from][]{EuclidSkyOverview}. One of these EAFs is the Cosmic Evolution Survey (COSMOS) field~\citep{Cosmos_scoville_07}, centred at RA (J2000) = \ang{150.119;;}, Dec (J2000) = +\ang{02.21;;}, covering an area of approximately $2.0\deg^2$, which has been extensively observed by several instruments over the last two decades from X-ray to radio frequencies, by ground- and space-based telescopes. A dedicated observing program on the COSMOS field has been carried out with the \textit{James Webb} Space Telescope~\citep[\JWST,][]{jwstoverview2023}, which resulted in the creation of the COSMOS2025~\citep[C25,][]{COSMOS2025} catalogue of more than \num{700000} sources.

While the aim of the first Euclid Quick Data Release \citep[Q1,][]{Q1-TP001} was to give a glimpse of the \Euclid capabilities on a small $63.1\,{\deg}^2$ footprint, the Euclid Data Release 1 (DR1) extends the covered sky area to roughly $1900\,{\deg}^2$~\citep{DR1-TP001}. The released data not only cover a large sky area, but have also been processed using updated and improved versions of both the software and the calibration products. The next data releases will continue to improve both quantitatively and qualitatively, as our knowledge of the telescope and of its instruments get better, together with our ability to analyse increasingly large volumes of data and assess their quality.

A pivotal step in the reduction chain of the data coming from \Euclid is the creation of photometric catalogues, involving the detection, deblending, and multi-wavelength flux measurements of the sources.  This task is carried out by the MERge Processing Function (MER-PF), developed by the MER Organisational Unit (OU-MER, hereafter MER). The images acquired by the Visible Camera~\citep[VIS,][]{EuclidSkyVIS} onboard \Euclid are processed by OU-VIS~\citep{DR1-TP002} to generate the calibrated frames, while those acquired with the Near-Infrared Spectrometer and Photometer~\citep[NISP,][]{EuclidSkyNISP} are processed by OU-NIR~\citep{DR1-TP003}. The stacked frames from external surveys are instead produced by OU-EXT~\citep{DR1-TP006} and together with the calibrated images from OU-VIS and OU-NIR, they are used as input to the MER-PF.

The MER data products are then used by downstream PFs to calculate the photometric redshifts, as well as to perform cosmic shear analyses and spectroscopic redshift measurements. A technical description of the full processing chain is presented in~\citet{DR1-TP033}, while the changes in the MER-PF since the Q1 release~\citep[hereafter \citetalias{Q1-TP004},][]{Q1-TP004} are extensively described in the MER DR1 paper~\citep[hereafter \citetalias{DR1-TP009},][]{DR1-TP009}. Together with this work and \citetalias{DR1-TP009}, the accuracy of the OU-MER DR1 photometry is also assessed using a synthetic source injection pipeline
(Euclid Collaboration: Kong et al., in prep.): the comparison of this work with our results is shown in the following sections.

This paper offers a complementary approach to the routine photometry validation performed in MER on \gaia-matched stars (described in \citetalias{Q1-TP004} and \citetalias{DR1-TP009}), focusing instead on extended extragalactic objects, which are the main target of the \Euclid core science. Since there is no analogue of the all-sky \gaia survey that can be used as a reference for extended sources, it is necessary to use data from extragalactic surveys; we chose to focus on the COSMOS field, exploiting the extensive multiwavelength data available in C25 (which combines \JWST observations with ancillary ground-based and \HST data) to predict the fluxes of individual sources in the four \Euclid bands, namely \IE, \YE, \JE, and \HE. To this end, we employ a spectral energy distribution (SED) fitting technique~\citep[see e.g.,][]{iyer2025SEDFitting} to derive a model spectrum for each source. Specifically, using a library of synthetic galaxy templates, we fit the observed fluxes and redshifts of the sources in C25 through a $\chi^2$ minimisation procedure, focussing on sources with high signal-to-noise ratio (S/N) in some selected \JWST bands. This allows us to determine the best-fitting SED for each source and, from the corresponding model spectrum, to compute predicted magnitudes in the four \Euclid bands. These predictions can then be compared with the photometric measurements from the MER catalogues. 

The paper is organised as follows. In Sect.~\ref{sect:datasets}, we describe the sub-selection of the MER catalogue and of the C25 data sets that have been used. In Sect.~\ref{sect:method}, we describe the SED-fitting procedure, including both the ingredients adopted to create the reference synthetic galaxy library and the software used. In Sect.~\ref{sect:analyses}, we report on the analyses performed, including the cross-matching between the catalogues and the various selection criteria we applied. In Sect.~\ref{sect:results}, we discuss the accuracy of the fitting procedure, comparing C25 with the forecast in the \Euclid and Hyper Suprime-Cam (HSC) bands; we also perform a direct comparison between C25 and MER for the HSC bands. We use MER catalogues for deep and wide tiles, searching for photometric offsets both for spectroscopic and photometric redshifts; in Sect.~\ref{sect:discussion} we discuss these results. In Sect.~\ref{sect:future} we present the conclusions of the work and give some perspectives on future improvements. Similarly to the C25 catalogue, we adopt a standard $\Lambda$CDM cosmology, with $H_0=70\,\si{\kilo\meter}\,\si{\s}^{-1}\,\si{\mega\parsec}^{-1}$, ${\Omm} = 0.3$, and ${\OmLa} = 0.7$.

\section{Data sets}\label{sect:datasets}
In this section we present the MER and COSMOS2025 data sets, focusing on the information that is most relevant for our goals. We refer the reader to \citetalias{Q1-TP004}, \citetalias{DR1-TP009}, and C25 for a more comprehensive and exhaustive description.
 
\begin{figure*}
    \centering
    \includegraphics[height=0.37\textheight]{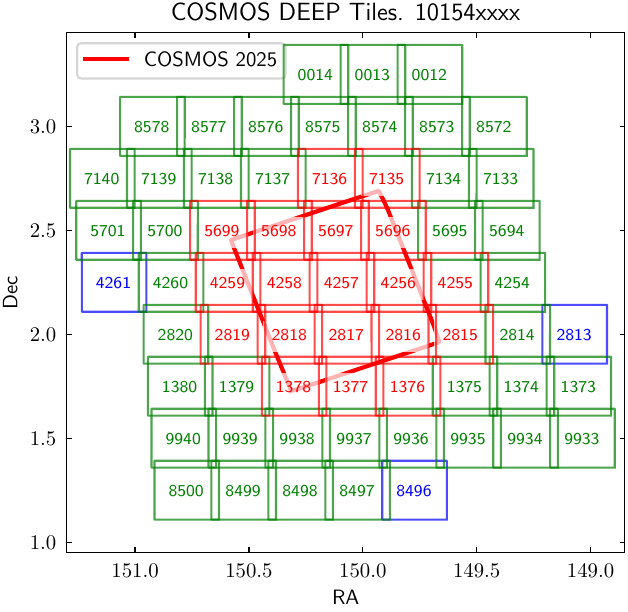}
    \includegraphics[height=0.37\textheight, trim=30 0 0 0, clip]{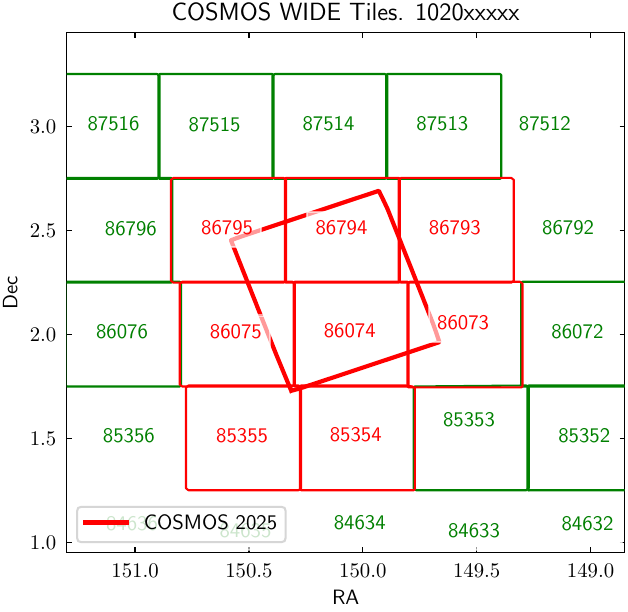}
    \caption{Tile IDs and tile area in the COSMOS field. \textit{Left}: outer tile area for the deep tiles (prefix is 10154) where 19 tiles (red) overlap with the C25 footprint (thick red rectangle), among the total of 60 tiles of the field (green polygons). The tile IDs in blue at the borders are not in DR1. \textit{Right}: inner tile area for the eight (red) wide tiles (prefix 1020) overlapping with the C25 footprint, while the other wide tiles (green) are outside it. There are 477\,600 objects in the MER deep catalogue in the C25 footprint, while this number is reduced to 200\,201 objects for the wide catalogue.}
    \label{fig:cosmos_tile_ids}
\end{figure*}

\subsection{MER Catalogue}\label{sect:mercat}

As described in \citetalias{Q1-TP004}, while the calibrated images provided by OU-VIS and OU-NIR depend on the telescope pointing following the survey strategy, the MER-PF is run on sky areas called `tiles', that are not tied to the pointing pattern and whose positions and size have been defined in \cite{2022ASPC..532..329K}. All the available OU-VIS, OU-NIR, and OU-EXT calibrated frames intersecting the tile area are used as input to the MER-PF, together with some additional configuration files, creating at the end a list of products, namely:
\begin{itemize}
\item a list of MER background-subtracted mosaics ({\tt MerBksMosaic}) per band (the starting point to measure the source photometry);
\item the \texttt{MerDetectionMosaic} used to get the list of VIS and NIR detected sources;
\item the segmentation map ({\tt MerSegmentationMap}) containing the spatial pixels associated with each individual source;
\item the final catalogue ({\tt MerFinalCatalog}) storing all the source information measured by the pipeline, complemented by a photometric deep catalogue in case of processing deep tiles.
\end{itemize}In this paper, we focus only on the \IE, \YE, \JE, and \HE photometry, with some additional comparisons done on the HSC~$g$, $r$, $i$, and $z$ bands. The details on the bands that are in each field both for the EWS and the EDS can be found in \citetalias{DR1-TP009}.

The data for COSMOS in DR1 consist of a series of repeated observations obtained between 22 and 28 November 2024, which were jointly processed to enable the detection of fainter sources. 60 deep tiles are released for DR1 over the COSMOS field, 19 of them overlapping, totally or partially, with the C25 footprint. The wide data for COSMOS correspond to just one of the observations carried out in this period and validated by MER;\footnote{The data are stored in the Euclid Archive System with \texttt{Header.DataSetRelease==COSMOS-WIDELIKE-DR1}.} a total of eight wide tiles overlap with the C25 footprint. The single catalogues from each of the 19 (eight) tiles are stacked together in order to obtain one single deep (wide) catalogue. While the extended tile area of the wide tile is $\ang{;32;}\times\ang{;32;}$, the deep ones have a $\ang{;17;}\times\ang{;17;}$ area (see \citetalias{DR1-TP009}).

The sources in the MER catalogues are detected either in the VIS mosaic (they are flagged with \texttt{VIS\_DET}=1 in the catalogue) or in a stack of the three NIR bands (\texttt{VIS\_DET}=0, see section~2.2 from \citetalias{DR1-TP009}). 
The deep tiles contain a total of 477\,600 MER sources within the C25 footprint (left panel in  Fig.~\ref{fig:cosmos_tile_ids}), approximately $91\%$ of which are detected in \IE (\texttt{VIS\_DET}=1). The wide tiles overlapping with the C25 footprint contain a total of 200\,201 MER sources (right panel in Fig.~\ref{fig:cosmos_tile_ids}), of which approximately $74\%$ are detected in \IE.
The percentage of sources detected in \IE is larger in the deep case because the objects detected in the NIR stack in the wide data can now be detected in \IE due to the increased image depth. Moreover, a non-negligible fraction of the NIR-detected sources in the wide catalogue are spurious, with no match in the deep catalogues (detailed discussion in section~6.1 of \citetalias{DR1-TP009}).

In this work, we focus exclusively on sources detected in \IE (\texttt{VIS\_DET}=1), since they constitute the majority of the sample and provide a sufficiently large data set for meaningful statistical analysis (Sect.~\ref{subsect:debl_selection}). Including NIR-detected sources in the photometric comparison would likely degrade the quality of our results (section~6.1 of \citetalias{DR1-TP009}).

\subsubsection{Photometric data}

The MER catalogues provide photometry measurements for all the detected sources in all the available bands. Three methods are used to estimate photometry, namely: (i) aperture photometry, obtained using a custom version of the software \texttt{A-PHOT}~\citep{Merlin2019}; (ii) template-fitting photometry, with a customised version of the \texttt{T-PHOT} code~\citep{Merlin2016}; and (iii) model-fitting photometry using analytical \sersic 2D profiles, performed using \texttt{SourceXtractor++}~\citep[{\tt SE++},][]{2022arXiv221202428K}.
A description of how these three methods work is given in \citetalias{Q1-TP004}, and more information on the content of the MER catalogues can be found in the Photometry Cookbook produced for the Q1 Data Release (PCQ1 hereafter).\footnote{\url{https://euclid.esac.esa.int/dr/q1/dpdd/merdpd/merphotometrycookbook.html}} All the fluxes used in this work are total fluxes, computed as described in PCQ1. The labels `APHOT', `TPHOT', and `SERSIC' are used in all the figures and in the text to indicate the photometry measurements coming from aperture photometry, template-fitting, and model-fitting photometry, respectively.

\subsection{COSMOS2025 sample}\label{subsect:cosmoscat}

COSMOS2025 is the COSMOS-Web~\citep{Cosmosweb23} catalogue of more than \num{700000} sources, released in June 2025, based on the \JWST 255 hours COSMOS-Web programme. The observations in four \JWST Near-InfraRed Camera~\citep[NIRCam,][]{nircam2023} filters, namely F115W, F150W, F277W, and F444W, and with filter F770W from \JWST Mid-InfraRed Instrument~\citep[MIRI,][]{miri2023} are complemented with ground- and space-based observations. SED-fitting between $0.3\,\micron$ and $8\,\micron$ is used to derive, among other properties, photometric redshifts and physical parameters, using both {\tt LePHARE}~\citep{lephare1, lephare2} and {\tt CIGALE}~\citep{cigale}. While the $0.54\deg{^2}$ footprint of COSMOS2025 is smaller (figure~1 from C25) with respect to COSMOS2020~\citep{Cosmos2020}, which covers roughly $2\deg^2$, the uncertainty in the photometric determination improves by a factor of 2 at magnitudes fainter than 26 with respect to the previous catalogue (figure~19 from C25); both the fraction of sources with photo-$z$s far from the spectroscopic redshift and the width of the distribution of such outliers are halved between COSMOS2020 and C25.

We retrieve from the COSMOS-Web website\footnote{\url{https://cosmos2025.iap.fr/}}  most of the data for this work, namely the photometric catalogues, the redshift measurements, and the SEDs. We use only two of the extensions available in the main catalogue FITS file: the first one, namely `PHOTOMETRY HOTCOLD AND SE++' (we refer to it as `Photometry'); and the second one, `LePhare'.\footnote{In this work we use version 1.1 of the COSMOS2025 catalogue and LePhare SEDs.} The reference catalogue was built using the fluxes reported in the columns \texttt{flux\_model\_\textasteriskcentered} of the `Photometry' extension (with the associated errors in \texttt{flux\_err-cal\_model\_\textasteriskcentered}) which are \sersic model total fluxes expressed in \si{\micro Jy}. We chose this instead of other flux measurements since \sersic model total fluxes is the only data set that has consistently provided photometry for all the C25 filters, and there is no need for combining different photometric measurements for different bands (in Fig.~\ref{fig:zphot_magauto_vs_sersic2d_photoz_0_1_nosrcmatch} we compare the results obtained using these fluxes against a combination of different flux measurements).

As suggested in the tutorial webpage of the COSMOS-Web website,\footnote{\url{https://cosmos2025.iap.fr/tutorials/catalog\_usage\_tutorial.html}} we reduce the initial \num{784016} sources to a robust and pure sample of \num{576143} galaxies, by requiring:
\begin{itemize}
    \item \texttt{type}=0 in `LePhare', to select only galaxies (\texttt{type}=1 and \texttt{type}=2 correspond to stars and QSOs, respectively);
    \item \texttt{warn\_flag}=0 in `Photometry', to keep only objects with no warning flag (i.e., far from bright stars);
    \item \texttt{abs\_val(mag\_model\_f444w)}<30 in `Photometry', to exclude extremely faint objects;
    \item \texttt{flag\_star\_hsc}=0 in `Photometry', to exclude sources that are outside bright stars (but still masked by the HSC star mask), whose photometry may be affected by stellar contamination.
\end{itemize}

The redshift of each source is used as an input to the SED-fitting procedure (Sect.~\ref{sect:method}) to determine the best-fitting SED and the fluxes in each band. In other words, the SED-fitting is not used to estimate photometric redshifts; instead, the redshift is fixed to previously estimated values. This choice avoids introducing an additional source of uncertainty arising from possible differences between the input redshifts and those that would be derived from our SED-fitting procedure.
    
We consider two subsamples of sources: one consisting of objects with available spectroscopic redshifts (spec-$z$) from~\cite{khostovan2025};\footnote{In this work we use release 1.1 of the spec-$z$ compilation from \url{https://github.com/cosmosastro/speczcompilation}.} and the other consisting of objects with photometric redshifts (photo-$z$) from C25.  For the spec-$z$ case, we chose to keep only the sources with confidence level ${\rm CL}>95\%$ in the catalogue (${\tt quality\_flag} >= 3$); there are 187\,549 of them, but only 16\,006 appear in C25 (they have an \texttt{Id\_COSMOS25} value in the spec-$z$ compilation, and a spec-$z$ greater than 0).\footnote{There are 48 sources appearing twice in the spec-$z$ compilation with the same \texttt{ID\_COSMOS25}: their redshifts are the average of the two numbers, the difference being less than 0.01 for most of the sources.} 

The photo-$z$ for each source is read from the \texttt{zfinal} column of the `LePhare' extension and the starting sample of \num{576143} sources is reduced to \num{576035} sources with a determined photo-$z$ (greater than 0). A comparison for the sources having both a spectroscopic redshift $z_{\rm spec}$ and a photometric redshift $z_{\rm phot}$ shows that only 5\% of the sources have $|z_{\rm spec}-z_{\rm phot}|/(1+z_{\rm spec})>0.1$. 

Our SED-fitting is performed on the C25 sources with high S/N, since the aim of this work is not to compare the full population of galaxies in the two catalogues, but rather to focus on sources with well-constrained flux measurements to find systematic biases or trends. We therefore select sources in C25 with ${\rm S/N}>10$ in each of the four \JWST bands covering the entire footprint, namely F115W, F150W, F277W, and F444W. Combining all the previous selection criteria, we perform the SED-fitting for \num{15268} sources for the spec-$z$ case and for \num{120008} sources for the photo-$z$ case. 

\begin{figure*}
    \centering
    \includegraphics[width=0.95\textwidth]{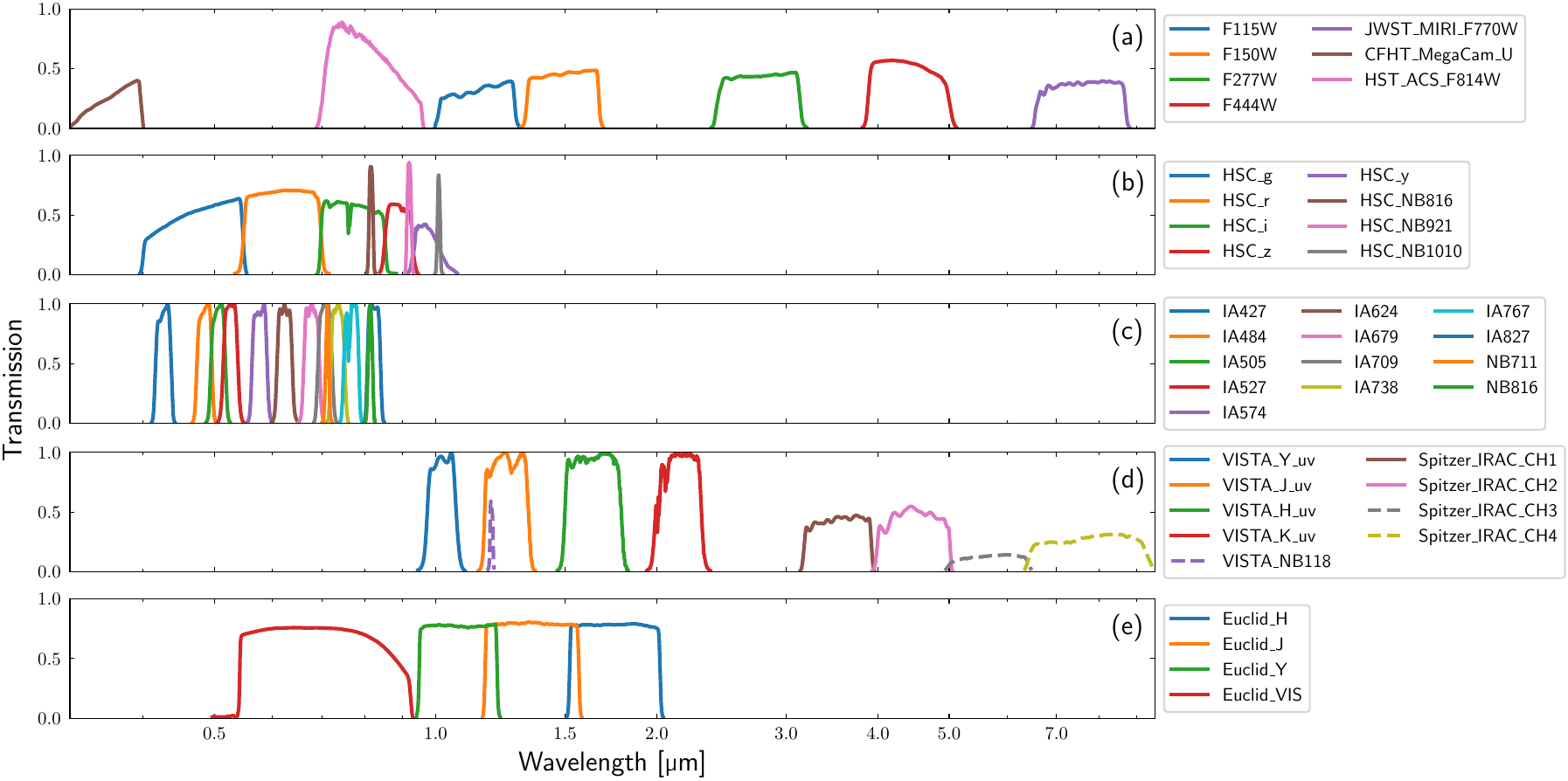}
    \caption{Transmission curves used for the SED-fitting procedure. (a) The five \JWST filters, \textit{U}-band from CFHT-MegaCam, and \HST-ACS F814W. (b) The five broadband filters, and three narrowband Hyper Suprime-Cam (HSC) filters. (c) 11 mediumbands filters and two narrowbands from the Subaru/Suprime-Cam. (d) The four broadband filters from VISTA-VIRCAM, the narrowband NR118 filter, and the four Spitzer-IRAC ones: the filters not used in the SED-fitting are represented with dashed lines. (e) The four \Euclid passbands. Each transmission curve is normalised to one at the peak during the SED-fitting.}
    \label{fig:zphot_filters}
\end{figure*}

\section{SED-fitting procedure}\label{sect:method}

The C25 photometric catalogue is complemented by an SED for each source. Since we construct our own SEDs, we perform a thorough comparison with those provided in C25 to make sure that no evident bias is introduced. A possible issue arises from the fact that the redshift of the C25 SED is the one that maximises the redshift probability distribution function PDF($z$), while the redshift (and fluxes) in the C25 catalogue are from the PDF($z$) median. The two redshifts can be very different, so we check that for the objects with a well constrained PDF($z$) no bias is introduced when using our SED instead of the C25 ones (more details in Appendix~\ref{app:lephare_sed_vs_zphot_sed}). After this check, we decide to adopt our SEDs as the reference to estimate the expected fluxes in the C25 and \Euclid bands.

We use the software \zphot~\citep[][]{Fontana2000, Grazian2006} to fit the observed fluxes from C25, estimate the predicted fluxes in the \Euclid bands, and obtain the \IE and NIR expected fluxes using the best-fit SED for each source, which is computed with a $\chi^2$ minimisation procedure. The details of the library used in the fitting procedure are given in Appendix~\ref{app:library}.

The 37 filters from C25 and the four \Euclid transmission filters used to create the synthetic galaxy library cover the wavelength range between $0.32\,\micron$ and approximately $10\,\micron$, using both narrow and broadband filters. Figure~\ref{fig:zphot_filters} shows all the filters: 
\begin{itemize}
    \item \JWST F115W, F150W, F277W, F444W from NIRCam and F770W from MIRI;
    \item CFHT \textit{U}-band imaging using MegaCam;
    \item \HST Advanced Camera for Surveys (ACS) using filter F814W;
    \item Subaru/Hyper Suprime-Cam (HSC) with $g$, $r$, $i$, $z$, $y$, NB0816, NB0921, and NB1010;
    \item Subaru/Suprime-Cam (SC) in 11 mediumbands (IB427, IA484, IB505, IA527, IB574, IA624, IA679, IB709, IA738, IA767, IB827), and two narrowbands (NB711, NB816);
    \item VISTA VIRCAM with \textit{Y}, \textit{J}, \textit{H}, \textit{Ks}, and NB118;
    \item \Spitzer channels 1, 2, 3, and 4 of IRAC.
    \item \Euclid VIS (\IE) and the three NISP (\YE, \JE, and \HE).
\end{itemize}
We use the official JWST database version 7.0 for the \JWST NIRCam filters (mean-system-throughput\footnote{\url{https://jwst-docs.stsci.edu/jwst-near-infrared-camera/nircam-instrumentation/nircam-filters\#gsc.tab=0}}); the remaining throughput curves were either fetched from the SVO Filter Profile Service\footnote{\url{https://svo2.cab.inta-csic.es/theory/fps/}} or from the COSMOS2020 website.\footnote{\url{https://cosmos.astro.caltech.edu/page/filterset}} The \Euclid \IE passband was taken from SVO, while the NISP photometric filters are from~\cite{Schirmer-EP18}. Filters from space-based telescopes combine the average detectors response with all the other telescope subsystems, while filters of ground-based instrument also include the average atmospheric transmission (1.2 airmass).

The VISTA NB118d narrowband, and the ch3 and ch4 IRAC bands, are too shallow to provide meaningful information to the SED-fitting procedure; for this reason, they were not used, as in C25. We include the ch1 and ch2 bands from IRAC, that were not used in C25, since they cover the range between 3 and 5 \micron, although they are not as deep as the \JWST F444W band and their fluxes have larger errors with respect to the \JWST bands. 

The $\chi^2$ minimisation procedure is performed on objects with ${\rm S/N} > 10$ in F115W, F150W, F277W, and F444W, as mentioned in Sect.~\ref{subsect:cosmoscat}. The \zphot runs provide, among other quantities, the modelled fluxes in each band and the reduced $\chi^2$ of the best-fit model. The effect of the S/N selection in the C25 \JWST bands is not predominant in the spec-$z$ case, but it can be seen for the photo-$z$ case in Fig.~\ref{fig:photoz_deep_wide_histograms_zphot_3} from the different distribution of the \zphot forecast magnitudes in the \Euclid bands.

\section{Analysis of the data set}\label{sect:analyses}
A comparison between the \Euclid forecast obtained from C25 and the MER DR1 final deep and wide catalogues is performed after a further selection based on the reduced $\chi^2$ values from \zphot (described in Sect.~\ref{subsect:selection}), a cross-matching between MER and C25 (Sect.~\ref{subsect:cross}), and a final cut based on the {\rm S/N} value in the MER catalogues (Sect.~\ref{subsect:high_snr_mer}).

\subsection{Selection criteria of \zphot \texorpdfstring{$\chi^2$}{chi2}}\label{subsect:selection}

\begin{figure}
    \centering
    \includegraphics[width=\linewidth]{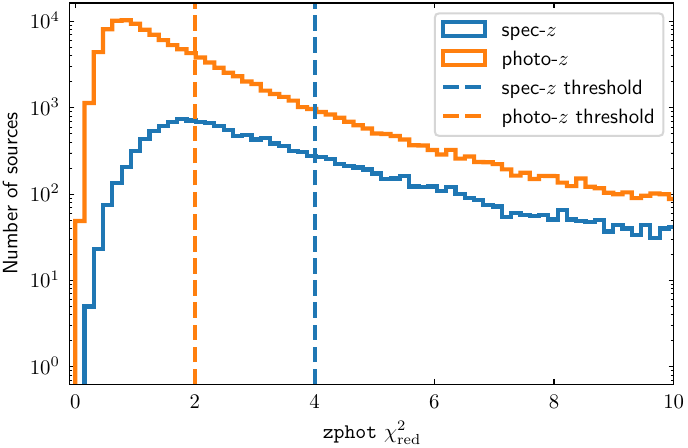}
    \caption{Histogram of the reduced $\chi^2$ values of the SED-fitting runs for the photo-$z$ (orange) and spec-$z$ (blue) samples. The distributions peak at $\chi_{\rm red}^2 \sim 2$ for the spec-$z$ sources, and at $\chi^2 \lesssim 1$  for the photo-$z$ ones. The number of degrees of freedom used to compute the reduced $\chi^2$ is 34, that is the number of filters used for the SED-fitting in \zphot.}
    \label{fig:photoz_specz_chi2}
\end{figure}

\begin{figure*}
    \centering
    \includegraphics[width=\linewidth]{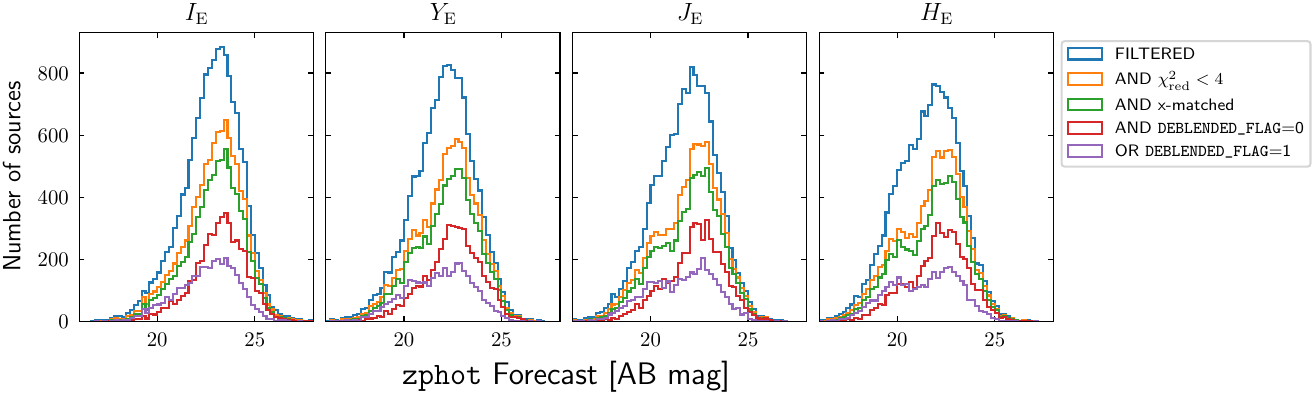}
    \caption{Distribution of sources in each \Euclid filter as a function of the \zphot forecast magnitude, for the spec-$z$ case. The histograms show the full initial sample of fitted sources (cyan), the sample after applying the $\chi_{\rm red}^2<4$ cut (orange), and the sample after cross-matching with the deep catalogue (green). Cross-matched sources are further divided into non-deblended (red) and deblended (purple) sources. Histograms for the very few NIR-detected sources are omitted, as they account for only a handful of objects. A similar plot for the photo-$z$ case is shown in Fig.~\ref{fig:photoz_deep_histograms_zphot}.}
    \label{fig:specz_deep_histograms_zphot}
\end{figure*}

\begin{figure*}
    \centering
    \includegraphics[width=\linewidth, trim=0 0 0 20, clip]{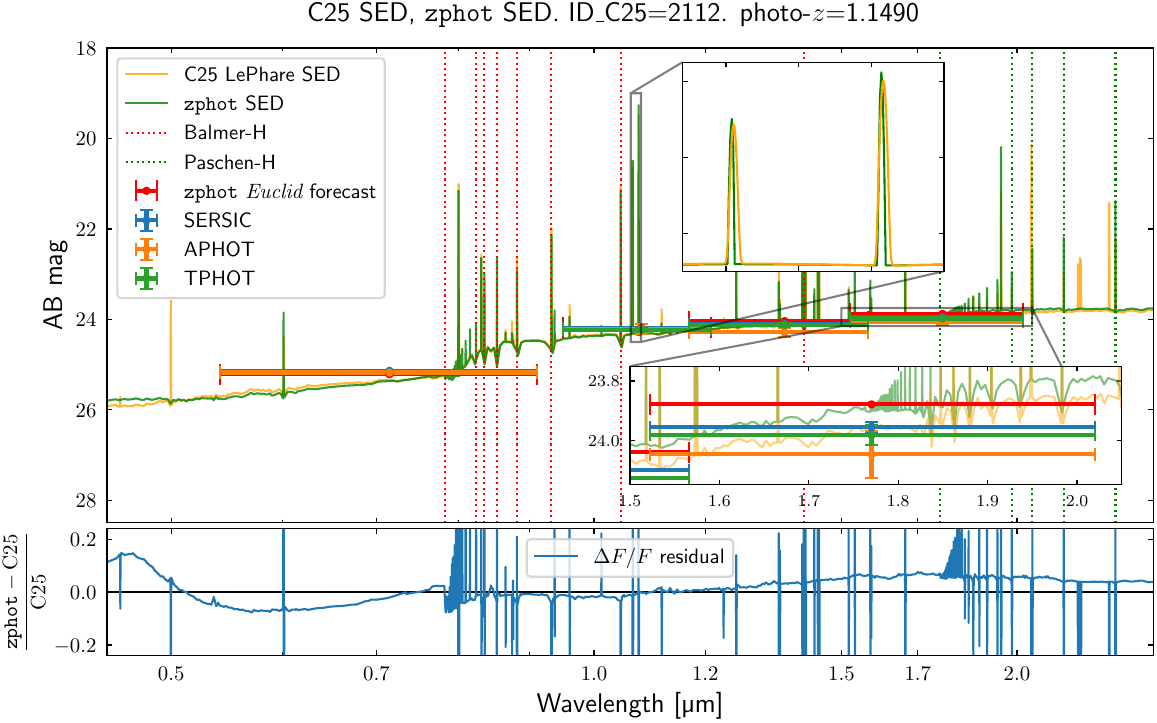}
    \caption{SEDs of a source (ID 2112 from C25, {\tt zfinal}=1.149) with good photometric redshift determination in COSMOS (i.e., with a well-constrained PDF($z$), see Sect.~\ref{sect:method}). The SED from C25 is in yellow, while the \zphot SED is in green, and the relative difference between the two SEDs is shown in the bottom panel; the upper-right corner shows a zoom-in on two emission lines ([\ion{O}{iii}]$\lambda$4958 and [\ion{O}{iii}]$\lambda$5007). The predicted magnitudes for the \Euclid filters from \zphot are in red. The error bars for the magnitudes along the $x$-axis show the wavelength ranges of the filter FWHM. The vertical dotted lines mark the redshifted Balmer (red) and Paschen (green) hydrogen emission lines. The zoomed-in window in the lower-right corner shows the \zphot forecast magnitude (red) for the \HE filter, and the values from the three MER photometric methods, namely model-fitting (blue, `SERSIC'), aperture (orange, `APHOT'), and template-fitting photometry (green, `TPHOT'), all of them plotted with their respective errors.} 
    \label{fig:SED_cosmos_zphot}
\end{figure*}

We initially considered applying pre-selection cuts to exclude stellar and faint sources; however, extensive testing demonstrated that this introduced spurious selection biases. We therefore retained the full catalogue to extend our analysis down to the faintest accessible magnitudes, relying on a stringent cross-matching between the catalogues (see Sect.~\ref{subsect:cross}) to ensure high sample purity and effectively filter out spurious detections.

The $\chi^2$ values resulting from the \zphot fitting are reduced $\chi^2$ values, $\chi_{\rm red}^2$. The initial $\chi^2$ is divided by the number of degrees of freedom, that is the number of filters used in the fitting (34 in our case). 

Since the fixed redshift in the spec-$z$ sample naturally leads to higher residual $\chi_{\rm red}^2$ compared to photo-$z$ fits where an additional degree of freedom allows the fits to artificially minimise the residuals, applying a single threshold would introduce an uneven selection. We therefore adopted $\chi_{\rm red}^2<4$ for spec-$z$ and $\chi_{\rm red}^2<2$ for photo-$z$ (Fig.~\ref{fig:photoz_specz_chi2}); these values represent the best compromise between fit reliability and sample completeness, retaining a homogeneous fraction of roughly 70\% of the best-fitted sources in each regime.

We checked that small changes in the value of the $\chi_{\rm red}^2$ threshold used for the selection cut has no impact on the final results; applying no cut at all on the $\chi_{\rm red}^2$ values increases the total number of sources used for the cross-match, especially the brighter ones (see Fig.~\ref{fig:specz_deep_histograms_zphot}), but the final results are only slightly worse.

Combining these selections with the previous ones we end up with \num{78880} photo-$z$s and \num{10443} spec-$z$s. The numbers of fitted sources and selected ones for the spec-$z$ case, against the expected \zphot magnitude, are reported in Fig.~\ref{fig:specz_deep_histograms_zphot}: a similar plot in which the ${\rm S/N} > 10$ selection in the \JWST bands from C25 is not performed would just add few hundred more sources at the fainter end. The photo-$z$ case is presented in Fig.~\ref{fig:photoz_deep_histograms_zphot}, where the case in which no S/N selection on the C25 \JWST bands is also applied.

\subsection{Cross-matching}\label{subsect:cross}
We finally cross-match the MER catalogues with C25 using \texttt{stilts}~\citep{topcat}, allowing for a maximum separation of \ang{;;0.1} to ensure a robust association. Since we select only galaxies from C25, and we use a small cross-match radius, we exclude most of the stars, transients, and spurious detections in the MER catalogues, while the number of underdeblended or overdeblended sources is only marginally reduced, since a positional match does not automatically guarantee a good deblending. The final samples comprise 8631 (\num{69729}) objects for the spec-$z$ (photo-$z$) MER deep catalogue, and \num{9081} (\num{54146}) for the spec-$z$ (photo-$z$) MER wide catalogue. The higher number of sources in the wide spec-$z$ case is discussed in Sect.~\ref{sect:discussion}. The distribution of sources after the cross-match for the spec-$z$ case is shown in Fig.~\ref{fig:specz_deep_histograms_zphot} (the photo-$z$ case is shown in Fig.~\ref{fig:photoz_deep_histograms_zphot}, where the selection cut on the S/N in the \JWST bands from C25 is both applied and not).

\subsection{Exclusion of low {\rm S/N} MER sources}\label{subsect:high_snr_mer}
We apply a final a-posteriori selection cut to the cross-matched sources, keeping only the objects with ${\rm S/N}>3$ in the band and photometric method we are considering. This should not be confused with the previous ${\rm S/N}>10$ cut which was performed on \JWST fluxes from C25: this additional ${\rm S/N}>3$ cut is performed on fluxes in the MER catalogues, to ensure an even more stringent quality of the selection.

This final sample results in slightly different numbers of objects across the various bands and photometric methods, reflecting both the differences between the algorithms and the source flux variation across the bands arising from the intrinsic shape of the SEDs of the sources.
An example of the resulting data set for a source with a good fit from \zphot, whose photometric redshift was well determined (PDF($z$) well constrained, see Sect.~\ref{sect:method}), is given in Fig.~\ref{fig:SED_cosmos_zphot}, where the predicted photometry and modelled SEDs from both \zphot and C25 are shown. 

\section{Results}\label{sect:results}

In this section we present the outcome of this study, comparing the fluxes in the \Euclid bands as predicted by the SED-fitting with the photometry measurements in the MER catalogues for the final selections of sources. The main results are presented in Sects.~\ref{sect:res_zphot}~and~\ref{sect:res_ext}, where we inspect the photometry measurements in the \Euclid bands and in the HSC bands, respectively, checking for biases between MER sources detected in \IE that are either affected or unaffected by the deblending algorithm (Sect.~\ref{subsect:debl_selection}). In Sect.~\ref{sect:zphot_goodness_checks} we present an analysis of the reliability of the \zphot fits, before the MER catalogues are considered, exploring different input fluxes from C25 and checking for any bias introduced by a sub-optimal modelling of some spectral features. 

In the main text we only present the analysis of the spec-$z$ case. Additional plots for the photo-$z$ case are shown in Appendix~\ref{app:res_zphot_MER_photoz_widedeep}. A complementary comparison between the \zphot forecast magnitudes and the convolution of the C25 SEDs with the \Euclid filters is discussed in Appendix~\ref{app:lephare_sed_vs_zphot_sed}. 


\subsection{Assessment of the MER photometry on the \Euclid filters}\label{sect:res_zphot}

\begin{figure*}
    \sidecaption
    \includegraphics[width=0.7\linewidth]{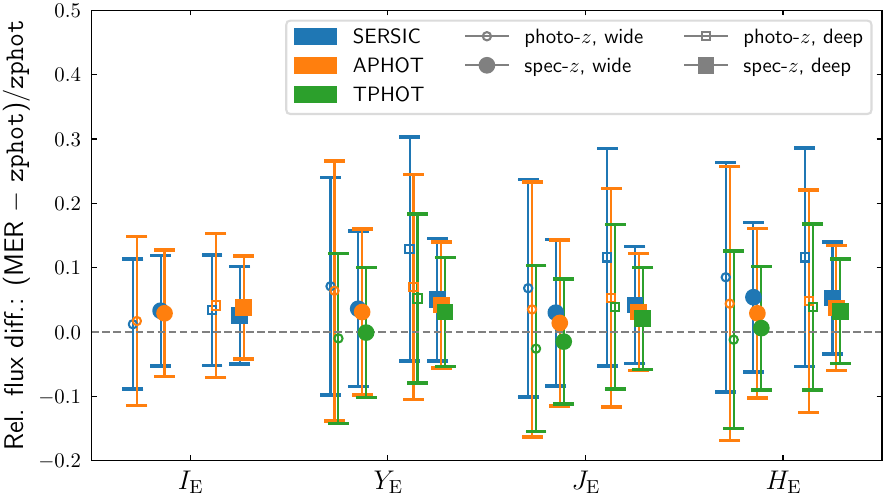}
    \caption{Median and normalised median absolute deviations (NMADs) of the relative flux differences between the MER fluxes and the \zphot forecast. These values are computed for all the methods (model-fitting (`SERSIC', blue), aperture photometry (`APHOT', orange), and template-fitting (`TPHOT', green)), depths (circle for wide, square for deep), redshift types (photo-$z$ small and empty, spec-$z$ larger and full), and \Euclid bands, for the sources detected in \IE, with {\tt DEBLENDED\_FLAG}=0, and ${\rm S/N}>3$ in the MER catalogue. The offset are computed for \zphot forecast magnitudes ranging from 18 to a different value for each band and catalogue depth, namely 25 (27) and 24 (25) for \IE and NIR bands  respectively, in wide (deep) mode. Figure~\ref{fig:offset_euclid_ALL_0_1_ALL_cumulative} shows the offset with respect to the \zphot magnitudes.}
    \label{fig:summary_plot}
\end{figure*}

Here we present the main result of this work: the comparison between the forecast from \zphot and the MER fluxes for the spec-$z$ sample, in each \Euclid filter, and for the three photometric methods (total fluxes from aperture photometry, model-fitting, and template-fitting). As mentioned, we only considered sources detected in \IE (\texttt{VIS\_DET}=1), with ${\rm S/N}>3$ in the MER bands, and that are not deblended in detection (\texttt{DEBLENDED\_FLAG}=0).\footnote{As described in \citetalias{Q1-TP004}, each object in the catalogue is assigned a \texttt{DEBLENDED\_FLAG} value of either 0 or 1. A value of 0 indicates that the source was not affected by the deblending algorithm, whereas a value of 1 indicates that the object was initially contaminated by nearby objects in the detection image, and was subsequently separated from them by the deblending algorithm. The MER pipeline employs {\tt SE++} for source detection, while deblending is performed using {\tt ASTErIsM}~\citep{Tramacere2016}.}

Since all the objects with spec-$z$ are bright enough to be detected in \IE, only 11 NIR-detected sources (\texttt{VIS\_DET}=0) are left after the cross-match with the deep catalogue (95 in the cross-match with the wide one). Since they are too few for the purpose of this work, and many are flagged as spurious, we chose to exclude them from the analysis. Similarly, for the photo-$z$ case we only focus on VIS-detected objects, as the NIR-detected ones left after the selection cuts are less than 1\% in the deep cross-match (less than 15\% in the wide).

We point out that the \IE flux in the `APHOT' row of the plots is \texttt{FLUX\_DETECTION\_TOTAL}, that is the flux within an ellipse with semimajor axis equal to 2.5 Kron radii (computed using equation 2 from \citep{Q1-TP004}). To obtain the flux in each NIR band, we scale the flux within a 2 FWHM aperture in the NIR band with the ratio between the Kron flux in \IE and the flux within an aperture of 2 FWHM in \IE (with both 2 FWHM fluxes measured on images PSF-matched to the worst PSF of the data set, typically a ground-based band from EXT). Furthermore, by construction there is no TPHOT measurement for \IE, and the related plot is intentionally left empty. More details are given in PCQ1.

A summary of the medians and normalised median absolute deviations (NMADs) of the resulting relative flux biases (measured versus predicted by \zphot) for the \Euclid bands, and for all methods, depths, and redshift types, are shown in Fig.~\ref{fig:summary_plot}. The magnitude ranges considered to compute the plotted values vary for depending on the bands and the depths, the minimum set to 18 and the maximum equal to 25 (27) and 24 (25) for \IE and NIR bands respectively, in wide (deep) mode . Significantly, the values are well consistent for all methods and bands. In almost all cases we find small positive offsets, with typical values of a few percent and in all cases well below 10\%. The medians are also reported in Table~\ref{tab:summary_table}, while the median offsets for all the methods and band as a function of the \zphot predicted magnitudes are shown in Fig.~\ref{fig:offset_euclid_ALL_0_1_ALL_cumulative}.

\begin{figure*}
    \centering
    \includegraphics[width=\linewidth]{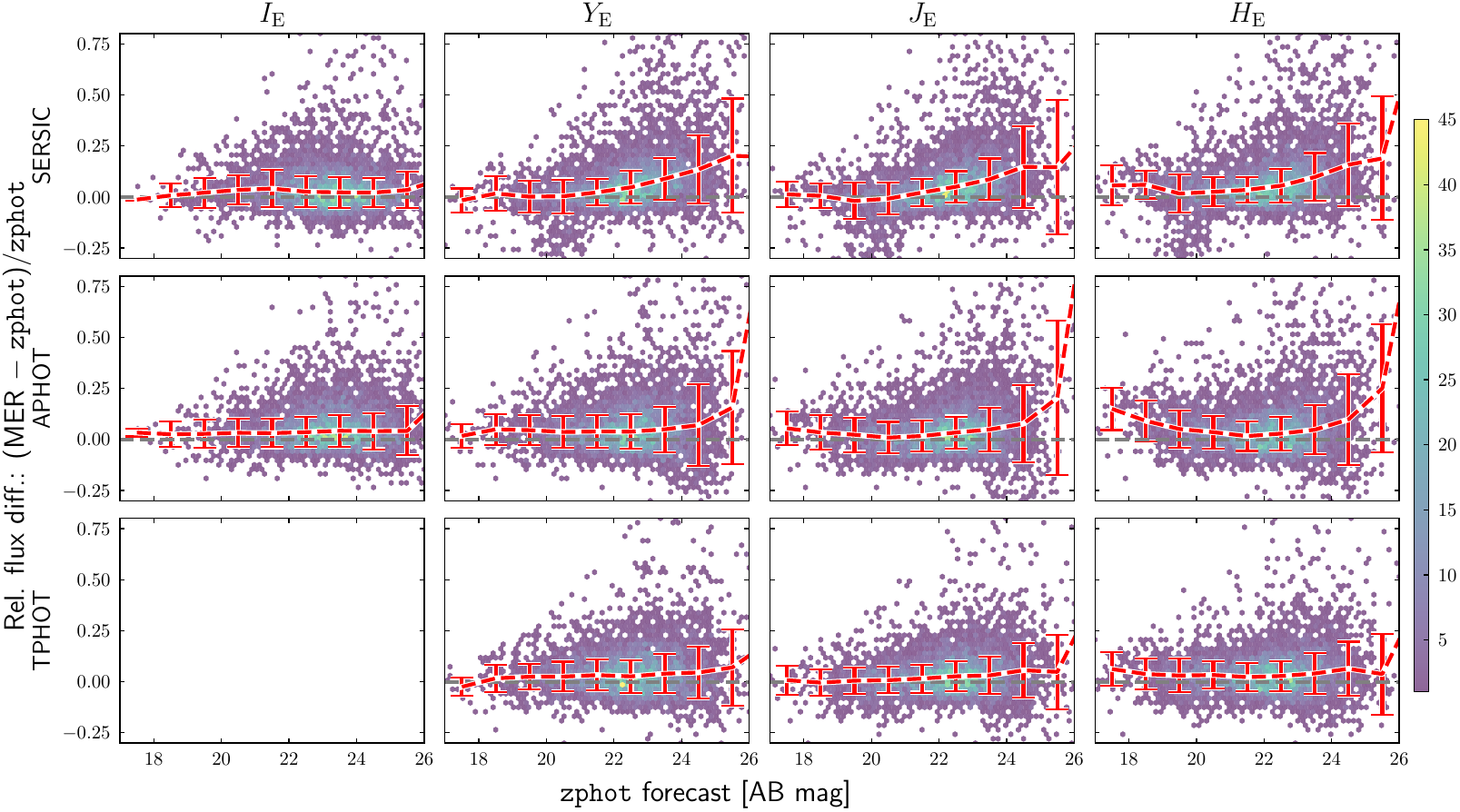}
    \includegraphics[width=\linewidth]{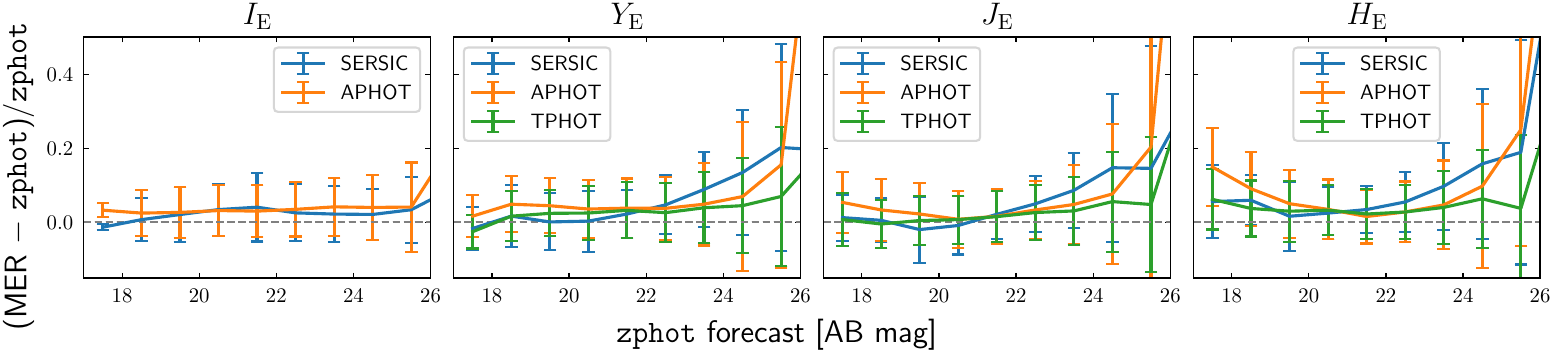}
    \includegraphics[width=\linewidth]{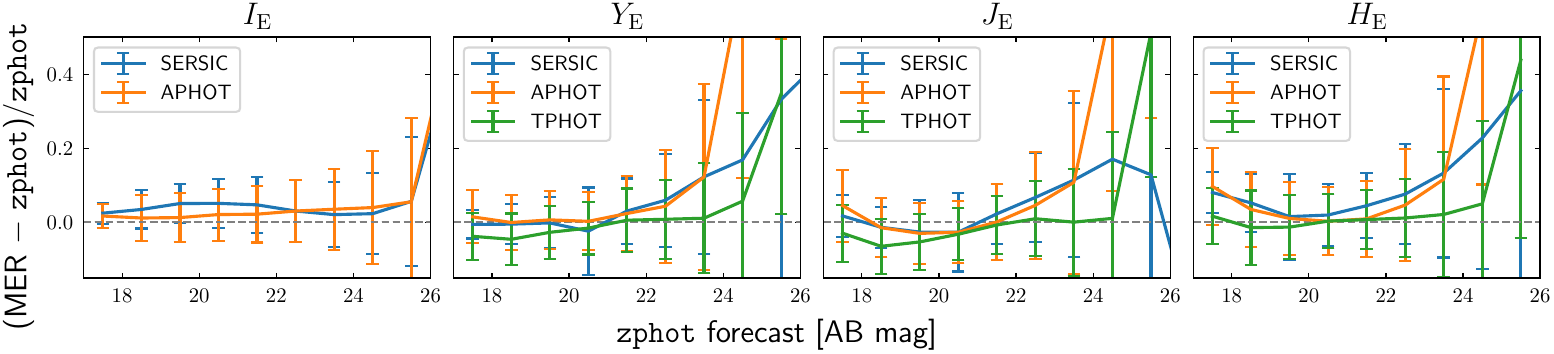}
    \caption{\textit{Top}: Relative flux differences between the MER fluxes and the \Euclid forecast from \zphot for the spec-$z$ sample. The following MER measurements from the deep catalogue are shown: model-fitting (`SERSIC'), aperture photometry (`APHOT'), and template-fitting (`TPHOT'). Only sources with \texttt{VIS\_DET}=1, \texttt{DEBLENDED\_FLAG}=0, and ${\rm S/N}>3$ in MER are shown in these plots. The medians and the normalised median absolute deviations (${\rm NMAD}=1.4826\times{\rm median}\lbrack|{\Delta F/F}_i - {\rm median}({\Delta F/F}_i)|\rbrack$, ${\Delta F/F}_i$ being the relative flux difference for each magnitude bin) are also reported for each method and band in red. By construction there is no TPHOT measurement for \IE, and the related plot is intentionally left empty. \textit{Middle}: Summary version of the upper plot, in which only the median (red) lines and the NMADs are reported. \textit{Bottom}: Comparison between the \Euclid forecast and the wide MER catalogue provided for the same bands and methods and with the same selection cuts. The medians in the middle and bottom panels are also reported in the top and middle panel of Fig.~\ref{fig:specz_deep_wide_histograms_zphot_3}. The plots obtained for the photo-$z$ case are shown in Fig.~\ref{fig:offset_euclid_wide_0_1}.}
    \label{fig:offset_euclid_0_1_specz}
\end{figure*}

\subsubsection{MER deep photometry}\label{subsect:res_specz_deep}
The detailed comparison for the spec-$z$ sample on the MER deep catalogue, for individual bands and photometric methods, is shown in the upper and middle panels of Fig.~\ref{fig:offset_euclid_0_1_specz}. The overall consistency is good, especially at bright magnitudes, for all three methods and in all bands. As anticipated, we find a positive average bias of the MER fluxes with respect to the \zphot forecast, by up to 5\% at magnitudes brighter than 23. Almost all the methods and bands show no average trend over a large magnitude range, with model-fitting presenting a larger bias at the faint end of the NISP bands with respect to aperture and template-fitting. Aperture photometry yields a small positive bias at the very bright end. While SERSIC has the smallest offset in \IE, TPHOT yields the lowest in the three NIR bands, with almost no trend with magnitudes.

Only roughly $6\%$ of sources in \IE and around $2\%$ in the NIR bands are fainter than AB magnitude 25 (the distribution for deblended and not deblended objects for the spec-$z$ deep case is shown in Fig.~\ref{fig:specz_deep_histograms_zphot}, while the photo-$z$ case is shown in Fig.~\ref{fig:photoz_deep_histograms_zphot}).  
All sources in the spec-$z$ case have ${\rm S/N}>3$ in \IE both for SERSIC and APHOT, and in the NIR bands most of the sources have ${\rm S/N}>3$ for all the methods as well. Only a handful of sources have ${\rm S/N}<3$ at magnitudes brighter than 25, the worst case being APHOT in \YE with 27 objects, their flux not being well determined (see also Fig.~\ref{fig:specz_deep_wide_histograms_zphot_3}). The medians of these plots are also reported in the top panel of Fig.~\ref{fig:specz_deep_wide_histograms_zphot_3}.

\subsubsection{MER wide photometry}\label{subsect:res_specz_wide}
The results for the MER wide catalogue are shown in the bottom panel of Fig.~\ref{fig:offset_euclid_0_1_specz} (the medians of these plots are also reported in the middle panel of Fig.~\ref{fig:specz_deep_wide_histograms_zphot_3}). The average offsets are flat for \IE in the whole magnitude range, with model-fitting showing a 5\% offset at the brighter magnitudes. 

However, all three methods show an upturn at the faint end of  the NISP bands, much more pronounced than in the deep case. This difference between wide and deep can be seen also in the photo-$z$ case shown in Appendix~\ref{app:res_zphot_MER_photoz_widedeep}. We attribute this to an Eddington-like bias, which we further discuss in Sect.~\ref{subsect:sel_bias_eddington}.

Template-fitting magnitudes show no bias down to about AB 24, while the other methods can reach a positive 10\% median difference with respect to the \zphot forecast. On the other hand, template-fitting slightly underestimates the flux of the brightest sources, where the other two methods show an offset compatible with zero. Importantly, both results (which are also evident in the photo-$z$ case, Fig.~\ref{fig:offset_euclid_wide_0_1}) are consistent with those found when using synthetic source injection on Euclid data (Euclid Collaboration: Kong et al., in prep). 

It is interesting to note that the flux bias we find in this analysis, obtained from photometry validation of extragalactic sources, has the opposite trend with respect to that from the routine MER validation performed on \gaia stars: while here the fluxes appear overestimated with respect to SED-fitting predictions, the fluxes of the \gaia stars are typically lower than expected, especially for \YE, \JE, and \HE (see e.g., the photometry overview plot in figure~10 from \citetalias{DR1-TP009}). The two methods are radically different; we further investigate this issue in the following subsection, while the effect of the applied selections is discussed in Sect.~\ref{sect:selection_effect}.

\subsection{Assessment of the MER photometry in the HSC bands}\label{sect:res_ext}

\begin{figure*}
    \centering
    \includegraphics[width=\linewidth]{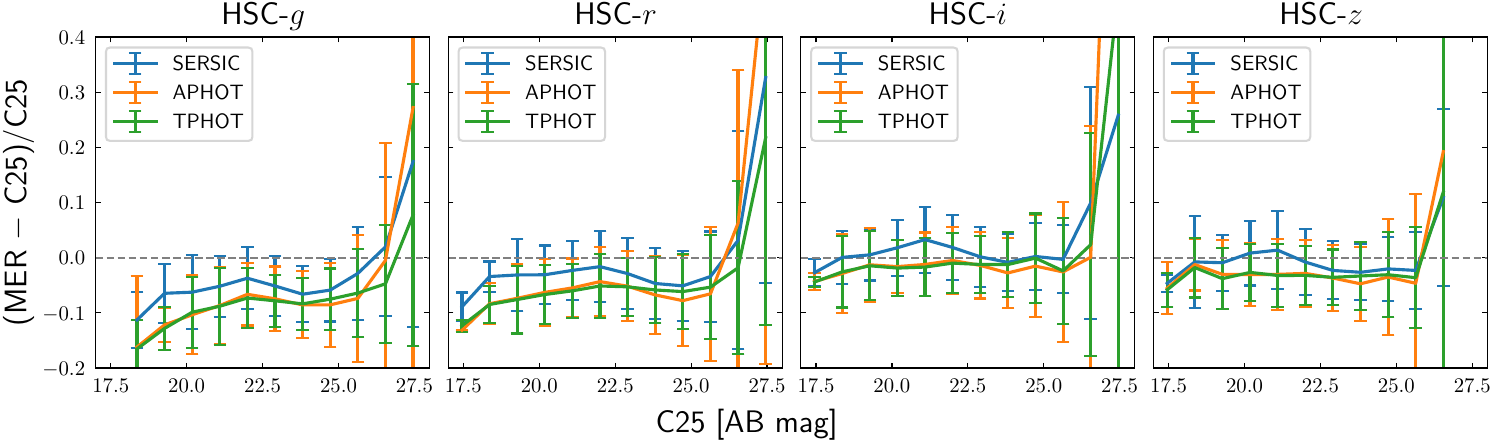}
    \includegraphics[width=\linewidth, trim=0 0 0 20, clip]{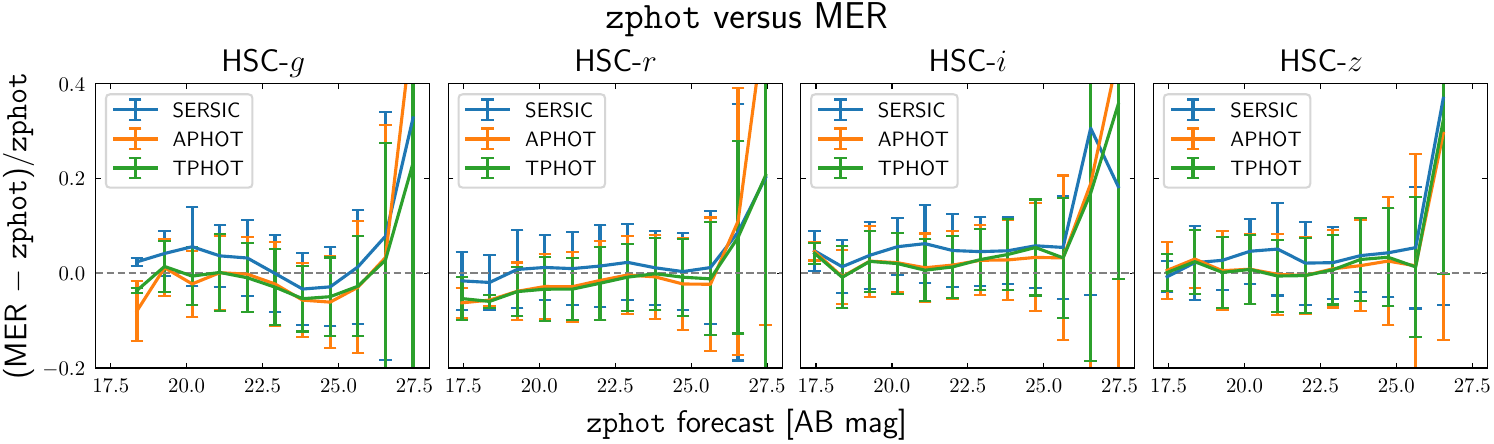}
    \caption{Same as Fig.~\ref{fig:offset_euclid_0_1_specz} for the HSC bands, namely HSC~$g$, $r$, $i$, and $z$. \textit{Top}: Relative flux differences between the MER fluxes and the C25 ones in the four HSC bands for model-fitting (`SERSIC'), aperture photometry (`APHOT'), and template-fitting (`TPHOT'). Only the median lines with the NMADs for the sources with \texttt{VIS\_DET}=1, \texttt{DEBLENDED\_FLAG}=0, and with ${\rm S/N} > 3$ in MER are shown in these plots. 
    \textit{Bottom}: While the upper panel directly compares MER and C25, these plots show a comparison between the MER fluxes and the \zphot forecast. The largest difference is seen in the HSC~$g$ band at the brighter end, while the other bands behave in the same way, within the errors, in the two panels. More details in Sect.~\ref{sect:hsc_discussion}. The photo-$z$ case is shown in Fig.~\ref{fig:offset_HSC_cosmos_mer_photoz}.}
    \label{fig:offset_HSC_cosmos_mer}
\end{figure*}

A direct comparison between C25 and MER can be performed for the four HSC bands that the two catalogues have in common, namely HSC~$g$, $r$, $i$, and $z$, only for the deep observations. Here we consider the same sample of sources used for the \Euclid bands comparison. In all the plots the objects are detected in \IE (\texttt{VIS\_DET}=1), not deblended (\texttt{DEBLENDED\_FLAG}=0), and have ${\rm S/N} > 3$ in the MER catalogue for the considered band (upper panel in Fig.~\ref{fig:offset_HSC_cosmos_mer}). 

Noticeably, with all the photometric methods, the MER fluxes (the only exception being the SERSIC fluxes in HSC~$i$) appear to be underestimated with respect to C25, at the level of 5--8\% for HSC~$g$ and HSC~$r$ between magnitudes 21 and 25. The offsets are less pronounced in the HSC~$i$ and HSC~$z$ bands, with a particularly good agreement in HSC~$i$ for all methods. No evident trend between magnitudes 21 and 25 is seen, for all methods and bands.

We then compared the predictions from \zphot with the MER catalogue in the same four HSC bands, in order to assess any bias introduced by the SED-fitting procedure (lower panel of Fig.~\ref{fig:offset_HSC_cosmos_mer}). The comparison reveals a clear excess at bright magnitudes in the HSC~$g$ band, whereas the other bands show a reasonably flat behaviour up to magnitude 25. Overall, the offsets that are below or consistent with zero in the direct comparison between C25 and MER, become consistent with zero or positive in the comparison between \zphot and MER. We interpret this as an evidence that, at least in the four HSC bands, \zphot tends to yield slightly underestimated fluxes. If this is the case, and the effect can be extended to all bands, then the positive bias discussed in Sects.~\ref{subsect:res_specz_deep} and \ref{subsect:res_specz_wide} might be mainly due to this effect. We further discuss this issue in Sect.~\ref{sect:hsc_discussion}. 

As we already mentioned in Sect.~\ref{subsect:res_specz_wide}, as we move to the faint end in Fig.~\ref{fig:offset_HSC_cosmos_mer} the MER fluxes appear to be more and more overestimated with respect to the \zphot forecast; we further discuss this feature in Sect.~\ref{subsect:sel_bias_eddington}.

\begin{figure*}
    \centering
     \begin{subfigure}[][][b]{\textwidth}
        \centering
        \includegraphics[width=0.2\textwidth]{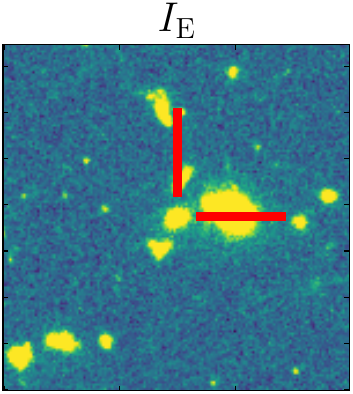}
        \includegraphics[width=0.2\textwidth]{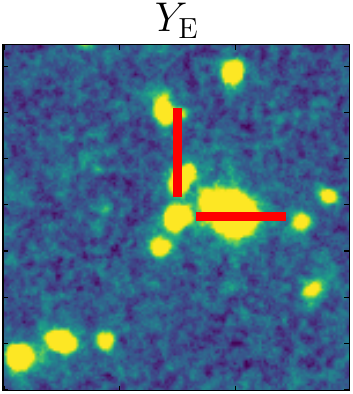}
        \includegraphics[width=0.2\textwidth]{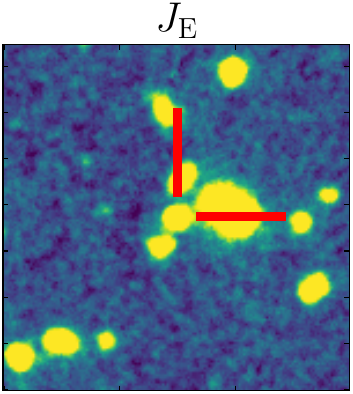}
        \includegraphics[width=0.2\textwidth]{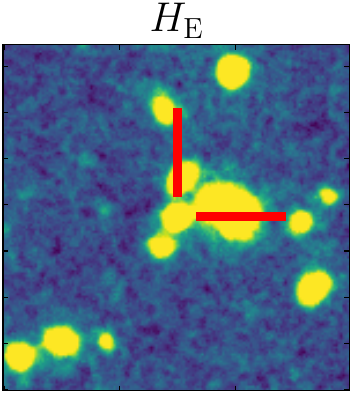}
    \end{subfigure}

     \begin{subfigure}[c][][]{0.32\textwidth}
        \centering
        \begin{picture}(140,140)
        \put(15,0){\includegraphics[width=0.95\textwidth, trim=45 52 45 40, clip]{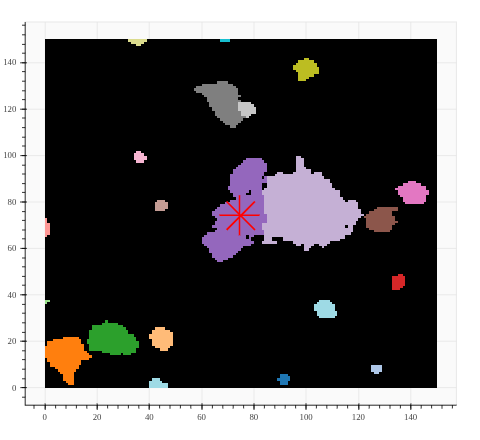}}
        \put(0,5){\rotatebox{90}{DR1 $-$ deep}}
        \end{picture}
     \end{subfigure}
     \hspace{15mm}
     \begin{subfigure}[c][][]{0.5\textwidth}
        \includegraphics[width=\textwidth]{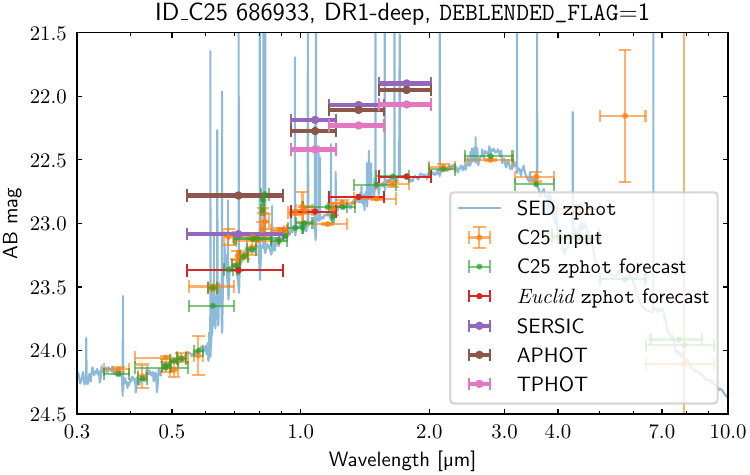}
     \end{subfigure}
     
     \begin{subfigure}[c][][]{0.32\textwidth}
         \centering
        \begin{picture}(140,140)
        \put(15,0){\includegraphics[width=0.95\textwidth, trim=45 52 45 40, clip]{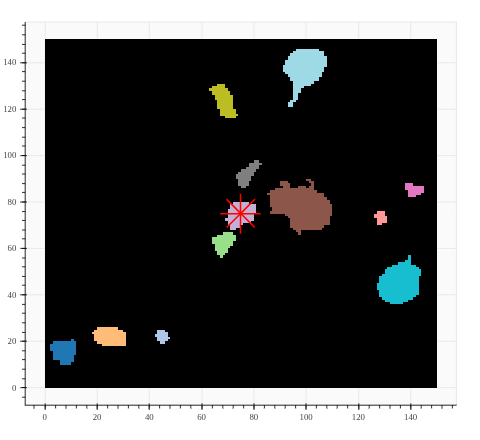}}
        \put(0,5){\rotatebox{90}{DR1 $-$ wide}}
        \end{picture}
     \end{subfigure}
     \hspace{15mm}
     \begin{subfigure}[c][][]{0.5\textwidth}
        \includegraphics[width=\textwidth]{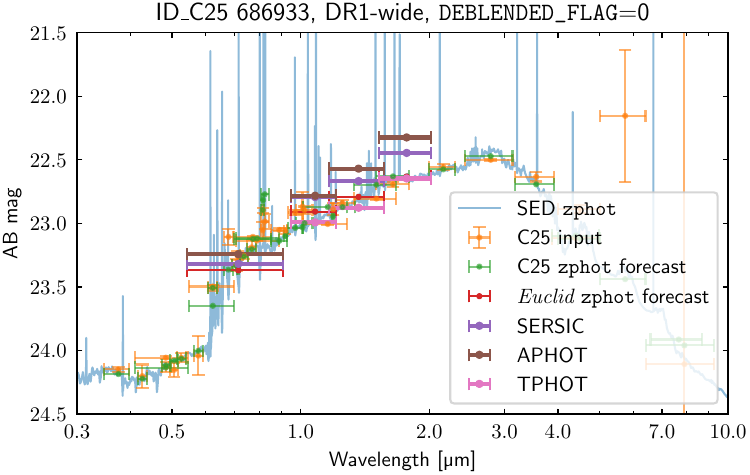}
     \end{subfigure}

     \begin{subfigure}[c][][]{0.32\textwidth}
        \centering
        \begin{picture}(140,140)
        \put(15,0){\includegraphics[width=0.95\textwidth, trim=45 52 45 40, clip]{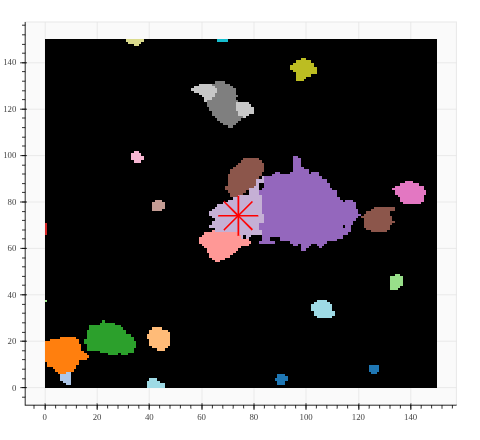}}
        \put(0,5){\rotatebox{90}{DR1 $-$ deep, Deblending update}}
        \end{picture}
     \end{subfigure}
     \hspace{15mm}
     \begin{subfigure}[c][][]{0.5\textwidth}
        \includegraphics[width=\textwidth]{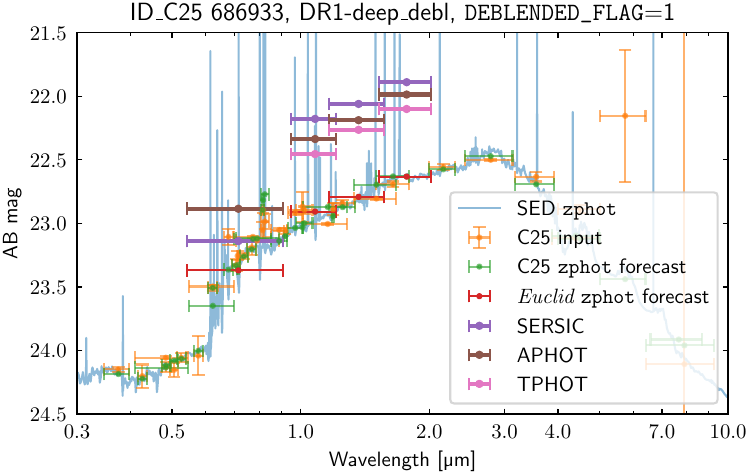}
     \end{subfigure}    
    \caption{Cutouts, segmentation maps, and photometry measurements for one example object (ID~\num{686933} from C25). {\it First row}: cutouts ($150\times150$ pixels) from the deep \IE, \YE, \JE, and \HE MER mosaics centred on the source. {\it Second row}: data from the Euclid DR1 deep imaging with the segmentation map (left, $150\times150$ pixels) and the various input values from C25 (orange), the spec-$z$ \zphot fit for \Euclid (red), the MER measurement from `SERSIC', `APHOT', `TPHOT', and the SED model (light blue line) from \zphot. The error bars along the $x$-axis for the photometric point is the FWHM of the filters. {\it Third row}: same as above, but for wide data, where the source was not blended with other nearby objects ({\tt DEBLENDED\_FLAG}=0). {\it Fourth row}: same as above, but data come from the reprocessing of COSMOS DR1 deep where we used the same code as for DR1 but with a different deblending configuration for {\tt ASTErIsM}.}
    \label{fig:deep_vs_wide_crossmatch}
\end{figure*}

\begin{figure}
    \centering
    \includegraphics[width=\linewidth]{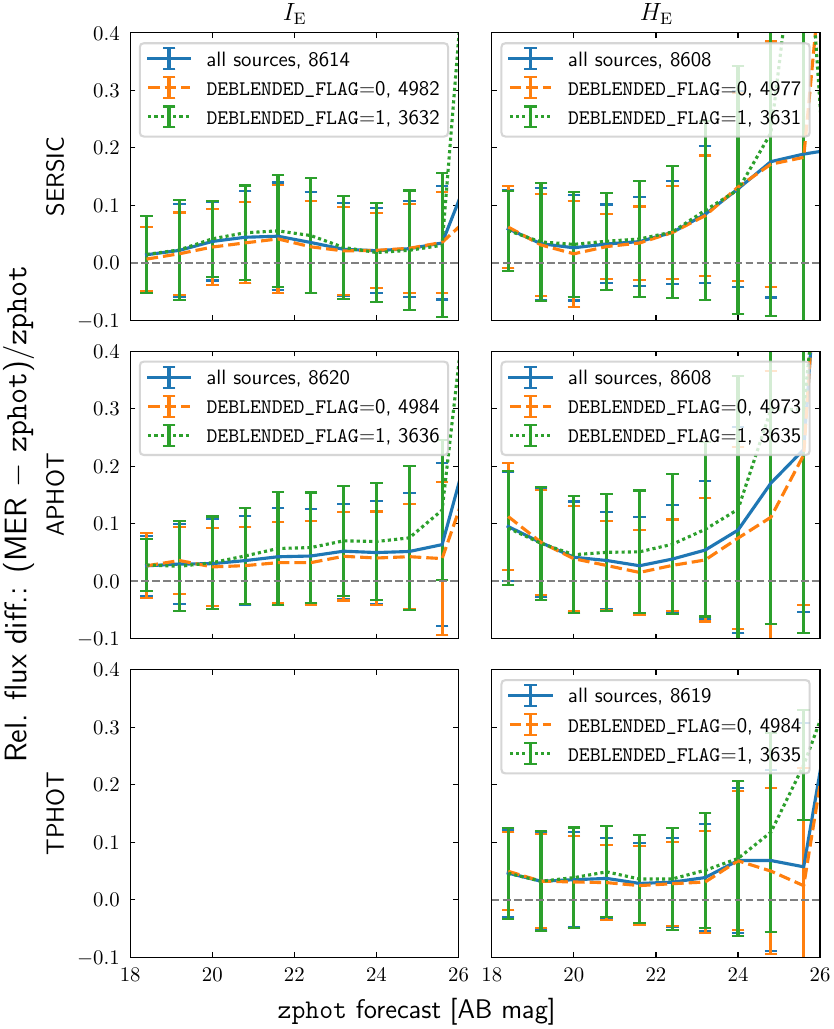}
    \caption{Relative flux differences between the MER fluxes and the \zphot fluxes as a function of the \zphot magnitude for \IE and \JE for VIS-detected sources, either for all sources, or for those with \texttt{DEBLENDED\_FLAG} equal to 0 or 1. The relative differences for \YE and \HE are similar to those shown for \JE and are not shown. The numbers in the legends of the plots show the numbers of sources that are used for each plot. By construction there is no TPHOT measurement for \IE, and the related plot is intentionally left empty. The photo-$z$ case is shown in Fig.~\ref{fig:MER_visdet_debl_photoz_deep}.}
    \label{fig:MER_visdet_debl_specz_deep_VIS_J}
\end{figure}

\subsection{Effect of blending}\label{subsect:debl_selection}

As shown in Sect.~\ref{subsect:cross}, in the spec-$z$ case we cross-match \num{9081} sources for the wide catalogue and \num{8631} sources for the deep one; in principle, however, one could expect more matches in the deeper images. While the faintest sources are obviously only present in the deep catalogue, many more objects turn out to be missing from it. After a thorough investigation, we found that this is due to a suboptimal deblending configuration used for the processing of the deep fields in DR1, which is a known issue within OU-MER (Sect.~\ref{sect:discussion}). This results both in a reduced number of sources in the strict cross-match, and in an overestimation of their flux (because the source was not deblended and its segmentation map includes pixels from neighbouring objects). As an example, Fig.~\ref{fig:deep_vs_wide_crossmatch} shows cutouts, segmentation maps, and photometry plots for an isolated and well-behaved object in wide data, which appears blended to a close neighbour in the deep data because of the more extended halos, and was indeed not deblended by the MER-PF. 

\begin{figure}
    \centering
    \includegraphics[width=\linewidth]{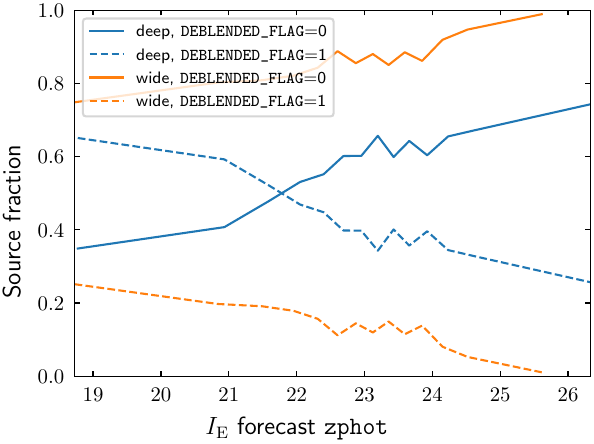}
    \caption{Fraction of sources with spectroscopic redshift, cross-matched with both the deep and wide catalogues, that are either not deblended ({\tt DEBLENDED\_FLAG}=0) or have been deblended ({\tt DEBLENDED\_FLAG}=1). The fraction of deblended sources is higher at the bright end and decreases towards fainter magnitudes.}
    \label{fig:specz_debl_fraction}
\end{figure}

The reduced number of cross-matched sources between deep and wide is observed only in the spec-$z$ case and not in the photo-$z$ case (\num{69729} and \num{54146} objects respectively). Moreover, the percentage of matched galaxies with {\tt DEBLENDED\_FLAG}=1 in the MER deep catalogue is approximately 40\% for the spec-$z$ case, and it is reduced to around $20\%$ for the photo-$z$ case. The effect of the non-optimal deblending is present but not dominant in the photo-$z$ case because non-spectroscopic sources are typically faint and small, and therefore less likely to overlap to other objects if they are detected; conversely, brighter and larger sources end up blending more frequently (see Fig.~\ref{fig:specz_debl_fraction}).

In Fig.~\ref{fig:MER_visdet_debl_specz_deep_VIS_J} we show a comparison between the \zphot forecast and the MER measurements for \IE and \JE (photo-$z$ case in Fig.~\ref{fig:MER_visdet_debl_photoz_deep}, with larger statistics and all the \Euclid bands), with different \texttt{DEBLENDED\_FLAG} selections, namely \texttt{DEBLENDED\_FLAG}=1, \texttt{DEBLENDED\_FLAG}=0, and no selection. Objects with \texttt{DEBLENDED\_FLAG}=1 generally show larger offsets, although the effect is strongly method-dependent: it is most evident for APHOT and substantially smaller for SERSIC and TPHOT, the latter being the least biased at the fainter magnitudes.

\subsection{Comparison of COSMOS2025 and \zphot photometry}\label{sect:zphot_goodness_checks}

We also performed a safety check to assess the reliability of the fit, comparing the input fluxes from C25 with the \zphot predicted fluxes for all the C25 bands. The median relative flux differences for the 37 C25 bands are within 10\% for most of the bands (Fig.~\ref{fig:offset_cosmos_bands}), with no clear wavelength trend, meaning that our SED-fitting method is not biased in any given wavelength range.

Strong spectral features, such as breaks and emission lines, might introduce biases in the \zphot estimates, because they can be challenging to model with photometric SED-fitting (Fig.~\ref{fig:hsc_y_redshift_offset}). Characterising the emission-line properties and their impact on \zphot is beyond the scope of this work, which instead focuses on assessing the quality and consistency of the DR1 MER photometry; however, we checked whether the exclusion of redshift ranges corresponding to some selected spectral features in the considered band would change the overall comparison. We found that excluding the redshift ranges with strong spectral features within the observed bands has a negligible effect on the final results (Fig.~\ref{fig:offset_z_filter_lines_photoz_0_1_deep}), so the comparison that we presented in this paper use the full redshift range.

\section{Discussion}\label{sect:discussion}

On top of the broad analysis of the accuracy of photometric data in the DR1 catalogues, the results presented in the previous section provide a basis for an assessment of the main limitations of this work, and their impact on the achievable precision. In this section we review some of the choices and assumptions we made, and their impact on the final results.

\subsection{COSMOS25 input}
A first important caveat is that all the results presented in this work are based on the C25 model-fitting photometry obtained with 2D \sersic profiles, which provides self-consistent flux measurements across all 37 bands. It is therefore worthwile stressing that there is no real `ground-truth' against which we can compare to actually assess how good a measurement is.

We choose to use C25 as a reference for our comparisons since it includes measurements from several instruments, and in particular exquisite data from \JWST observations. Their depth and resolution enable an accurate characterisation of individual sources and mitigate the effects of source blending. However, the photometry measurements performed on the \JWST images can still be affected by photometric calibration uncertainties, sub-optimal background subtraction and local sky brightness variations, modelling assumptions, and other observational systematic effects. 

\subsection{\zphot bias from HSC photometry}\label{sect:hsc_discussion}
The results from the assessment of the photometry in the EXT HSC bands presented in Sect.~\ref{sect:res_ext} show that moving from a direct comparison between C25 and MER to that between \zphot and MER, the measured offsets increase by 5\% to 10\%, suggesting that \zphot is underestimating the fluxes and that it is introducing a bias. This is observed not only in the spec-$z$ case in Fig.~\ref{fig:offset_HSC_cosmos_mer}, but also in the photo-$z$ case from Fig.~\ref{fig:offset_HSC_cosmos_mer_photoz} which consistent with Fig.~\ref{fig:offset_cosmos_bands}, where the {\tt zphot} forecast in the HSC bands is 3\% to 5\% below the C25 input fluxes.

While we only see that \zphot is underestimating the flux for the HSC bands, we can try to understand its effect in the \Euclid bands, assuming that \zphot is underestimating the expected fluxes for them as well: this argument is more valid in \IE, since this band overlaps with the common HSC bands we inspected, while is less demonstrable for the NISP bands. That said, we have shown that comparing the forecast from \zphot with the actual measurements reported in the MER deep catalogue for the \Euclid bands (middle panel in Fig.~\ref{fig:offset_euclid_0_1_specz}) the fluxes in MER are overestimated by up to 5\% between magnitude 21 and 25. We conclude that assuming \zphot is underestimating the flux in the \Euclid bands such positive offsets could be reduced substantially reduced. 

\subsection{Selection effects}\label{sect:selection_effect}
The selections criteria presented in Sect. \ref{sect:analyses} can potentially affect the trends we found. We discuss them here, checking whether they are actually introducing any selection bias. In Fig.~\ref{fig:specz_deep_histograms_zphot} we show the sources distributions after the different selection cuts are applied, for the spec-$z$ case (the photo-$z$ case is shown in Fig.~\ref{fig:photoz_deep_histograms_zphot}), including the distribution of objects cross-matched with the deep catalogue which are deblended ({\tt DEBLENDED\_FLAG}=1) or are left unaltered by the deblending algorithm ({\tt DEBLENDED\_FLAG}=0). We see that the ${\rm S/N} > 10$ selection on the \JWST bands for the spec-$z$ case only removes few hundreds of sources at the faint end of the \zphot forecast magnitude distributions (not plotted because the effect is hardly visible), and thus does not affect the overall results. The $\chi^2_{\rm red} < 4$ cut mostly affects the brighter sources, while the cross-match removes objects across the full magnitude range with no evident selection effect.

\begin{figure*}
    \includegraphics[width=\linewidth]{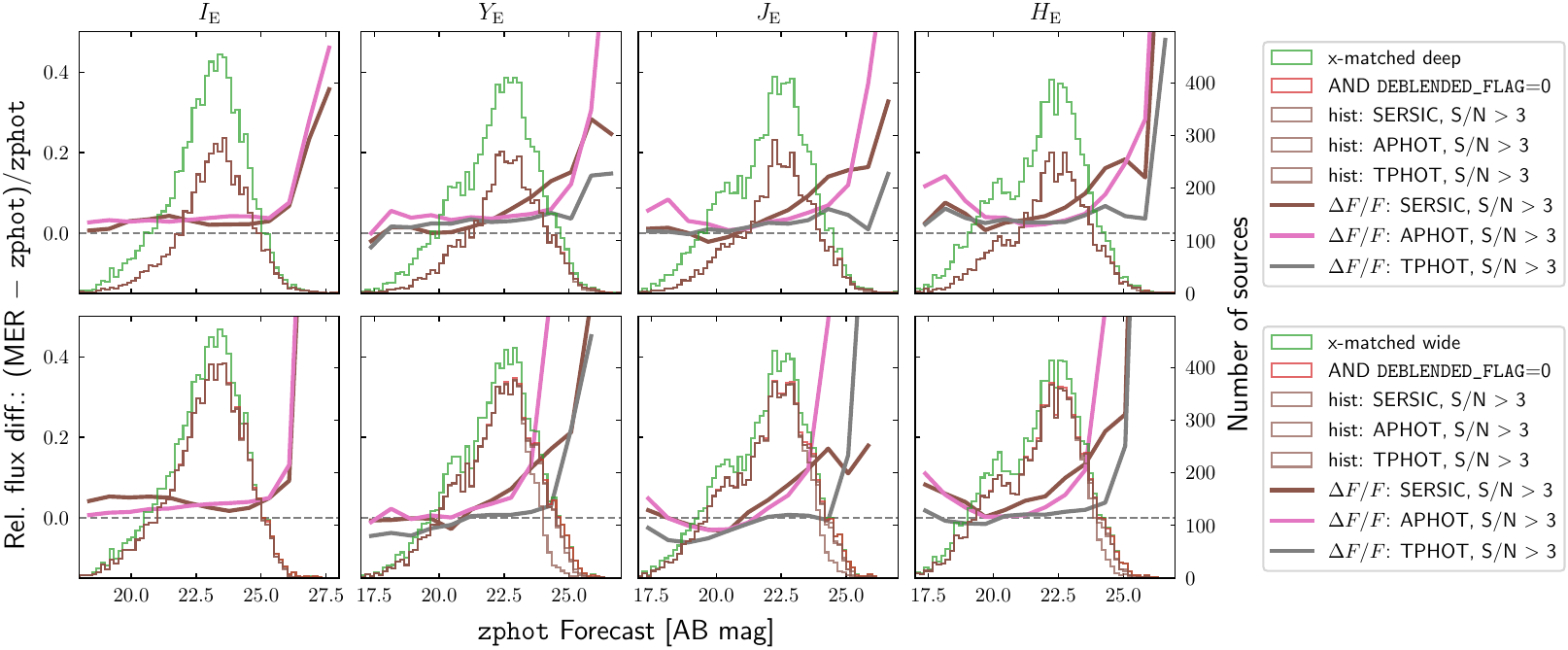}
    \includegraphics[width=\linewidth]{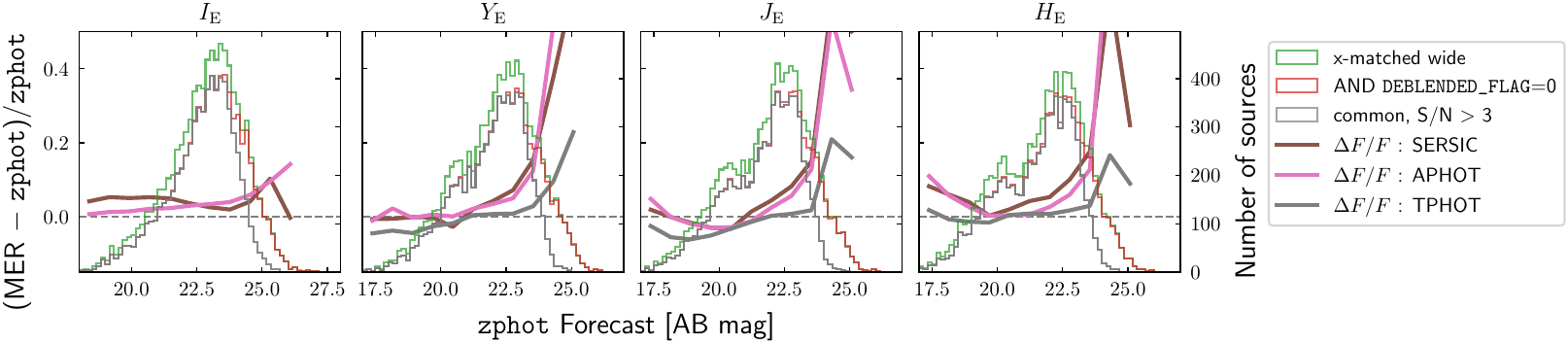}
    \caption{Number of spec-$z$ sources and distribution of the median relative flux difference $\Delta F/F$  as a function of the predicted \zphot magnitude in the four \Euclid filters. After the $\chi^2_{\rm red} < 4$ cut is applied, the sources are cross-matched either with the deep catalogue (top panel) or the wide one (middle panel), shown in green. Then we only consider the sources with {\tt DEBLENDED\_FLAG}=0 (shown in red) and with ${\rm S/N} > 3$: this last cut selects a slightly different number of sources for each method and band (as seen in the histograms). The distributions are shown for SERSIC (brown), APHOT (pink), and TPHOT (gray) measurements. A reference horizontal line at $\Delta F/F=0$ is plotted as a dashed gray line. The bottom panel shows the same plot as per the wide case, but we keep only the common sources that have ${\rm S/N} > 3$ in all the bands and for all the three photometric methods, meaning the same objects are being compared (`common' histogram shown in brown). The same plots for the photo-$z$ sample are shown in Fig.~\ref{fig:photoz_deep_wide_histograms_zphot_3}.}
    \label{fig:specz_deep_wide_histograms_zphot_3}
\end{figure*}

Figure~\ref{fig:specz_deep_wide_histograms_zphot_3} shows the population of sources in the spec-$z$ case, after the $\chi^2_{\rm red} < 4$ cut, which are cross-matched with the deep catalogue (top panel) and with the wide one (middle panel); we select those sources which are not deblended ({\rm DEBLENDED\_FLAG}=0), whose distribution is also shown. The distribution of those sources that are selected using a ${\rm S/N} > 3$ cut in the \Euclid bands, for the three photometric methods, are also shown.

In the deep case, almost all the cross-matched sources pass the S/N criteria and are selected (as mentioned in Sect.~\ref{subsect:res_specz_deep}). In the wide case, there is little difference in \IE both for SERSIC and APHOT; in the NIR bands only a few sources are lost for SERSIC and TPHOT, while APHOT is more affected by the ${\rm S/N} > 3$ cut, especially at magnitudes fainter than 23. The \HE band is the less affected, because the bulk of galaxies are brighter there, so more objects survive the selection. 

Since the ${\rm S/N} > 3$ criterion selects a different number of sources in different bands, the resulting comparisons are not fully consistent, because they are performed on slightly different subsamples of sources. To check the overall effect of this issue, in the bottom panel of Fig.~\ref{fig:specz_deep_wide_histograms_zphot_3} we show the distributions of all the sources in the spec-$z$ wide case that have ${\rm S/N} > 3$ in all the bands and methods. Only a few minor differences with respect to the previous case appear: SERSIC and APHOT show a small positive offset in \IE, which extends only to the brightest sources when compared to the previous case. In the NIR bands, SERSIC offsets are now larger at the faint end, closely following APHOT, while TPHOT is again the method with the smaller offsets and with the smallest trends down to magnitude 23. 

We performed the same test for the photo-$z$ case, both for the deep and the wide case, reaching similar conclusions (see Fig.~\ref{fig:photoz_deep_wide_histograms_zphot_3}).
After these tests we can conclude that the performances of \IE SERSIC and APHOT are comparable, TPHOT is still the method retaining the best accuracy in the NIR bands down to the faint end of the distribution, and the selection procedure described in Sect.~\ref{sect:analyses} does not substantially affect the results.

\subsection{Deblending accuracy}

In Sect.~\ref{subsect:debl_selection} we analysed how the performance of different photometric methods changes as a function of the {\tt DEBLENDED\_FLAG} parameter.
As shown in Fig.~\ref{fig:MER_visdet_debl_specz_deep_VIS_J}, there are clear trends across the various bands and as a function of the deblending parameter. Unsurprisingly, APHOT shows larger offsets for sources that have been deblended in detection, because of flux contamination within the aperture by neighbouring objects.\footnote{We remind that when estimating the flux of a source with {\tt APHOT}, the pixels assigned to different objects are excluded from the summation, but the segmentation map usually does not cover the full extension of the faint wings of galaxies; they can therefore contaminate the measurement.} Instead, both SERSIC and TPHOT show very small differences between deblended and not deblended sources, TPHOT showing an overall smaller offset at the faintest magnitudes compared to SERSIC.

We find similar results for the photo-$z$ case in Fig.~\ref{fig:MER_visdet_debl_photoz_deep}, where the numbers are close to the spec-$z$ ones. Since around 80\% of the sources in the photo-$z$ case are not affected by the deblending algorithm, the difference between the offsets calculated using all the sources ({\tt deblended\_flag}=None) and using only the non-deblended ones ({\tt deblended\_flag}=0) is much smaller in the photo-$z$ case than in the spec-$z$ case (which includes roughly $40\%$ of deblended sources).

Since the deblending parameters were not optimised for the deep processing in DR1 (see Fig.~\ref{fig:deep_vs_wide_crossmatch} in Sect.~\ref{subsect:debl_selection} and \citetalias{DR1-TP009} for DR1 discussion and DR2 plans), we tested a reprocessing of the deep DR1 tiles in COSMOS using a new deblending configuration, tuned within OU-MER for previous calibration purposes. While a more in-depth analysis is under way, we have checked that the source in Fig.~\ref{fig:deep_vs_wide_crossmatch}, which was blended in the deep data in DR1, is now successfully deblended.

\subsection{Eddington-like bias}\label{subsect:sel_bias_eddington}
\begin{figure}
    \centering
    \includegraphics[width=\linewidth]{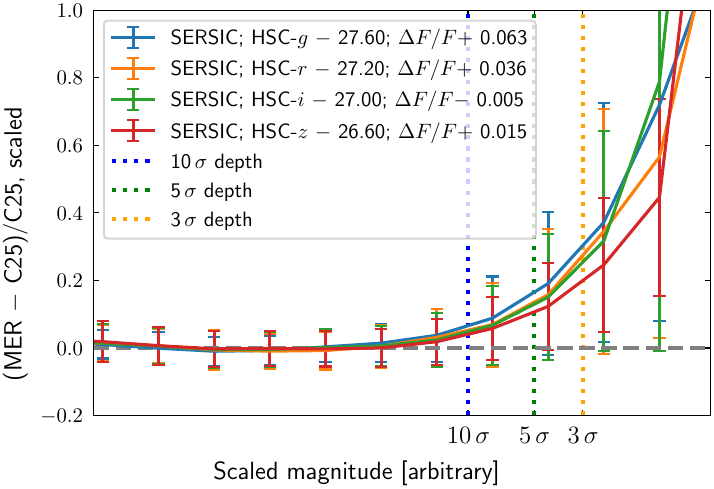}
    \caption{Relative flux differences between the MER model-fitting photometry and the C25 one in the four common HSC bands. The upturn at faint magnitudes has the same behaviour once the magnitudes are shifted to the $5\,\sigma$ limiting magnitude in each HSC band and corrected for any constant bias. The positions of the 10\,$\sigma$, 5\,$\sigma$, and 3\,$\sigma$ magnitude limits are shown as vertical lines.}
    \label{fig:mag_lim_HSC_5sigma}
\end{figure}

In all the analysed cases, when comparing the measured fluxes to the expected ones we found an evident upturn at the faint end of the distribution. The exact position of the feature changes from case to case (with TPHOT typically having it at the faintest magnitudes in all the scenarios). This upturn is less visible in the spec-$z$ deep case, because of the larger number of faint sources, but becomes clear in the spec-$z$ wide case, and even more so in the photo-$z$ case both for the deep and wide scenarios. 

We interpret this feature as an Eddington-like selection effect~\citep{Eddington}. Due to random noise fluctuations in the image, faint sources close to the detection limit can be detected, therefore ending up in the selected sample of sources, only if they happen to fall on a positive noise fluctuation.  This bias is not due to a real physical motivation, but just to a correct measurement of a randomly increased flux. Since the magnitude limit of the EWS is at brighter magnitude than the one of the EDS, the upturn in the wide case appears at brighter magnitude with respect to the deep case. For similar reasons, the upturn is also more visible in the photo-$z$ deep case with respect to the spec-$z$ scenario. 

The upturn is also visible in the HSC bands, presented in Sect.~\ref{sect:res_ext}, both for the spec-$z$ (Fig.~\ref{fig:offset_HSC_cosmos_mer}) and for the photo-$z$ case (Fig.~\ref{fig:offset_HSC_cosmos_mer_photoz}). To make sure that we are not misinterpreting this issue, in Fig.~\ref{fig:mag_lim_HSC_5sigma} we plot the relative flux differences between C25 and MER for model-fitting photometry, scaling all the magnitudes to the nominal $5\,\sigma$ magnitude limit for each band as reported in~\cite{COSMOS2025}, and correcting for any zero-point difference using the median between magnitudes 20 and 24: the upturns are perfectly consistent in all bands relatively to their detection limit.

\section{Summary and conclusions}\label{sect:future}

We presented an assessment of the photometric accuracy of DR1 OU-MER catalogues focusing on distant galaxies, the core business of the \Euclid mission.
We used the COSMOS2025 photometric catalogue and a SED-fitting technique to predict the fluxes expected in the \Euclid \IE and NISP bands, for a sample of bright sources with spectroscopic redshifts, and a larger sample of sources with photometric redshifts to extend the analysis to fainter sources. We compared the SED-fitting predicted fluxes with those in the MER catalogues, for both the deep and the wide cases, for a total of 8631 (\num{69729}) objects for the spec-$z$ (photo-$z$) MER deep catalogue, and \num{9081} (\num{54146}) for the spec-$z$ (photo-$z$) MER wide catalogue.

We found a good agreement between the predicted and measured values in all four \Euclid bands and for all three photometric methods that are used for the MER catalogues, namely aperture photometry with {\tt A-PHOT}, template photometry with {\tt T-PHOT}, and model-fitting photometry using analytical \sersic 2D profiles with \texttt{SE++}. MER measured fluxes are typically larger than the \zphot predictions by a few percent, and within about 10\% over the considered magnitude ranges, with the deep catalogue showing a larger positive offset with respect to the wide catalogue. While the photometry measurements from model-fitting and aperture photometry in \IE show similar offsets, template-fitting photometry yields overall better result in the NIR bands down to fainter magnitudes in all the considered scenarios; this is in line with the results from synthetic source injection reported in Euclid Collaboration: Kong et al. (in prep.).

As pointed out in Sect. \ref{sect:discussion}, the methodology adopted in the present work is subject to several unavoidable limitations, including (but not limited to) the accuracy of the C25 fluxes and redshifts adopted as references, the reliability of the parameters used to construct the synthetic galaxy library for SED-fitting, and the uncertainties associated with the photometric measurements in the MER catalogues. As a result, the intrinsic accuracy of our tests is likely of the same order of magnitude as the measured offsets. For example, comparing the MER deep catalogue both with the C25 fluxes and with the \zphot predictions using the common HSC bands we find that both the MER fluxes and the \zphot forecasts have negative average offsets; if we speculate that the bias introduced by \zphot is also affecting the \Euclid bands in the same direction, the positive offset of the MER photometry with respect to \zphot forecasts in the \Euclid bands might be substantially reduced. We also discuss the effect of an Eddington-like selection bias at the faint-end of the distributions, where the sources included in our selections are only the ones having their fluxes artificially increased by random local noise image fluctuations, causing a spurious upturn of the median bias.

Future studies focussed on the \Euclid deep fields might strengthen the conclusions obtained on COSMOS, ensuring that they are not affected by systematic effects caused by the sky position or by the Galactic latitude. Furthermore, improved calibrations and future updates to the OU-MER pipeline, which is expected to undergo further development in the coming years, will likely provide additional insights and hence smaller uncertainties.

\begin{acknowledgements}
The research activities in this paper were carried out with the contribution of the NextGenerationEU funds within the National Recovery and Resilience Plan (PNRR), Mission 4 -- Education and Research, Component 2 -- From Research to Business (M4C2), Investment Line 3.1 -- Strengthening and creation of Research Infrastructures, Project IR0000034 – “STILES -- Strengthening the Italian Leadership in ELT and SKA”, CUP C33C22000640006.
\AckEC
\AckDRone See \cite{DR1cite}.
Based on data from UNIONS, a scientific collaboration using three Hawaii-based telescopes: CFHT, Pan-STARRS, and Subaru (\url{www.skysurvey.cc}). This work uses results from the ESA mission {\it Gaia}, whose data are being processed by the \gaia Data Processing and Analysis Consortium \url{https://www.cosmos.esa.int/gaia}.
\end{acknowledgements}

\bibliographystyle{aa}
\bibliography{Euclid, dr1_cosmos}

\begin{appendix}

\section{Dependence on SED-fitting method}\label{app:lephare_sed_vs_zphot_sed}

\begin{figure*}
    \centering
    \includegraphics[width=0.9\linewidth]{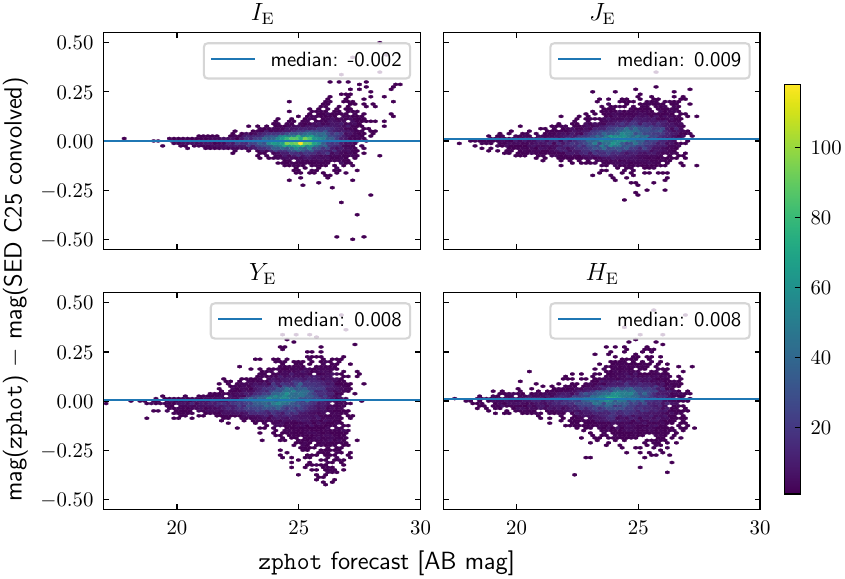}
    \caption{Density plot of the difference between the magnitude predicted by \zphot and the one estimated after convolving the COSMOS SEDs with the \Euclid filters, as a function of the \zphot magnitude. This check has been performed only for the 8546 sources with well-constrained PDF($z$). The median of the difference is 5 to 10 mmags, with a large deviation from the median appearing at faint magnitudes.}
    \label{fig:sed_convolve_cosmos_zphot}
\end{figure*}

The COSMOS2025 {\tt LePhare} SEDs\footnote{\url{https://cosmos2025.iap.fr/catalog\_download.html\#lephare-seds-and-pdfz}} are provided as an {\tt h5} file including best-fit galaxy SEDs, source IDs, and photo-$z$ for all the sources.  We compared the SEDs from C25 against the ones we created in our work with \zphot, for internal consistency. While the redshift we used in the photo-$z$ catalogue is the \texttt{zfinal} from `LePhare', the redshift for the SED is the \texttt{zchi2} from the catalogue.
These two values can be significantly different, because the PDF($z$) might not be well constrained; therefore, we re-computed the fluxes in the \Euclid filters from the SED only for the sources where \texttt{abs}(\texttt{zfinal}$-$\texttt{zchi2})/\texttt{zfinal} < 0.001. The selection of sources with ${\rm S/N}>10$ in the four \JWST bands, with reduced $\chi^2<2$ from \zphot, and cross-matched with the MER deep catalogue, leaves 8546 sources.

As a safety check, we computed the \Euclid fluxes of these well-behaved objects using the C25 SEDs. We then compared the fluxes with the forecasts from \zphot, confirming that the two sets of SEDs give consistent results, especially at the bright end of the considered magnitude range (Fig.~\ref{fig:sed_convolve_cosmos_zphot}). Given this reassuring result, we decided to use the SEDs and magnitude forecast from \zphot throughout this work.

\section{SED-fitting library}\label{app:library}
To perform the SED-fitting runs with \zphot we create a grid of synthetic models spanning a redshift interval from 0 to 10, on a 0.025 step grid. We use the \citet{bc03} stellar population synthesis library, and we adopted delayed exponential star formation histories [${\rm SFR}\propto t^2\exp(-t/\tau)$] with $\tau$ parameters ranging from 0.1 to 7~Gyr. We chose a \citet{Chabrier2003} initial mass function, \citet{calzetti2000} extinction law with $E(B-V)$ from 0 to 1.1, and \citet{fan06} IGM absorption. We include models with metallicity from 2\% to 250\% $Z_{\odot}$, and age from 0.01 to 13.8~Gyr. A similar library was used for the estimates of the physical parameters in the ASTRODEEP-GS43 catalogue~\citep{Merlin2021}.

\section{Photometric redshift sample}\label{app:res_zphot_MER_photoz_widedeep}

While the analysis in the main text was performed on the spec-$z$ sample (except where explicitly mentioned), here we summarise the results from the analysis of the photo-$z$ sample, both with the deep and wide catalogues. Compared to the spec-$z$ case, the sample is substantially larger and extends to fainter magnitudes. 

\begin{figure*}
    \centering
    \includegraphics[width=\linewidth]{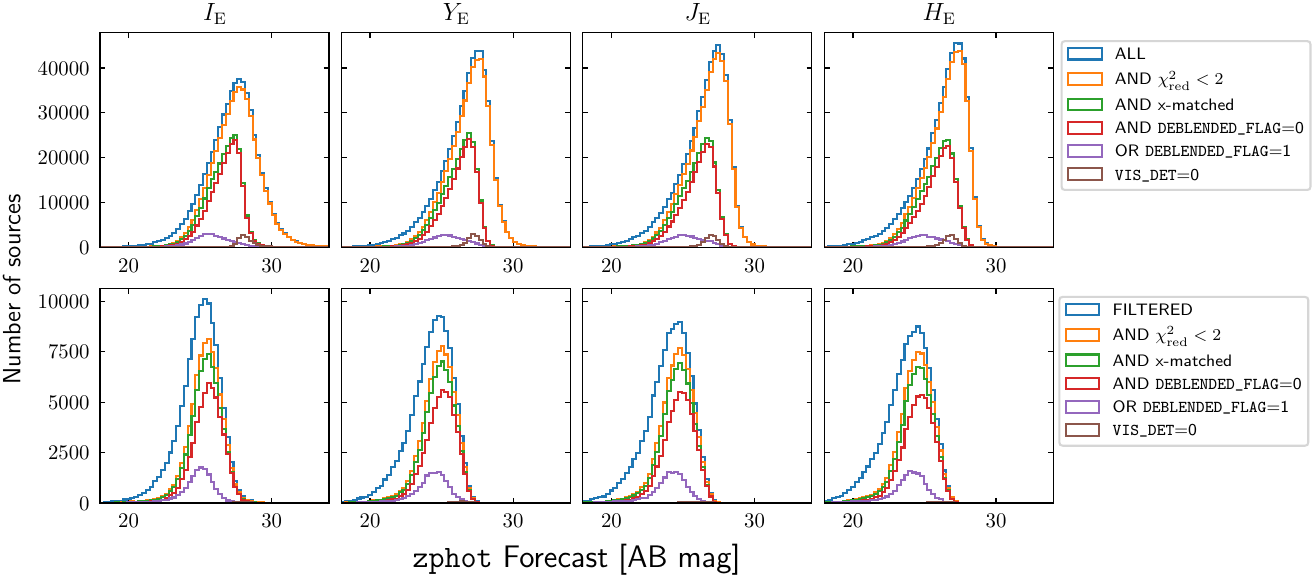}
    \caption{Distribution of sources in each \Euclid filter as a function of the \zphot forecast magnitude for the photo-$z$ case. Shown are the histograms
for the full initial sample of fitted sources (cyan), after the $\chi_{\rm red}^2<2$ cut (orange), and after the cross-match with the deep catalogue (green). The distribution of cross-matched sources is either not deblended (red) or deblended (purple). The NIR-detected sources ({\tt VIS\_DET}=0) are also reported (brown). \textit{Top}: No S/N selection for the \JWST bands is applied. \textit{Bottom}: Only the C25 sources with ${\rm S/N} > 10$ in the \JWST bands are considered. The specz-$z$ case is shown in Fig.~\ref{fig:specz_deep_histograms_zphot}.}. 
    \label{fig:photoz_deep_histograms_zphot}
\end{figure*}

\begin{figure*}
    \includegraphics[width=\linewidth]{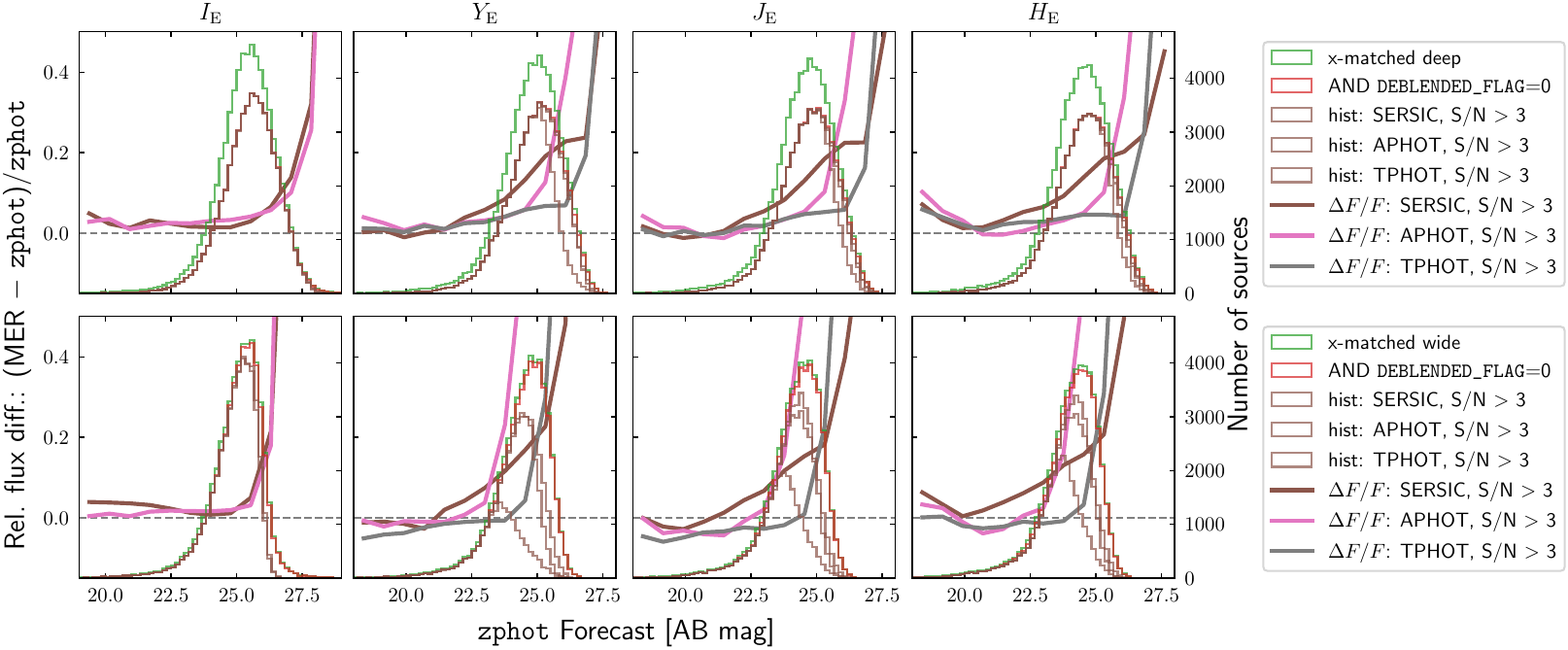}
    \includegraphics[width=\linewidth]{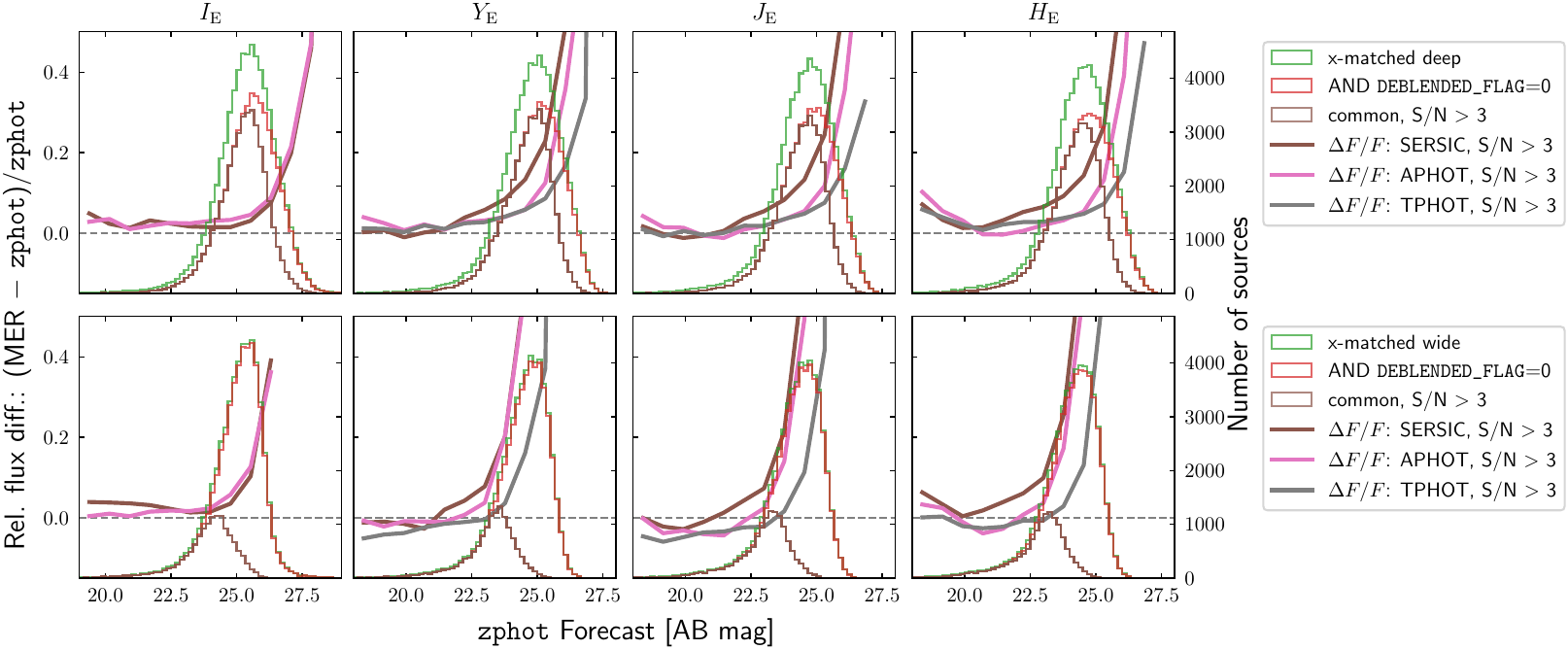}
    \caption{Number of sources and median relative flux difference $\Delta F/F$ for the photo-$z$ sample, as a function of the \zphot forecast magnitudes in the four \Euclid filters. After the $\chi^2_{\rm red} < 2$ cut is applied, the sources are cross-matched either with the deep catalogue (top panel) or the wide one (second panel), represented in green. 
    Then we only consider the sources with {\tt DEBLENDED\_FLAG}=0 (shown in red) and with ${\rm S/N} > 3$: this last cut selects a slightly different number of sources for each method and band (as seen in the histograms). The distributions are shown for SERSIC (brown), APHOT (pink), and TPHOT (gray). A reference horizontal line at $\Delta F/F=0$ is shown as a dashed gray line. The same plots are shown for the deep case (third panel) and for the wide case (bottom panel), but only for sources that have ${\rm S/N} > 3$ in all the bands and for all the three photometric methods (`common' histogram shown in brown). The same plots for the spec-$z$ sample are shown in Fig.~\ref{fig:specz_deep_wide_histograms_zphot_3}.}
    \label{fig:photoz_deep_wide_histograms_zphot_3}
\end{figure*}

\begin{figure*}
    \centering
    \includegraphics[width=\linewidth]{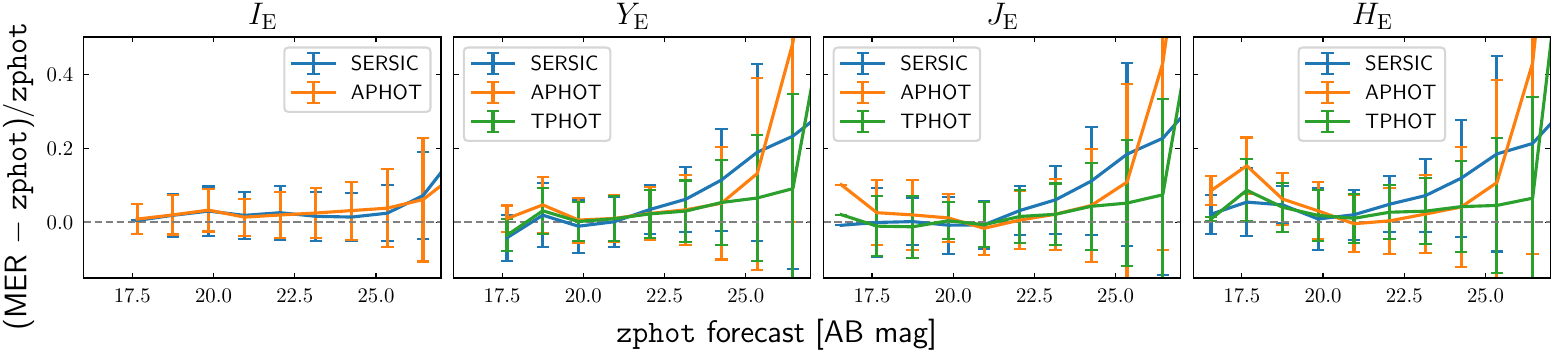}
    \includegraphics[width=\linewidth]{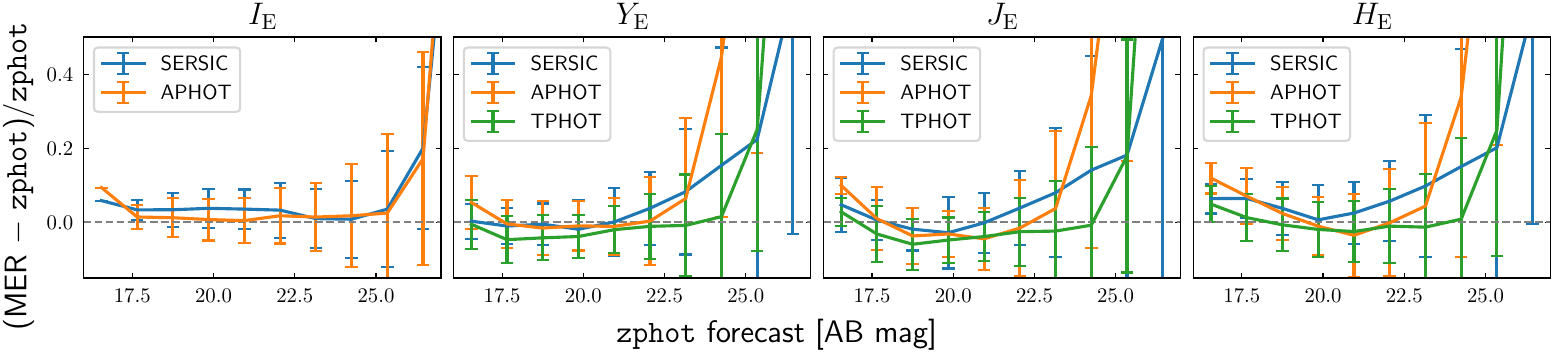}
    \caption{Relative flux differences between MER and the forecast from \zphot for the deep photo-$z$ case (upper panel) and for the wide photo-$z$ case (lower panel), for model-fitting (`SERSIC'), aperture photometry (`APHOT'), and template-fitting (`TPHOT'). Only the sources with \texttt{VIS\_DET}=1, \texttt{DEBLENDED\_FLAG}=0, and ${\rm S/N}>3$ are shown in these plots. The medians in these panels are also reported in the top and middle panel of Fig.~\ref{fig:photoz_deep_wide_histograms_zphot_3}.}
    \label{fig:offset_euclid_wide_0_1}
\end{figure*}

\subsection{Deep observations}

Figure~\ref{fig:photoz_deep_histograms_zphot} shows the distribution of sources as a function of the forecast \zphot magnitude, with the various selection cuts taken into account (analogous to Fig.~\ref{fig:specz_deep_histograms_zphot} for the spec-$z$ case), and with both the full data set (tagged as `ALL') and the one filtered according to ${\rm S/N} > 10$ in the \JWST bands (`FILTERED'). While the $\chi_{\rm red}^2$ selection cut is mostly affecting the brighter sources, the S/N selection on the \JWST bands affects the fainter ones. The \num{12544} NIR-detected objects only appear at the faint end in the upper panel (`ALL'), while in the `FILTERED' data set only 425 sources are NIR-detected (they are barely visible in the figure, because having lower S/N in the \JWST bands, most of them are filtered out when applying this cut).

Focusing on the objects detected in \IE and with {\tt DEBLENDED\_FLAG}=0 in the `FILTERED' sample, we checked whether the ${\rm S/N} > 3$ cut in the MER catalogue for the different bands and photometric methods introduces any further bias. In Fig.~\ref{fig:photoz_deep_wide_histograms_zphot_3} we show the distribution of the considered sources, cross-matched with the deep catalogue (upper panel) or with the wide one (lower panel), as a function of the magnitude predicted by \zphot, for all the methods and bands. The deep case has more objects than the wide case, and includes a larger fraction of isolated galaxies {\tt DEBLENDED\_FLAG}=0. The deep photo-$z$ case turns out to behave similarly to the spec-$z$ wide case, where APHOT is loosing some sources at magnitudes fainter than some threshold (23 in the spec-$z$ case, 25 here); the effect is more visible in \YE than in \HE. In the photo-$z$ wide case, \IE shows almost no difference between APHOT and SERSIC, while in the NISP bands APHOT is loosing sources at magnitudes fainter than 23, the effect being more visible in \YE than in \HE. Finally, TPHOT and SERSIC behave in a similar way in the NISP bands.

The relative flux differences with the \zphot forecasts for the photo-$z$ deep catalogue is shown in the upper panel of Fig.~\ref{fig:offset_euclid_wide_0_1}. The positive relative offset between MER and the \zphot forecast found in the spec-$z$ case is also present here, meaning that the MER catalogue is on average brighter than C25. The median offsets show similar trends to the spec-$z$ case, and again they become larger at fainter magnitudes, where the Eddington-like selection bias becomes important. TPHOT consistently stays below a 10\% offset at magnitudes brighter than 27, with a basically flat trend in the NISP bands. The \IE flux from model-fitting behaves comparably to the APHOT measurement, whereas it reaches a 20\% difference around magnitude 26 in the NISP bands; however, it is important to stress that the number of sources at the faint end is very small. The medians of these plots are also reported in the top and middle panel of Fig.~\ref{fig:photoz_deep_wide_histograms_zphot_3}.

In the bottom panel of Fig.~\ref{fig:photoz_deep_wide_histograms_zphot_3} we compare the MER fluxes with the \zphot predictions selecting those with ${\rm S/N} > 3$ in all the bands and in all the methods, similarly to what done in Sect.~\ref{sect:selection_effect}. While APHOT and TPHOT median curves do not change much, the latter still yielding the smallest bias, SERSIC has now larger offsets at bright magnitudes. In general, the differences are not dramatic, confirming that the selection effects have a limited impact on our results.

\begin{figure*}
    \centering
    \includegraphics[width=\linewidth]{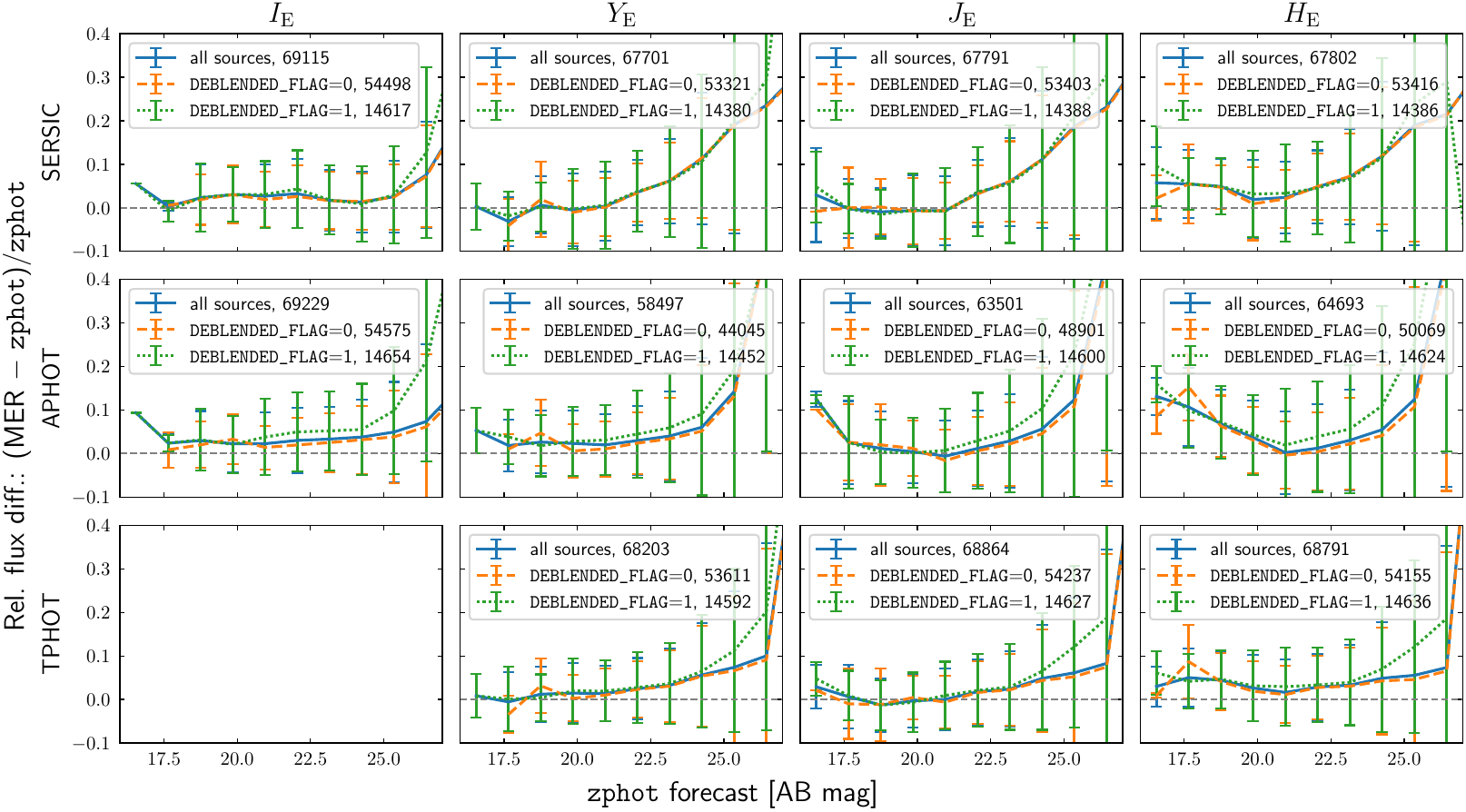}
    \caption{Relative flux differences between the MER fluxes and the \zphot fluxes as a function of the \zphot magnitude, for VIS-detected sources with \texttt{DEBLENDED\_FLAG} equal to 0 (orange dashed) or to 1 (dotted green), for model-fitting (`SERSIC'), aperture photometry (`APHOT'), and template-fitting (`TPHOT'). The solid blue line shows the trends for the full sample. In 'SERSIC' the deblending plays almost no role and in NISP the biases grow fast at magnitudes fainter than 20, while for APHOT and TPHOT non-deblended objects (i.e., well isolated in detection) have smaller offsets. The number of sources for each selection are reported in the legends. By construction there is no TPHOT measurement for \IE, and the related plot is intentionally left empty.}
    \label{fig:MER_visdet_debl_photoz_deep}
\end{figure*}

\begin{figure*}
    \centering
    \includegraphics[width=\linewidth]{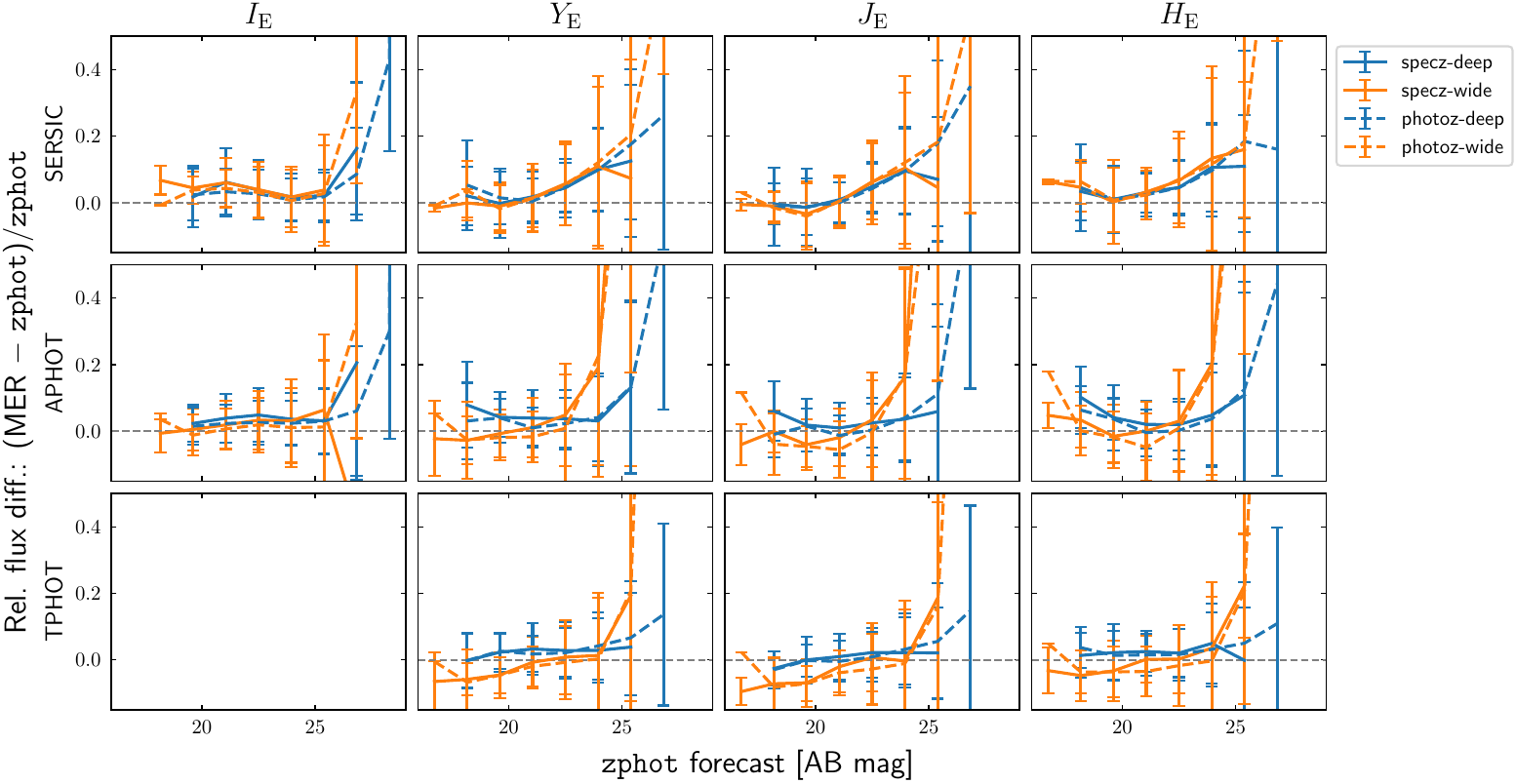}
    \caption{Comparison of relative flux differences between the MER fluxes and the predicted \zphot fluxes as a function of the \zphot magnitude for the photo-$z$ (dashed) and spec-$z$ (solid) samples. for VIS-detected sources with \texttt{DEBLENDED\_FLAG}=0 (i.e., well isolated in detection), for model-fitting (`SERSIC'), aperture photometry (`APHOT'), and template-fitting (`TPHOT'). Both the wide and the deep cases are shown here for all bands. By construction there is no TPHOT measurement for \IE, and the related plot is intentionally left empty.}
    \label{fig:offset_euclid_ALL_0_1_ALL_cumulative}
\end{figure*}

\begin{table}
\caption{Summary of the median offsets for all bands, methods, depths, and redshift types. We exclude the faint end of the distributions where Eddington-like selection effects become relevant, using a different limit for each band and catalogue depth: 25 (27) and 24 (25) for the \IE and NIR bands, in wide (deep) mode respectively. Figure~\ref{fig:offset_euclid_ALL_0_1_ALL_cumulative} shows the trends used to calculate these values.}
\centering
\begin{tblr}{
  row{1} = {c},
  row{2} = {c},
  row{3} = {c},
  row{4} = {c},
  cell{1}{4} = {c=2}{},
  cell{3}{1} = {r=6}{c}, 
  cell{9}{1} = {r=6}{c},
  cell{15}{1} = {r=6}{c},
  cell{21}{1} = {r=6}{c},
  cell{3}{2} = {r=2}{c}, 
  cell{5}{2} = {r=2}{c},
  cell{7}{2} = {r=2}{c},
  cell{9}{2} = {r=2}{c},
  cell{11}{2} = {r=2}{c},
  cell{13}{2} = {r=2}{c},
  cell{15}{2} = {r=2}{c},
  cell{17}{2} = {r=2}{c},
  cell{19}{2} = {r=2}{c},
  cell{21}{2} = {r=2}{c},
  cell{23}{2} = {r=2}{c},
  cell{25}{2} = {r=2}{c},
  cell{3-26}{3-5} = {c},
  hline{3,9,15,21} = {-}{},
  hline{5,7, 11,13, 17,19, 23,25} = {2-5}{},
}
\hline\hline
Band   & Method    & Redshift    & Median offset &      \\
       &           & (spec-$z$/photo-$z$)     & wide   & deep \\
\IE    & SERSIC    & spec-$z$  &  0.033   &   0.027  \\
       &           & photo-$z$ &  0.012   &   0.034  \\
       & APHOT     & spec-$z$  &  0.029   &   0.038  \\
       &           & photo-$z$ &  0.017   &   0.041  \\
       & TPHOT     & spec-$z$  &  $\dots$ &   $\dots$  \\
       &           & photo-$z$ &  $\dots$ &   $\dots$ \\
\YE    & SERSIC    & spec-$z$  &  0.036   &   0.049  \\
       &           & photo-$z$ &  0.071   &   0.129  \\
       & APHOT     & spec-$z$  &  0.031   &   0.042  \\
       &           & photo-$z$ &  0.064   &   0.070  \\
       & TPHOT     & spec-$z$  &  $-0.002$   &   0.030  \\
       &           & photo-$z$ &  $-0.010$   &   0.052  \\
\JE    & SERSIC    & spec-$z$  &  0.030   &   0.042  \\
       &           & photo-$z$ &  0.068   &   0.116  \\
       & APHOT     & spec-$z$  &  0.014   &   0.031  \\
       &           & photo-$z$ &  0.035   &   0.053  \\
       & TPHOT     & spec-$z$  &  $-0.015$   &   0.021  \\
       &           & photo-$z$ &  $-0.026$   &   0.039  \\
\HE    & SERSIC    & spec-$z$  &  0.055   &   0.054  \\
       &           & photo-$z$ &  0.085   &   0.116  \\
       & APHOT     & spec-$z$  &  0.029   &   0.037  \\
       &           & photo-$z$ &  0.044   &   0.048  \\
       & TPHOT     & spec-$z$  &  0.006   &   0.032  \\
       &           & photo-$z$ &  $-0.012$   &   0.039  \\
\hline
\end{tblr}
\label{tab:summary_table}
\end{table}

In Fig.~\ref{fig:MER_visdet_debl_photoz_deep} we show the relative differences between the \zphot predictions and the MER fluxes for different deblending parameter values, similarly to what shown in Fig.~\ref{fig:MER_visdet_debl_specz_deep_VIS_J} for the spec-$z$ case.

\subsection{Wide observations}

The relative flux differences between the \zphot predictions and the MER fluxes for the photo-$z$ sample in the wide catalogue (which is produced using just one visit of the COSMOS field among the several ones available) is shown in the lower panel of Fig.~\ref{fig:offset_euclid_wide_0_1}. 
Since the catalogue is shallower, the faint-end upturn due to Eddington-like selection bias begins two magnitudes brighter than in the deep counterpart, consistently with the nominal difference in depth between wide an deep observations.
SERSIC has trends comparable to those in the deep case in the NISP bands, and the \IE flux is measured consistenly with SERSIC and APHOT. Again, TPHOT has a flat median compatible with zero down to magnitude 24. 

In the bottom panel of Fig.~\ref{fig:photoz_deep_wide_histograms_zphot_3} we repeat the exercise of comparing the MER fluxes in the wide catalogue with the \zphot forecast selecting the objects with ${\rm S/N} > 3$ in all the bands and in all the methods, again similarly to what done in Sect. \ref{sect:selection_effect}. In \IE, SERSIC and APHOT behave similarly to the previous case, the curves moving slightly at brighter magnitude, which is the same behaviour of APHOT and TPHOT in the NIR bands. While  TPHOT remains the method with the smallest offsets in the NIR bands, SERSIC becomes the method with largest offsets in the NIR bands for the photo-$z$ wide case. The summary offsets for all method, bands, redshifts modes, and survey depths are reported in Fig.~\ref{fig:summary_plot}, Table~\ref{tab:summary_table}, and in Fig.~\ref{fig:offset_euclid_ALL_0_1_ALL_cumulative}.

\subsection{HSC for photo-{\it z}}

\begin{figure*}
    \centering
    \includegraphics[width=\linewidth]{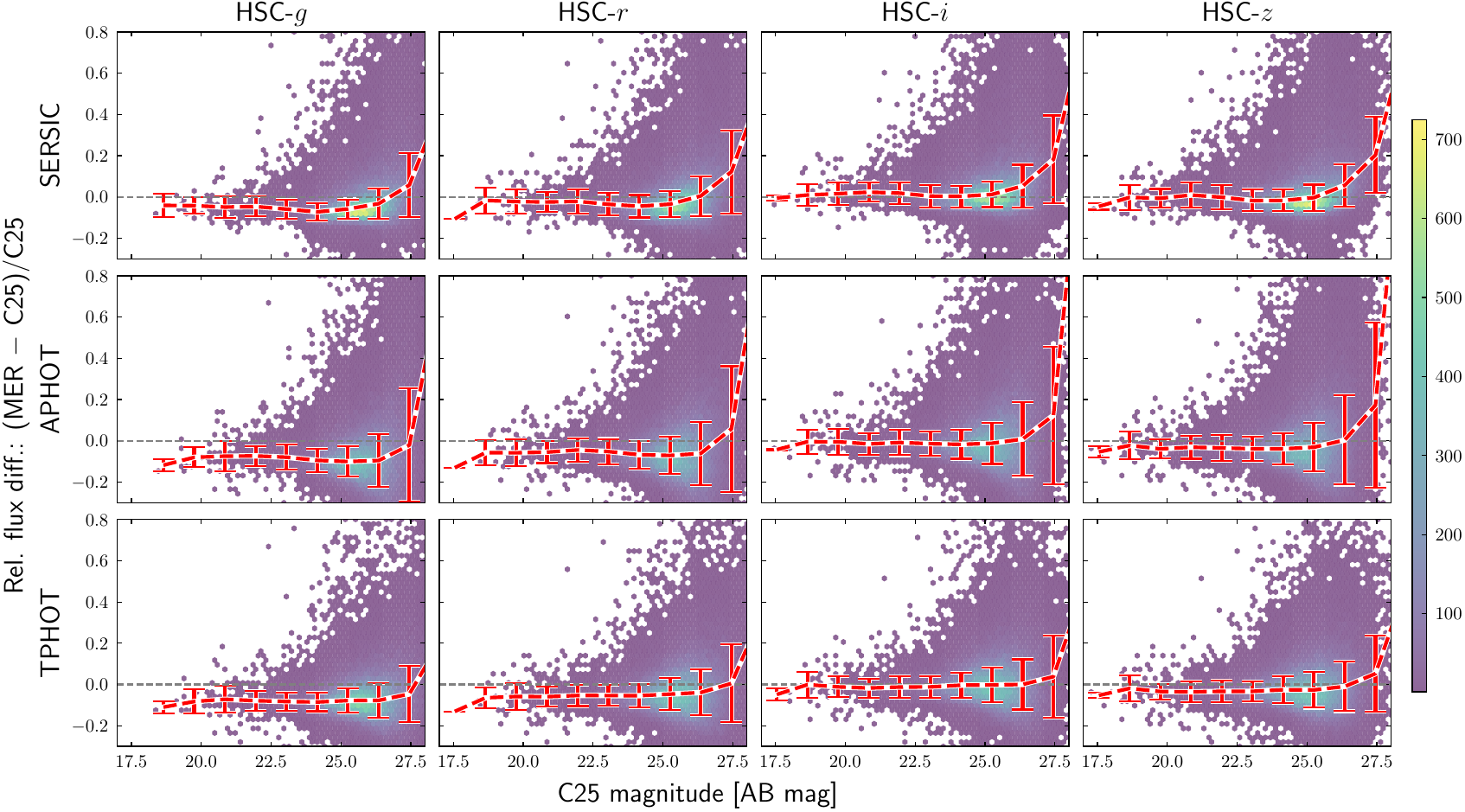}
    \includegraphics[width=\linewidth]{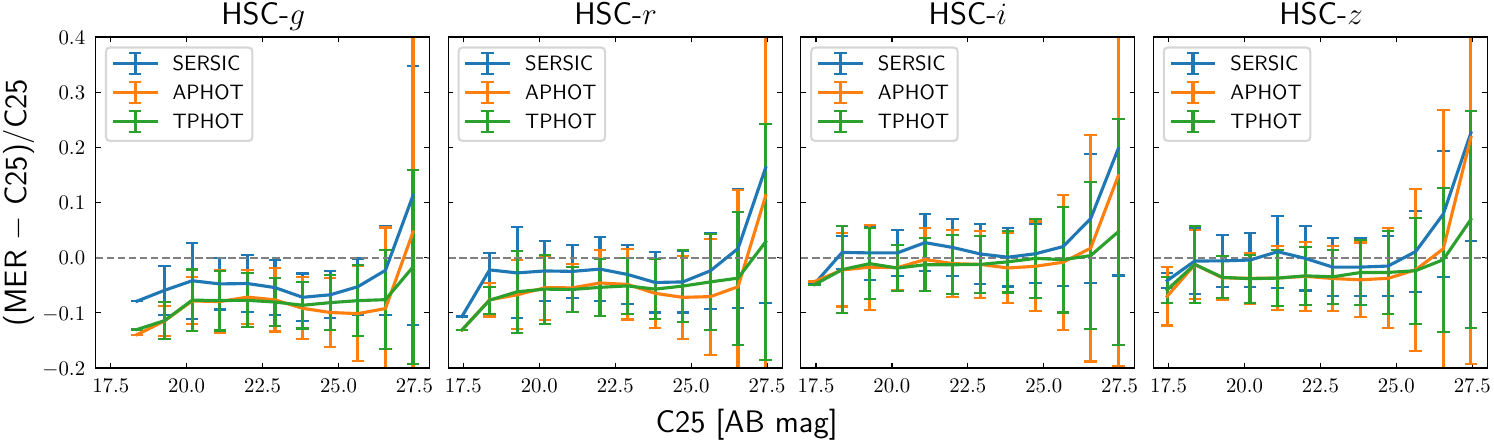}
    \includegraphics[width=\linewidth, trim=0 0 0 20, clip]{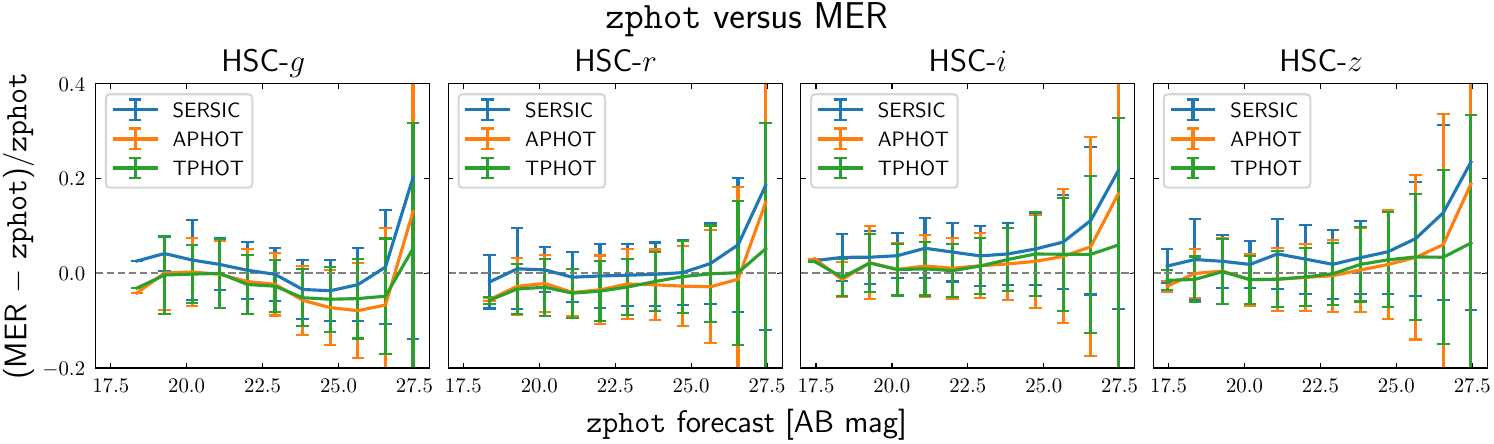}
    \caption{Comparison between the measured C25 fluxes, MER measured fluxes, and \zphot predicted fluxes for the photo-$z$ sample, for HSC~$g$, $r$, $i$, and $z$. \textit{Top}: scatter plots of relative flux differences between the MER and C25 fluxes. Only \texttt{VIS\_DET}=1 and \texttt{DEBLENDED\_FLAG}=0 sources are shown in these plots. \textit{Middle}: summary version of the upper plot, with median trends and  NMADs. \textit{Bottom}: relative flux difference between the \zphot forecast and MER. The largest difference is at the bright end of HSC~$g$, while the other bands have similar trends in the two panels. The analogous of this plot for the spec-$z$ case is Fig.~\ref{fig:offset_HSC_cosmos_mer}.}
    \label{fig:offset_HSC_cosmos_mer_photoz}
\end{figure*}

For completeness, in Fig.~\ref{fig:offset_HSC_cosmos_mer_photoz} we show the HSC bands comparison between C25, MER, and the \zphot forecasts, for the photo-$z$ sample. This is the analogous  of the spec-$z$ case shown in Fig.~\ref{fig:offset_HSC_cosmos_mer}.

\begin{figure*}
    \centering
    \includegraphics[width=\linewidth]{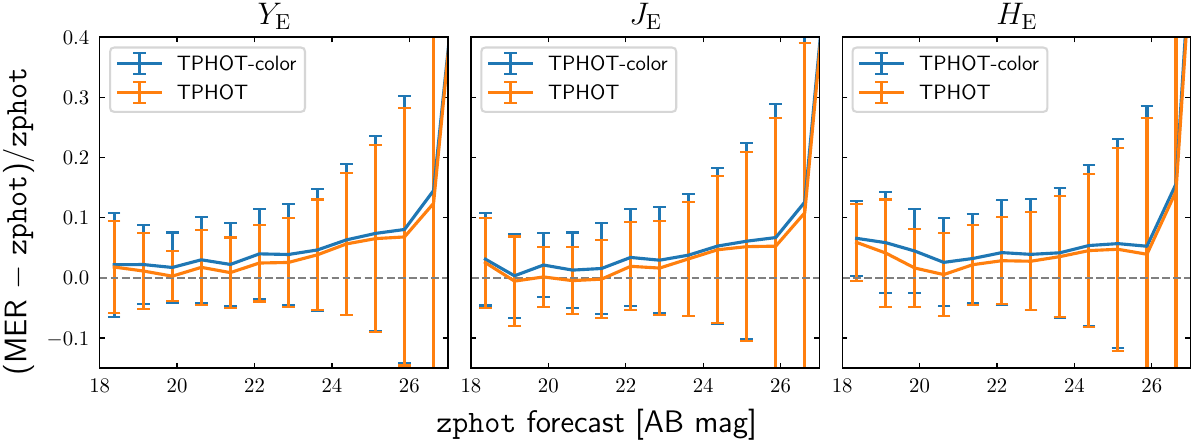}
    \caption{Relative flux differences between the MER fluxes and the \zphot ones, for template-fitting, for \texttt{FLUX\_\textasteriskcentered\_TEMPLFIT} directly from the MER catalogue (`TPHOT'), or using the ratio between \texttt{FLUX\_DETECTION\_TOTAL} and the \texttt{FLUX\_VIS\_TO\_\textasteriskcentered\_TEMPLFIT} colums as a corrective term to scale the detection (Kron) flux, as done for apertures (`TPHOT-color').}
    \label{fig:tphotcolor_deep_photoz}
\end{figure*}

\subsection{TPHOT versus TPHOTcolor}\label{app:tphotcolor}

As discussed in the Photometry Cookbook, there are two possible ways to use the template-fitting estimates: one can either take the \texttt{FLUX\_\textasteriskcentered\_TEMPLFIT} columns in the MER catalogue directly as a total flux, or use them to compute a colour term with \texttt{FLUX\_VIS\_TO\_\textasteriskcentered\_TEMPLFIT}, which can be then used to scale the Kron detection flux, as in the APHOT method:
\begin{equation}
\begin{split}
    {\rm Flux}_{\rm J, VIS-det}&= {\rm FLUX\_DETECTION\_TOTAL}\, \\
    &\times\frac{\rm FLUX\_J\_TEMPLFIT}{\rm FLUX\_VIS\_TO\_J\_TEMPLFIT}\,,
\end{split}
\end{equation}for the VIS-detected sources.
We compared these two approaches, and the relative difference between the MER estimates computed with them against the \zphot forecast is shown in Fig.~\ref{fig:tphotcolor_deep_photoz}. The difference between the two methods is very small (of the order of 1--2\%), in contrast with the results by Euclid Collaboration: Kong et al. (in prep.), who find some non-negligible differences which favour the `TPHOT-color' method.

\section{C25 -- \zphot difference for the 37 filters}

\label{appsect:offset_cosmos_bands}
\begin{figure*}
    \centering
    \includegraphics[width=1.0\linewidth]{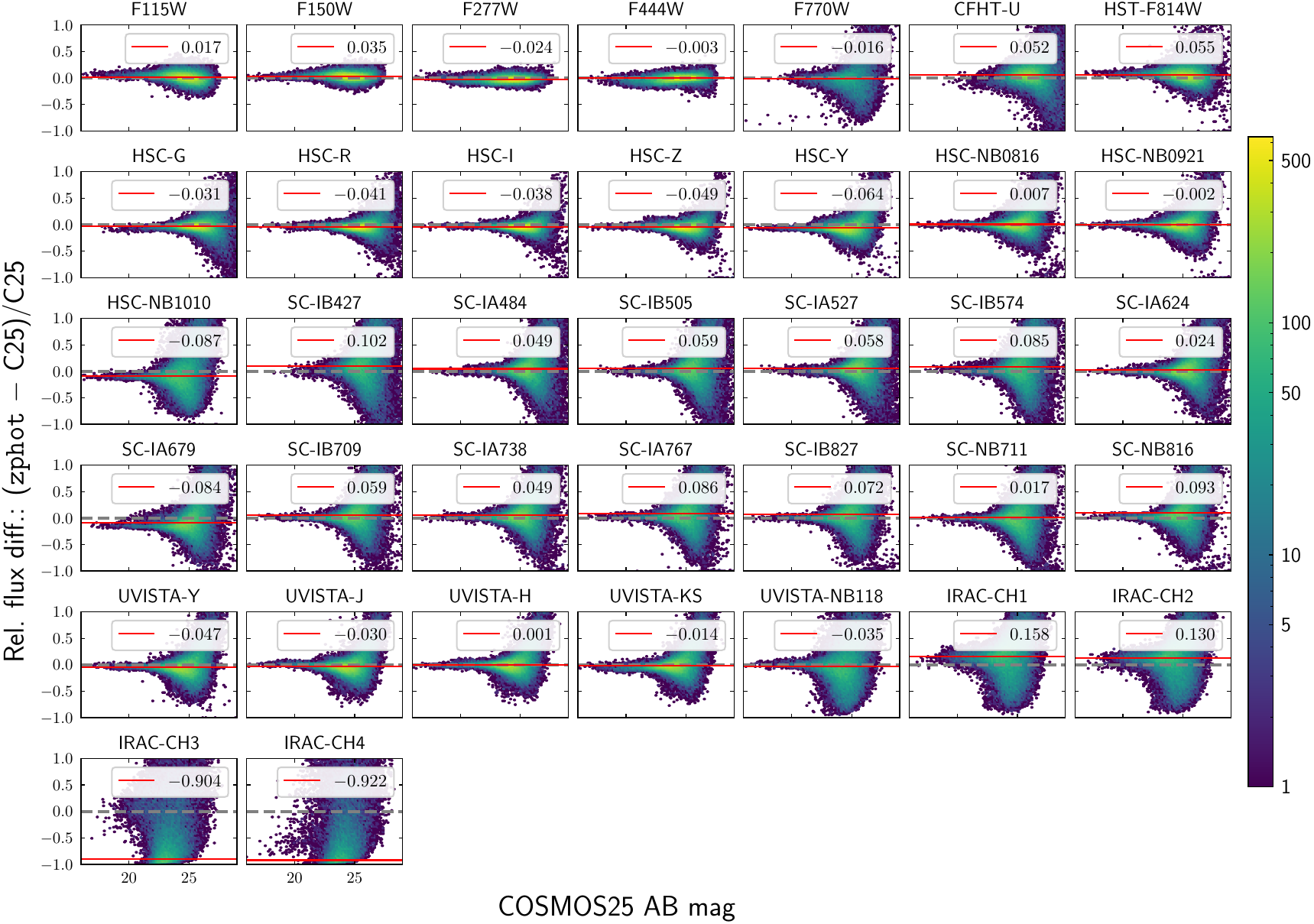}
    \includegraphics[width=1.0\linewidth]{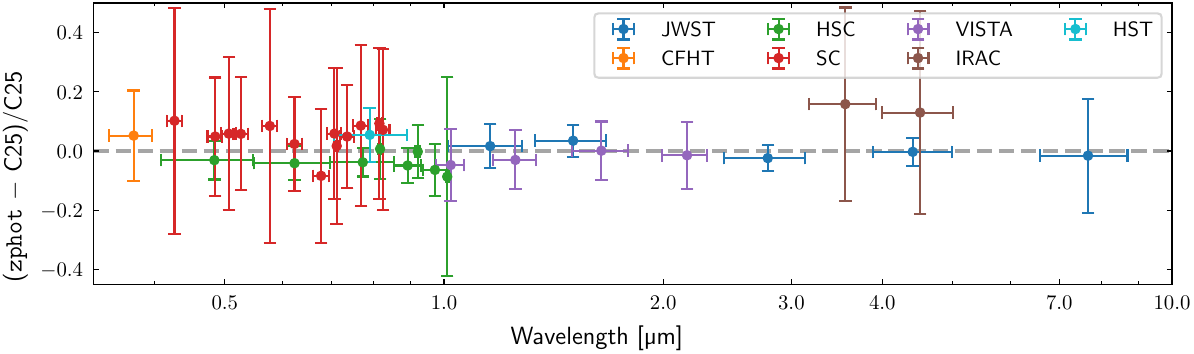}
    \caption{\textit{Top}: Density plot of the relative flux differences between the \zphot predicted fluxes and the input C25 fluxes plotted against the C25 AB magnitudes for all the sources with $\chi_{\rm red}^2 < 2$ and ${\rm S/N} > 10$ in \JWST F115W, F150W, F277W, and F444W (photo-$z$ sample). The red lines show the medians of the relative flux difference. \textit{Bottom}: Median and normalised median absolute deviations (NMAD) for all the 34 bands used in \zphot for the fit as a function of the central filter wavelength. The error bars along the $x$-axis represent the wavelength window of each filter. Different colours refer to different facilities, highlighting instrument-dependent systematics: JWST (blue), CFHT (orange), HSC (green), SuprimeCam (red), VISTA (purple), IRAC (brown), and HST (cyan).}
    \label{fig:offset_cosmos_bands}
\end{figure*}

\begin{figure}
    \centering
    \includegraphics[width=\linewidth 
    ]{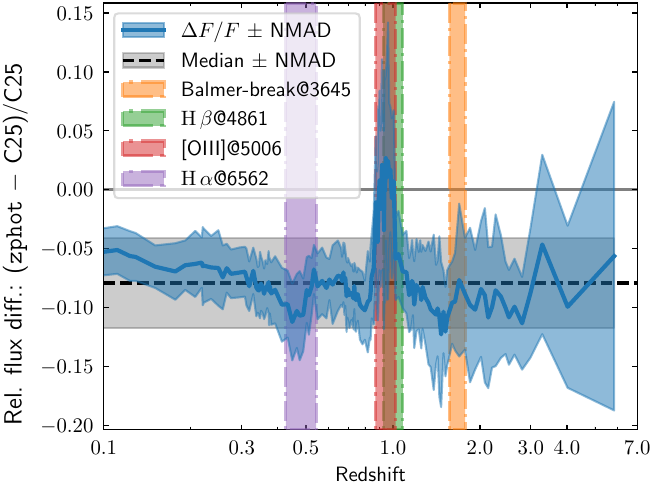}
    \caption{Relative flux difference between the \zphot fitted flux and the C25 flux for the HSC~$y$ filter, plotted against the spectroscopic redshift of sources with $\chi^2_{\rm red}<4$. The redshift range is divided in 150 bins with roughly the same number of sources per bin. The blue band is 2 NMADs wide around the median. ${\rm NMAD}_k=1.4626\times{\rm median}\lbrack|{\Delta F/F}_k - {\rm median}({\Delta F/F}_k)|\rbrack$, where ${\Delta F/F}_k$ are the relative flux differences in the $k$th redshift bin. The grey horizontal region represents the median and its NMAD across the whole redshift range. The vertical coloured areas correspond to the redshift ranges within which a certain spectral feature is observed with this filter. As an example, the [\ion{O}{iii}]$\lambda$5007 emission line for sources with redshift between 0.869 and 1.021 (red rectangle) falls within the HSC~$y$ band wavelength range 936.0--1012.0 nm. The modelling of this emission line in the SED-fitting procedure shows a large scatter for sources at redshift around 1.0, yielding an overestimation of up to 15\% above the median value.}
    \label{fig:hsc_y_redshift_offset}
\end{figure}

\begin{figure*}
    \centering
    \includegraphics[width=\linewidth]{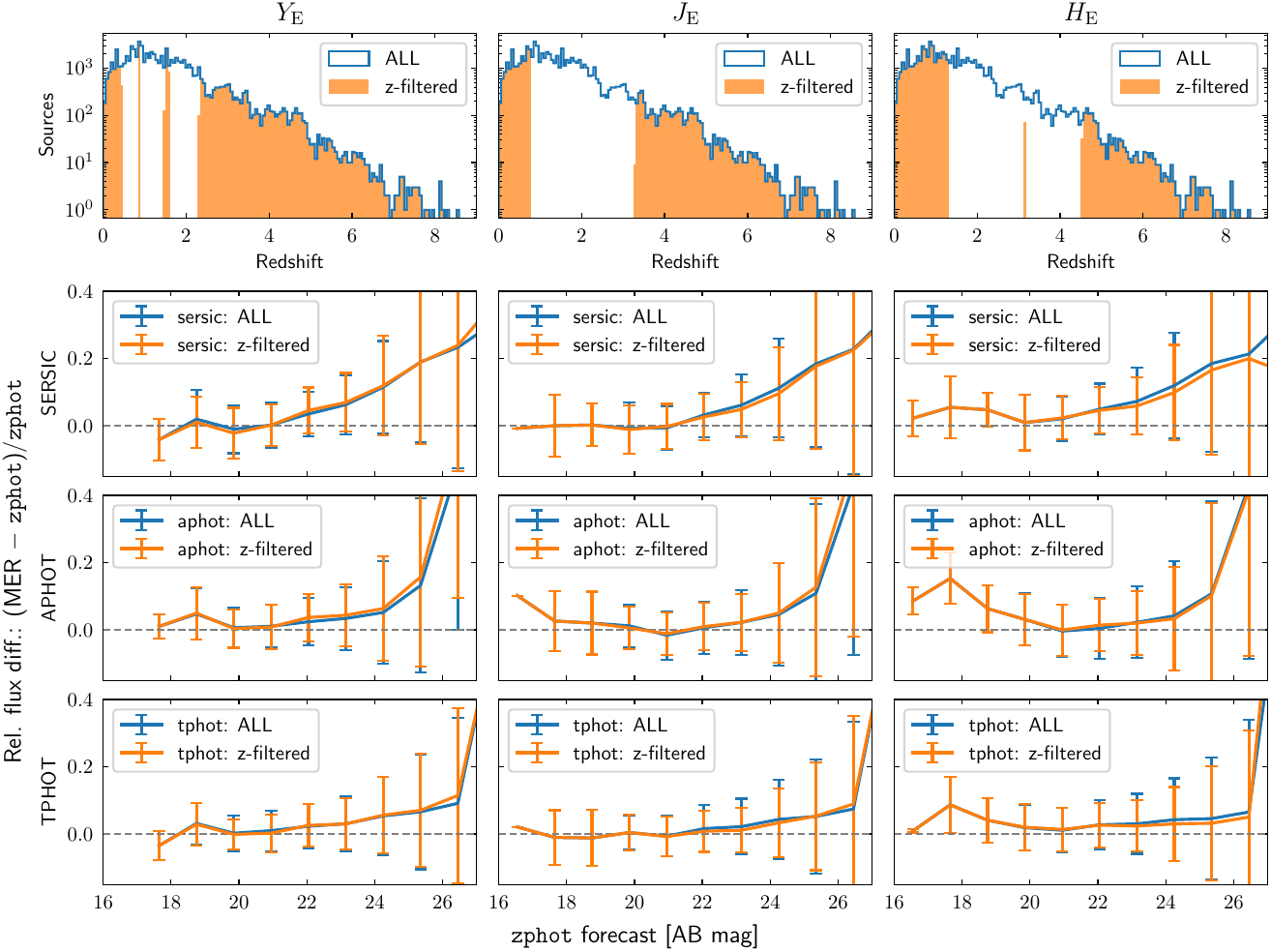}
    \caption{Effect on the photometry assessment of the sub-optimal modelling of strong emission lines in the SED-fitting process. `ALL' refers to sources from the photo-$z$ sample with $\chi^2_{\rm red} < 2$, {\tt DEBLENDED\_FLAG}=0, {\rm VIS\_DET}=1 and with ${\rm S/N} > 3$ in MER; `z-filtered' refers to sources  which are not affected by the presence of strong emission lines. {\it Top row}: Redshift distribution for the two selections. {\it Second to fourth rows}: Relative flux differences between the MER flux in the three NIR bands, for model-fitting (`SERSIC'), aperture photometry (`APHOT'), and template-fitting (`TPHOT'). }
    \label{fig:offset_z_filter_lines_photoz_0_1_deep}
\end{figure*}

We compared the C25 and \zphot predicted fluxes across all C25 bands to assess their consistency and identify any potential biases, especially for sources with high S/N in C25. The top panel of Fig.~\ref{fig:offset_cosmos_bands} shows the relative flux differences between the C25 and \zphot fluxes as a function of the C25 AB magnitudes for the sources from the photo-$z$ sample having ${\rm S/N} > 10$ in the four \JWST NIRCam bands. The median of the relative differences is within 5 to 10\%, with no particular trends within the magnitude range for most of the bands: IRAC ch1 and ch2  show the largest offsets (the other two IRAC bands are excluded from the fit, and therefore we do not expect their relative flux difference to be well constrained). In the bottom panel of Fig.~\ref{fig:offset_cosmos_bands} the median and NMAD for each filter are plotted as a function of its central wavelength. While we find no evident wavelength-dependent systematic, instrument-dependent biases up to a few percents seems to be present, possibly due to different calibration pipelines and reductions; however, all offsets are compatible with zero within their uncertainties.


To understand the effect of non-optimal modelling of the spectral features in SED-fitting, we checked the ones that are more likely to cause large differences between the input flux from C25 and the \zphot modelled one, focusing on the C25 bands that overlap with the \Euclid NIR bands. The four selected features are the Balmer break at $3645\,\si{\angstrom}$, \hb at $4861\,\si{\angstrom}$, \ionF{O}{iii}\,at $5006\,\si{\angstrom}$, and \ha at $6562\,\si{\angstrom}$; obviously, they correspond to different redshift ranges depending on the considered band.  
We checked whether the exclusion of redshift ranges corresponding to the four selected spectral features in the considered band would change the overall comparison. As an example, Fig.~\ref{fig:hsc_y_redshift_offset} shows the relative flux difference between the C25 values and \zphot forecasts for the spec-$z$ case in the HSC~$y$ band, plotted against the source redshifts, together with the redshift intervals at which the four spectral features must be redshifted in order to fall within HSC~$y$. For each \Euclid NIR band we discarded the sources falling in the relevant redshift ranges before computing the global relative flux differences shown in Fig.~\ref{fig:offset_z_filter_lines_photoz_0_1_deep} for the photo-$z$ MER deep case. Since the difference between the full and the reduced sample is well within the measurement uncertainties, we decided to keep the full sample.


\begin{figure*}
    \centering
    \includegraphics[width=1.0\linewidth]{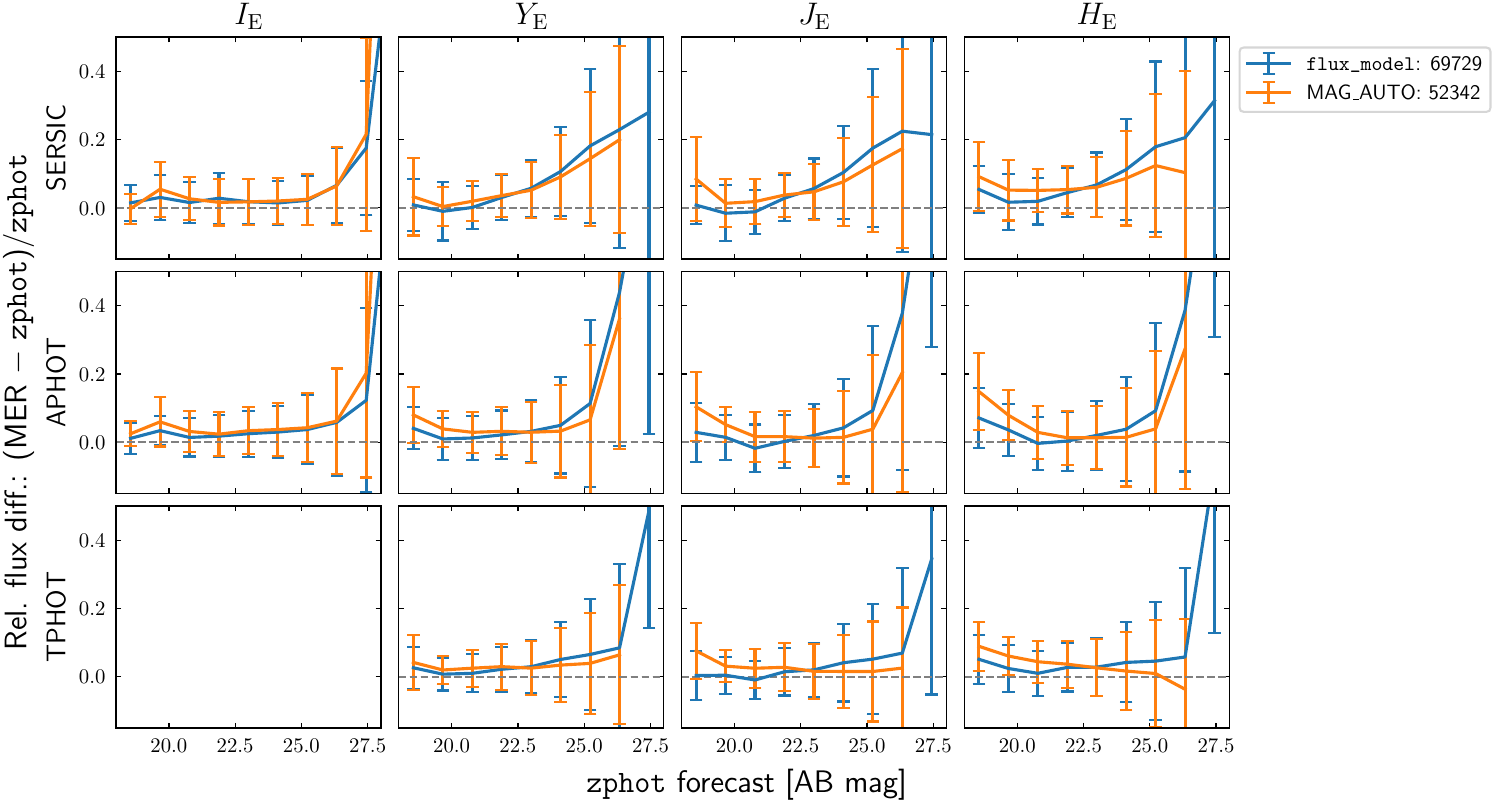}
    \includegraphics[width=1.0\linewidth]{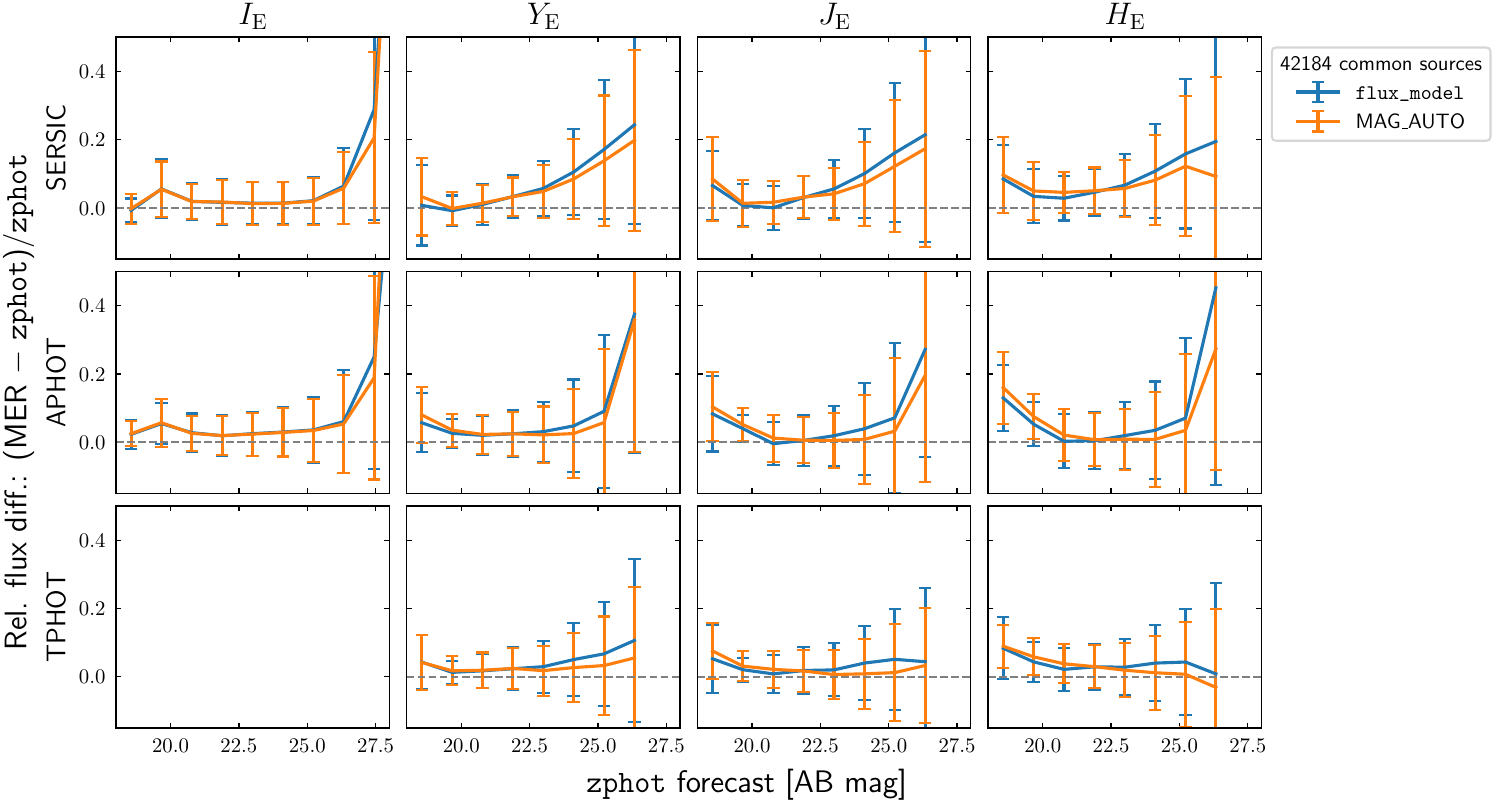}
    \caption{Comparison of the relative flux differences between MER and the \zphot predicted fluxes for two runs using (i) the \sersic for the 37 C25 bands (`{\tt flux\_model}') and (ii) the `MAG\_AUTO' magnitudes from the six available C25 bands (five \JWST and one HST, all the others being `{\tt flux\_model}' fluxes), for the photo-$z$ sources with {\tt DEBLENDED\_FLAG}=0, detected in \IE, and with ${\rm S/N}>3$ in MER. The three top panels show the relative flux differences for the full population of sources (\num{69729} for `{\tt flux\_model}' and \num{52342} for `MAG\_AUTO'). The three bottom panels are obtained using the sources in common between the two data sets, resulting in a total of 42184 sources. By construction there is no TPHOT measurement for \IE, and the related plot is intentionally left empty.}
    \label{fig:zphot_magauto_vs_sersic2d_photoz_0_1_nosrcmatch}
\end{figure*}

We checked whether using different input fluxes rather than the \sersic fit from C25 would change the results. In Fig.~\ref{fig:zphot_magauto_vs_sersic2d_photoz_0_1_nosrcmatch} we show the relative difference between the \zphot and MER fluxes for the photo-$z$ case, using either the \sersic fluxes for all the C25 bands (`{\tt flux\_model}' run) or the {\tt MAG\_AUTO} magnitudes available for the five \JWST and the additional HST bands, complemented again with \sersic fluxes for the remaining C25 bands (`MAG\_AUTO' run). In both runs we excluded the same bands from the fitting procedure. The final number of sources is \num{69729} for the `{\tt flux\_model}' run, and it is reduced to \num{52342} for the `MAG\_AUTO' run. In \IE the \sersic fluxes yield smaller average offsets at the very faint end and at the bright end, but very little differences are present between magnitudes 20 and 27. In the NIR bands, although the offsets are slightly reduced at faint magnitudes when switching from \sersic fluxes to {\tt MAG\_AUTO}, the two measurements always remain consistent within the uncertainties. 

\end{appendix}

\end{document}